\documentclass[preprint]{revtex4-2}
\nonstopmode
\usepackage{lmodern}
\usepackage{amsmath}
\usepackage{amsbsy}
\usepackage{pstricks,graphicx}
\RequirePackage{mathbbol}
\RequirePackage{amssymb}
\RequirePackage{mathtools}
\newcommand{\bra}[1]{\ensuremath{\langle #1|}}
\newcommand{\ket}[1]{\ensuremath{|#1\rangle}}
\newcommand{\braket}[2]{\ensuremath{\langle #1|#2\rangle}} 
\newcommand{\dirint}[3]{\ensuremath{\langle #1|#2|#3\rangle}}

\newcommand{\bs}{\boldsymbol}

\newcommand{\ul}{\underline}
\newcommand{\lo}{\overline}
\newcommand{\half}{{\textstyle\frac{1}{2}}}
\newcommand{\sixth}{{\textstyle\frac{1}{6}}}

\newcommand{\dul}[1]{\underline{\underline{#1}}}
\newcommand{\vtilde}[1]{\tilde{\raisebox{0pt}[0.85\height]{$\tilde{#1}$}}} 
\def\vcenterpdf#1{\mathrlap{\parbox{0pt}{\includegraphics[]{#1}}}\hphantom{\includegraphics[]{#1}}}
\def\vcpdf#1{\mathrlap{\parbox{0pt}{\includegraphics[width=11.0cm]{#1}}}\hphantom{\includegraphics[width=11.0cm]{#1}}}

\begin{document}

\title{The electron propagator for non-interacting particles: An alternative approach to quasi-degenerate perturbation theory (QDPT)}
\author{J. Schirmer}
\affiliation{Theoretische Chemie, Physikalisch-Chemisches Institut, Universit\"{a}t Heidelberg\\
Im Neuenheimer Feld 229, D-69120 Heidelberg, Germany}

\setcounter{tocdepth}{0}
\thispagestyle{empty}
\date{September 27, 2023}

\begin{abstract}
This article presents an alternative formulation of quasi-degenerate perturbation theory (QDPT). The development results by simplifying established many-body (MB) techniques to systems of non-interacting particles (NIP). While the physical information of a NIP system is fully equivalent with that of the underlying one-particle system, NIP adaptions of MB schemes can acquire useful new roles in the one-particle context.  
In particular, the algebraic-diagrammatic construction (ADC) approach to the particle-detachment part of the electron propagator, turns into a new formalism of QDPT for the original one-particle system, establishing a diagrammatic perturbation theory for the matrix elements of the effective hamiltonian and the associated effective eigenvector coefficients (EEC).
Using the NIP-ADC procedure explicit QDPT expansions are derived through third order.
Alternatively, the intermediate state representation (ISR), underlying the ADC approach, allows one to construct the effective secular quantities directly from  successive orders in the PT expansions of the NIP ground state and ground-state energy. Moreover, exact closed-form expressions can be derived for the effective hamiltonian and the EEC matrix in terms of 'target state' eigenvectors and eigenvalues of the one-particle system. These results are fully consistent with corresponding expressions derived within the QDPT context. The NIP approach to QDPT is not confined to actual particles but applies to arbitrary quantum systems. Any quantum system can be seen as formally constituting a "hyper-particle" giving rise to a system of non-interacting hyper-particles.  
The NIP adaptions of the biorthogonal coupled-cluster (BCC) and unitary coupled-cluster (UCC) methods are briefly discussed as well. 
\end{abstract}

\maketitle



\section{Introduction}

Perturbation theory (PT), based on a suitable partitioning $\hat H = \hat H_0 + \hat H_1$ of the hamiltonian into an "unperturbed" or zeroth-order part,  $\hat H_0$, and 
a perturbation part, $\hat H_1$, has proved a useful approach to the treatment of ground states of quantum systems, in particular, many-particle systems. 
In practice, PT works provided the zeroth-order ground state is energetically well separated from the higher-lying (zeroth-order) states and the coupling  
of the ground state to the higher-lying states, as effected by the perturbation part, 
$\hat H_1$, is small relative to the energy gap $\Delta E$. Specifically, the ratio 
$\gamma =|W|/\Delta E$, must be sufficiently small, where $|W|$ is a blanket measure for the coupling strength. Evidently, the PT approach breaks down if the criterion $\gamma < 1$ is not met. In particular, this is the case if the zeroth-order ground state is degenerate ($\Delta E = 0$) or nearly (quasi-) degenerate ($\Delta E \approx 0$) with one or more other zeroth-order states. 
 
If the conditions for a PT treatment do not apply to the ground state itself, but rather to a group of states (including the ground state), one may resort to an obvious extension of the 
strict PT approach referred to as quasi-degenerate perturbation theory (QDPT). Here, one constructs, as a representation of an 'effective' hamiltonian, a secular matrix for the particular group of states, also referred to as 'model states', where the coupling to the complementary (non-model) states is taken into account via perturbation theory. This amounts to a combination of matrix diagonalization and perturbation theory: diagonalization of the effective secular matrix featuring perturbation expansions of the secular matrix elements. 
Obviously, such a concept is not necessarily confined to a group of energetically lowest states, but can eventually be applied to states in a higher energy window as well. 

There is a rather extensive literature on QDPT or, more generally, effective hamiltonians, and the reader is referred to review 
articles~\cite{kle74:786,bra77:187,kva77:345,lin78:33,sha80:5711,dur87:321,fre88:1} and books~\cite{Lindgren:1982,Shavitt:2009} for accounts of the various developments and further references. It should be noted that parts of the QDPT literature are not easy to read, because often the formal level is not kept separate from the complexities encountered in the treatment of many-body systems~\cite{hos79:3827}. The latter developments are also referred to as multi-reference many-body perturbation theory (MR-MBPT) or QD-MBPT~\cite{svr87:625}.

Essentially, one may distinguish three approaches. 
The first approach, perhaps being the most direct one, was devised by Bloch~\cite{blo58:329} for the case of degenerate model states and subsequently generalized by Lindgren~\cite{lin74:2441} and Kvasni\v{c}ka~\cite{kva74:605}. 
The central notion here is the 'wave operator' relating the model states to the exact energy eigenstates. The wave operator satisfies the general Bloch 
equation~\cite{blo58:329,lin74:2441,kva74:605}, which allows one to generate its PT expansion and that of the associated effective hamiltonian.

The second approach sets out from Brioullin-Wigner (BW) perturbation theory, which derives from using partitioning~\cite{low62:969} in the solution of the secular equation. Accordingly,  the ground-state BWPT treatment can readily be extended to deal with a group of model states. However, the BWPT expansions are energy dependent, that is, they constitute implicit equations to be solved iteratively. To obtain energy-independent Rayleigh-Schr\"{o}dinger (RS) type PT expansions, one has to get rid of the energy dependence in the BW expressions, as was demonstrated by des Cloiseaux~\cite{clo60:321} and Brandow~\cite{bra67:771} in the case of degenerate model states. A generalization to non-degenerate model states was reported by Kvasni\v{c}ka~\cite{kva77:345}.

The third QDPT approach, being the most general one, goes back to Van Vleck~\cite{vle29:4467}, Jordahl~\cite{jor33:87}, and Kemble~\cite{Kemble:1937} almost a century ago. It is based on the concept of unitary or, more generally, similarity transformations (ST) that bring the hamiltonian into block-diagonal form. Essentially, any effective hamiltonian can be written in such a form, and here not even a PT-type partitioning of the 
hamiltonian needs to be supposed~\cite{ced89:2427}. For an instructive review of the ST approach, also referred to as Van Vleck perturbation theory (VVPT), the reader is referred to Shavitt and Redmon~\cite{sha80:5711}.

The purpose of this paper is to present an alternative formulation of QDPT, which - while corroborating the QDPT essentials - is completely independent of the previous standard derivations. This approach, envisaged some 30 years ago but not published until now, emerges as a surprising by-product of well-established many-body methods, such as the algebraic-diagrammatic construction (ADC) and the intermediate state representation (ISR) approaches~\cite{sch98:4734,mer96:2140,sch04:11449,Schirm:2018} related to the electron propagator. More specifically, this spin-off comes about when these
many-body methods are applied to the case of non-interacting particles (NIP). While the NIP concept does not convey new physics, the ADC and ISR formalisms adopt useful new roles in the design of QDPT for the underlying one-particle system. 

The physics of a NIP system is trivial, of course, since the solution of the underlying one-particle problem solves the many-particle problem as well, which pertains not only to the $N$-particle ground-state, but also to excitations of $N$,$N\pm1$,$\dots$ particles.    
The electron propagator, $\bs G(\omega)$, for example, reduces to a representation of the resolvent operator of the underlying one-particle problem. However, the $\bs G^-(\omega)$ part of the propagator, associated with particle detachment, is of interest, as it can directly be identified with the main block in a partitioning of the   
one-particle resolvent matrix according to the distinction of $N$ one-particle "model" states  and complementary states. Here, the number $N$ can be chosen as appropriate. As a consequence, the ADC and ISR schemes, devised for a separate treatment of the particle detachment part of $\bs G(\omega)$ (as well as the attachment part), turn into means of dealing with the partitioned one-particle resolvent matrix, that is, a particular form into which QDPT 
can be cast. In particular, this allows for an ADC-based diagrammatic generation of the PT expansions of the hermitian effective hamiltonian and the associated effective eigenvector coefficients of QDPT. Alternatively, these PT expansions can be obtained also via the ISR procedure, which, moreover, allows one to establish closed-form expressions for the relevant QDPT quantities in terms of the eigenvectors and eigenvalues of the one-particle secular equation.  

Any given one-particle problem can formally be conceived as a many-particle problem lacking particle-particle interaction, namely, as the problem of $1h$ (one-hole) excitations in a system of $N\!-\!1$ non-interacting particles. Thereby, the original many-body techniques become applicable in the one-particle context, allowing, in particular, for a stringent new approach to QDPT. As will be seen, the apparent restriction implied by the particle concept can be lifted. Any quantum system can be viewed as constituting a non-interacting "hyper-particle" at the level of a suitably defined zeroth-order hamiltonian. This, in turn, allows one to generalize the results based on the particle concept to any quantum system.

This article is organized as follows. 
Sec. II introduces the NIP concept, reviews briefly the electron propagator, and
describes its simplification in the NIP case. Sec. III treats the ADC procedure
for the particle detachment part of the electron propagator in the NIP specialization. Here, explicit second-order (ADC(2)) results are derived, while the extension to third-order is presented in App.~A. The ISR concept and its NIP adaption are the subject of Sec.~IV. In addition to the derivation of explicit PT expansions through 2nd order (partly presented in App.~C), exact closed-form expressions are obtained for the effective ADC/ISR secular matrices.  
In Sec.~V the NIP-ADC/ISR approach is generalized to an arbitrary quantum system, and in this form compared to the actual QDPT formulation applying here. Thereby the validity of the general NIP-ADC/ISR approach to QDPT is established. Some QDPT extensions are briefly addressed as well.
A brief summary and some conclusions are given in the final Sec.~VI.
 
Besides the already mentioned App.~A and App.~C, there are three more appendices.
App.~B inspects how the ground and excited states of a NIP system are treated within the biorthogonal coupled-cluster (BCC) framework. A proof is given of the block-diagonal structure of the NIP-ADC/ISR secular matrix, which is based on the findings for the BCC secular matrix.
There is also a look at the unitary coupled-cluster (UCC) approach.
App.~D reviews the derivation of PT expansions for the wave operator and effective hamiltonians according to conventional QDPT practices. A compilation of the explicit third-order PT expansions of the effective hamiltonian and the effective eigenvector amplitudes, as obtained via the present NIP-ADC procedure, is given in App.~E.

\section{The electron propagator for non-interacting particles}

The one-particle Green's function or electron propagator (see, e.g., Fetter and Walecka~\cite{Fetter:1971} and references therein) has proven a useful means 
in the treatment of interacting particles, such as electrons in an atom or molecule,
specifically providing a direct approach to ionization (particle detachment) and electron affinities (attachment) processes. 
In the following we will inspect the conversion of the electron propagator effected by eliminating the two-body interaction, that is, in its application to system of non-interacting particles (NIP). As an example, one may imagine an atom in which the unperturbed hamiltonian $\hat H_0$ only accounts for the kinetic energy of the electrons and the electron-nucleus attraction, whereas there is no Coulomb repulsion of the electrons, and    
the perturbation part of the hamiltonian is a one-particle operator, e.g., describing the effect of an outer electric field.

\subsection{A system of $N$ non-interacting particles} 
Let us consider a one-particle system, e.g., a particle subject to an external force field, 
where the hamiltonian is given by
\begin{equation}
\label{eq:H} 
\hat h = \hat h_0 + \hat w
\end{equation}
and the Hilbert space is spanned by one-particle wavefunctions (orbitals) $\psi(\xi)$, where $\xi$ denotes 
a suitable set of variables, e.g., the spatial coordinates plus a spin variable, $\xi \equiv \bs r \sigma$. 
We assume that the eigenvalue problem of $\hat h_0$,  
\begin{equation}
\label{eq:h0phi}
\hat h_0\, \phi_p(\xi) = \epsilon_p \phi_p(\xi), \;\;  p=1,2, \dots
\end{equation} 
has been solved, and the (energetically ordered) eigenfunctions $\phi_p(\bs \xi), p = 1,2,\dots$,   
establish an orthonormal basis of "unperturbed" one-particle states. 

Using the corresponding matrix representation of $\hat h$, 
\begin{equation}
\bs H = \bs \epsilon + \bs W
\end{equation}
where 
\begin{equation}
\label{eq:Hpq}
H_{pq} = \epsilon_p \delta_{pq} + w_{pq}, \;\; w_{pq} = \dirint{\phi_p}{\hat w}{\phi_q}
\end{equation}
the Schr\"{o}dinger equation for the full hamiltonian, 
\begin{equation}
\label{eq:hpsi}
\hat h \,\psi_n(\xi) = e_n \psi_n(\xi), \;\; n=1,2, \dots
\end{equation} 
can be written in a compact matrix form according to
\begin{equation}
\label{eq:seceq}
\bs H \bs X = \bs X \bs E, \; \bs X^\dagger \bs X = \bs 1
\end{equation}
Here $\bs X$ and $\bs E$ denote the matrix of eigenvectors, $\underline X_n$, and the diagonal matrix of eigenvalues, 
$e_n$, respectively. Again we suppose energetical order for the eigenvalues, $e_n \leq e_{n+1}$.
The eigenfunctions $\psi_n$ of $\bs H$ are given as linear combinations
\begin{equation}
\label{eq:psi}
 \psi_n = \sum_p X_{pn} \phi_p, \;\; n = 1, 2, \dots 
\end{equation}
of the orbitals $\phi_p$.

In the present context, the resolvent operator of the one-particle system, 
\begin{equation}
\label{eq:rop}
\hat R(\omega) = (\omega - \hat h)^{-1}
\end{equation}
is of relevance. 
The representation of $\hat R(\omega)$ in terms of the unperturbed states,  
\begin{equation}
R_{pq}(\omega) = \dirint{\phi_p}{(\omega - \hat h)^{-1}}{\phi_q}
\end{equation}
can be written in matrix form as
\begin{equation}
\label{eq:ropmat}
\bs R(\omega) = (\omega \bs 1 - \bs H)^{-1} = \bs X (\omega \bs 1 - \bs E)^{-1} \bs X^\dagger
\end{equation}
where the explicit expression in the second equation follows as a 
consequence the secular equations~(\ref{eq:seceq}).

After the exposition of the one-particle level, we now can turn to the case of many non-interacting particles (NIPs) 
Specifically, we consider a system consisting of $N$ particles of the given kind and, moreover, suppose that  
the particles behave as fermions with regard to the permutation symmetry in the $N$-particle wave functions.

The corresponding $N$-particle hamiltonian is given by the sum  
\begin{equation}
\label{eq:nparth}
\hat H = \sum_{i=1}^N \hat h(i)
\end{equation}
of one-particle hamiltonians,
where $\hat h(i)$ is the operator~(\ref{eq:H}) acting on particle $i$.
Using the formalism of second quantization, which can readily be adapted to the 
present case, the hamiltonian can be written in the form
\begin{equation}
\label{eq:Hmb}
\hat H = \hat H_0 + \hat W = \sum_p \epsilon_p c^\dagger_p c_p + \sum_{p,q} w_{pq}c^\dagger_p c_q
\end{equation}
which is independent of the respective particle number $N$.
Here $c^\dagger_p (c_p)$ denotes the creation (destruction) operator for a particle in the orbital $\phi_p$.
Likewise, one may introduce the fermion operators $\tilde c^\dagger_p$ ($\tilde c_p$) based on the eigenfunctions $\psi_p$ of the (one-particle) hamiltonian $\hat h$. The unitary transformation~(\ref{eq:psi}) between the two sets of orbitals 
translates into corresponding transformations of the fermion operators, such as 
the expansion
\begin{equation}
\label{eq:ctilde}
 c_p = \sum_s X_{ps} \tilde c_s
\end{equation}
of the destruction operators $c_p$ in terms of the operators $\tilde c_s$.

The Slater determinant 
\begin{equation}
\label{eq:Phi0}
\ket{\Phi_0^N} = |\phi_1 \phi_2 \dots \phi_N|
\end{equation}
of the energetically lowest orbitals $\phi_p, \, p = 1, \dots ,N$,
constitutes the "unperturbed" $N$-particle ground state, that is,
\begin{equation}
\hat H_0 \ket{\Phi_0^N} = E^{(0)}_0 \ket{\Phi_0^N}  
\end{equation}
with the ground-state energy
\begin{equation}
E^{(0)}_0 = \sum_{k=1}^N \epsilon_k
\end{equation}
For further use, we introduce "occupation numbers" with respect to the unperturbed ground state $\ket{\Phi^N_0}$:
\begin{equation}
n_p = 1- \overline n_p = \begin{cases}{1,}&p \leq N \;\; \text{- occupied}
\\{0,}& p > N \;\;\text{- 
unoccupied}\end{cases} 
\end{equation}
Moreover, we 
shall adopt the familiar notation in which the indices $a,b,c,\dots$ and $i,j,k,\dots$ refer to
unoccupied and occupied orbitals, respectively, while the indices $p,q,r,\dots$ are general. 

Obviously, the treatment of the NIP system is trivial inasmuch as
the solution of the one-particle secular equations~(\ref{eq:seceq}) solves 
the $N$-particle problem as well. The exact $N$-particle ground state
is simply given by the Slater determinant
\begin{equation}
\label{eq:Psi0}
\ket{\Psi_0^N} = |\psi_1 \psi_2 \dots \psi_N|
\end{equation}
in terms of the one-particle energy eigenstates $\psi_n, n =1, \dots, N$:
\begin{equation}
\hat H \ket{\Psi_0^N} = E^N_0 \ket{\Psi_0^N},  
\end{equation}
where the ground-state energy is given by
\begin{equation}
\label{eq:mbgsen}
E^N_0 = \sum_{n=1}^N e_n
\end{equation}

Moreover, the entire manifold of excited states is available as well. For example, 
the single excitations can be written as 
\begin{equation}
\ket{\Psi^N_{ak}} = \tilde c^\dagger_a \tilde c_k \ket{\Psi_0^N}, \,\, a > N, k \leq N  
\end{equation}
being eigenstates of $\hat H$ with energies $E_{ak} = E^N_0 + e_a - e_k$. 
This also applies to "generalized" excitations associated with the removal or attachment of a particle, such as the one-hole ($1h$) states of the ($N\!-\!1$)-particle system,
\begin{equation}
\label{eq:ex1h}
\ket{\Psi^{N-1}_k} = \tilde c_k \ket{\Psi_0^N}, \;\, k \leq N  
\end{equation}
or the one-particle ($1p$) states of the ($N$+$1$)-particle system,
\begin{equation}
\ket{\Psi^{N+1}_a} = \tilde c^\dagger_a \ket{\Psi_0^N}, \;\, a > N  
\end{equation}
Here the respective energies are $E^{N-1}_k = E^N_0 - e_k$ and $E^{N+1}_a = E^N_0 + e_a$.

The distinction of $1h$ and $1p$ states of $N\!-\!1$ and $N\!+\!1$ (non-interacting) particles, respectively,
entails a corresponding partitioning at the one-particle level. Here, 
the exact one-particle states, $\psi_p$, divide into two groups,
\begin{align}
\nonumber
(\bs 1) \equiv\{\psi_p, \, p\leq N\}\\
\label{eq:set12x} 
(\bs 2) \equiv\{\psi_p, \, p > N\}
\end{align} 
where the former and latter states are "occupied" and "unoccupied" in the exact ground state~(\ref{eq:Psi0}) of the NIP system. 
This mirrors the partitioning of 
the unperturbed one-particle states, $\phi_p$, with regard to their occupation in the unperturbed
ground state~(\ref{eq:Phi0}) according to
\begin{align}
\nonumber
(\bs 1) \equiv\{\phi_p, \, p\leq N\}\\
\label{eq:set12} 
(\bs 2) \equiv\{\phi_p, \, p > N\}
\end{align}

Noting that the particle number $N$ can be chosen arbitrarily, we may say that the pair $(\hat h_0, N)$
marks a partitioning of the one-particle Hilbert space into a space spanned by the unperturbed states of group $(\bs 1)$
and its orthogonal complement spanned by the unperturbed states of group $(\bs 2)$. In the QDPT context the former space is referred to as the "model space". In an analogous way, the pair $(\hat h, N)$ entails a partitioning of the one-particle Hilbert space 
into a "target space" and its orthogonal complement corresponding to the partitioning~(\ref{eq:set12x}) of the  
eigenstates of $\hat h$. These related partitionings are essential elements in the QDPT formulation, here of the one-particle system.

\subsection{The electron propagator}

In energy representation the components of the electron propagator 
take on the form
\begin{equation}
G_{pq}(\omega) = G^-_{pq}(\omega) + G^+_{pq}(\omega)
\end{equation}
where the two parts are given by 
\begin{align}
\label{eq:Gmin}
G^-_{pq}(\omega) &= \dirint{\Psi^N_0}{c_q^\dagger\left(\omega  + \hat H - E^N_0 - i\eta\right)^{-1}c_p}{\Psi^N_0}\\
\label{eq:Gplus}
G^+_{pq}(\omega) &= \dirint{\Psi^N_0}{c_p\left(\omega - \hat H + E^N_0  + i\eta\right)^{-1}c_q^\dagger}{\Psi^N_0}
\end{align}
The two parts $G^\pm_{pq}$ can be understood as being matrix elements of the many-body resolvent variants
$(\omega +
 \hat H - E^N_0)^{-1}$ and $(\omega -
 \hat H + E^N_0)^{-1}$ taken with respect to ($N\!-\!1$)-particle states and ($N\!+\!1$)-particle states, respectively. 
The infinitesimals $\pm i \eta$ are of mathematical significance, being required for the definiteness of the Fourier transformations between the time and energy representations. 

The physical content of the electron propagator parts becomes explicit in their spectral representations obtained by using the resolution of the identity in terms of the exact 
($N\mp 1$)-particle energy eigenstates. For $\bs G^-(\omega)$ the spectral representation reads   
\begin{equation}
\label{eq:specrep}
G^{-}_{pq}(\omega) = \sum_n \frac{x_{pn}x_{qn}^*}{\omega - \omega_n -i\eta}
\end{equation} 
where the pole positions $\omega_n$ can be identified with the negative ionization energies,
\begin{equation}
\label{eq:omega}
\omega_n = - (E_n^{N\!-\!1} - E^N_0)
\end{equation}
and
\begin{equation}
\label{eq:specamp}
x_{pn} = \dirint{\Psi_n^{N\!-\!1}}{c_p}{\Psi_0^N}
\end{equation}
denote the spectral amplitudes associated with the ($N\!-\!1$)-particle states $\ket{\Psi_n^{N\!-\!1}}$.

Now we turn to the NIP system and inspect the significant
simplification that comes with it.   
Let us consider the spectral representation for $G^-_{pq}(\omega)$, being of the general form of 
Eq.~(\ref{eq:specrep}) and  
recall that the $1h$ states of $N-1$ particles,
\begin{equation}
\label{eq:exstats}
\ket{\Psi_m^{N\!-\!1}} = \tilde c_m \ket{\Psi^N_0},\; m = 1, \dots, N
\end{equation}
are eigenstates of $\hat H$,
\begin{equation}
 \hat H \ket{\Psi_m^{N\!-\!1}} = E^{N\!-\!1}_m \ket{\Psi_m^{N\!-\!1}}
 \end{equation} 
with the energies
\begin{equation}
\label{eq:deten}
E^{N\!-\!1}_m = E^N_0 - e_m
\end{equation}
The corresponding pole positions~(\ref{eq:omega}) are given by $\omega_m = e_m$.
According to Eq.~(\ref{eq:ctilde}), the spectral amplitudes~(\ref{eq:specamp}) can be identified 
with the eigenvector components $X_{pm}$ of the one-particle secular equation~(\ref{eq:seceq}),
\begin{equation}
\label{eq:specampx}
x_{pm} = \dirint{\Psi_m^{N\!-\!1}}{c_p}{\Psi^N_0} = X_{pm}, \; m = 1, \dots, N
\end{equation}
Obviously, only $1h$ states do appear in the spectral representation for $G^-_{pq}(\omega)$,
because the spectral amplitudes for states of higher excitation classes vanish. 
For example, the spectral amplitudes for a two-hole-one-particle ($2h$-$1p$) state 
$\ket{\Psi_{aij}^{N\!-\!1}} = \tilde c_a^\dagger \tilde c_i \tilde c_j \ket{\Psi^N_0}$, are given by 
$x_{p,aij} = \dirint{\Psi_{aij}^{N\!-\!1}}{c_p}{\Psi_0^N}$. According to Eq.~(\ref{eq:ctilde}), 
$c_p \ket{\Psi_0^N}$ is a linear combination of $1h$ states, so that $x_{p,aij}$ vanishes.

As a result, the general spectral representation~(\ref{eq:specrep}) simplifies to  
\begin{equation}
\label{eq:sr-}
 G^-_{pq}(\omega) = \sum_{m\leq N} X_{pm} (\omega -e_m -i \eta)^{-1} X_{qm}^*
\end{equation}
In an analogous way, the $\bs G^+(\omega)$ part can be written according to
\begin{equation}
\label{eq:sr+}
 G^+_{pq}(\omega) = \sum_{m > N} X_{pm} (\omega - e_m + i \eta)^{-1} X_{qm}^*
\end{equation}
Here, the spectral amplitudes are given by
\begin{equation}
\label{eq:sa+}
x_{pm} = \dirint{\Psi_0^N}{c_p}{\Psi_m^{N\!+\!1}} = X_{pm}, \;\; m = N+1, N+2, \dots 
\end{equation}  
Note that only $1p$ states enter the spectral representation of  $G^+_{pq}(\omega)$.

Disregarding the infinitesimals $\pm i \eta$, which are not essential here, the expressions~(\ref{eq:sr-}) 
and (\ref{eq:sr+}) for the two parts can be combined, which allows us to write $\bs G(\omega)$ in a compact matrix form as follows: 
\begin{equation}
\label{eq:Gecf}
\bs G(\omega) = \bs G^-(\omega) + \bs G^+(\omega) = \bs X (\omega \bs 1 - \bs E)^{-1} \bs X^\dagger 
\end{equation}
Here $\bs X$ and $\bs E$ are the eigenvector and eigenvalue matrices of the one-particle
secular problem~(\ref{eq:seceq}), which can be used write $\bs G(\omega)$ as 
\begin{equation}
\label{eq:Gsol}
\bs G(\omega)= (\omega - \bs H)^{-1}
\end{equation}
where $\bs H = \bs \epsilon + \bs W$ is the matrix representation of the one-particle hamiltonian~(\ref{eq:H}).
This means that for a system of non-interacting particles the electron propagator $\bs G(\omega)$ is simply given by 
the representation~(\ref{eq:ropmat}) of the one-particle resolvent operator:
\begin{equation}
\bs G(\omega) = \bs R(\omega)
\end{equation}
Obviously, the electron propagator of a NIP system does not convey any physical content beyond
that of the underlying one-particle system.  

But then what is the meaning of the two parts $\bs G^-(\omega)$ and $\bs G^+(\omega)$?
According to partitioning specifications~(\ref{eq:set12x},\ref{eq:set12}), the eigenvector matrix $\bs X$ can 
be written in the partitioned form 
\begin{equation}
\label{eq:part}
\bs X = 
\left(
\begin{array}{cc}
\bs X_{11} & \bs X_{12} \\
\bs X_{21} & \bs X_{22} 
\end{array}
\right)
\end{equation}
Introducing corresponding projector matrices,
\begin{equation}
\label{eq:proms}
\bs P_0 = 
\left(
\begin{array}{cc}
\bs 1 & \bs 0 \\
\bs 0 & \bs 0 
\end{array}
\right)
,\;\;\; 
\bs Q_0 = 
\left(
\begin{array}{cc}
\bs 0 & \bs 0 \\
\bs 0 & \bs 1 
\end{array}
\right)
\end{equation}
the unit matrix can be written as
\begin{equation}
\bs 1 = \bs P_0 + \bs Q_0
\end{equation}
which in turn can be used in Eq.~(\ref{eq:ropmat}) to decompose the 
resolvent matrix into two parts, 
\begin{equation}
\bs R(\omega) = \bs R^I(\omega) + \bs R^{II}(\omega) = \bs x (\omega \bs 1 - \bs E_1)^{-1} \bs x^\dagger 
+ \bs y (\omega \bs 1 - \bs E_2)^{-1} \bs y^\dagger  
\end{equation}
where $\bs x$ and $\bs y$ denote the matrices formed by the eigenvectors of group $\bs 1$ and $\bs 2$, respectively:
\begin{equation}
\label{eq:X1} 
\bs x = 
\left(
\begin{array}{c}
\bs X_{11}  \\
\bs X_{21}  
\end{array}
\right), \;\;\;
\bs y = 
\left(
\begin{array}{c}
\bs X_{12}  \\
\bs X_{22}  
\end{array}
\right)
\end{equation}
and $\bs E_1$ and $\bs E_2$ denote the diagonal matrices of the energy eigenvalues $e_n,\, n \leq N$, and $e_n, \, n > N$, 
respectively.

According to Eq.~(\ref{eq:specampx}), the matrix of the spectral amplitudes~(\ref{eq:specamp}) is given by 
$\bs x$ as well, so that 
the $\bs G^-(\omega)$ part of the electron propagator can be written as 
\begin{equation}
\label{eq:srmat-}
\bs G^-(\omega)=  \bs x (\omega \bs 1 - \bs E_1)^{-1} \bs x^\dagger = \bs R^I(\omega)
\end{equation}
which establishes the identity $\bs G^-(\omega) = \bs R^I(\omega)$.

Let us emphasize the dual roles of the quantities $\bs x$ and $\bs E_1$:
On the one-particle level, $\bs x$ is the matrix formed by the first $N$ eigenvectors of the 
one-particle secular problem~(\ref{eq:seceq}), and $\bs E_1$ is the diagonal matrix of the 
corresponding energy eigenvalues; on the many-particle level, by contrast, these quantities refer  
to the matrix of spectroscopic amplitudes and the diagonal matrix of energies, respectively,
associated with the removal of a particle from the NIP system.   
 
An analogous expression applies to the $\bs G^+(\omega)$ part,
\begin{equation}
\label{eq:srmat+}
\bs G^+(\omega) = \bs y (\omega \bs 1 - \bs E_2)^{-1} \bs y^\dagger = \bs R^{II}(\omega)
\end{equation}
where $\bs y$ is the matrix of the spectroscopic amplitudes~(\ref{eq:sr+}), or, likewise, the matrix formed by the eigenvectors $\ul X_n, n > N$ of Eq.~(\ref{eq:seceq}).

The identification of $\bs G^-(\omega)$ with $\bs R^I(\omega)$ (and $\bs G^+(\omega)$ with $\bs R^{II}(\omega)$)  
shows again that the NIP concept does not go beyond the framework set by the underlying one-particle system.
However, the methods devised for the separate treatment of the ionization part $\bs G^-(\omega)$ (or $\bs G^+(\omega)$),
such as the ADC/ISR schemes, apply as well to the resolvent part, $\bs R^I(\omega)$ (or $\bs R^{II}(\omega)$),
thereby establishing an alternative approach to QDPT, as will be addressed in Sec.~V.

\subsection{Diagrammatic perturbation theory}

The absence of two-particle interactions radically simplifies the familiar diagrammatic perturbation theory (see Fetter and Walecka~\cite{Fetter:1971})
for the electron propagator.
\begin{figure}
\begin{equation*}
\bs G \equiv \quad \vcenterpdf{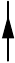} \quad + \quad
\vcenterpdf{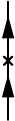} \quad + \quad \vcenterpdf{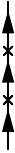} \quad +  \quad \dots
\end{equation*}
\caption{Diagrammatic PT expansion for a one-particle interaction.}
\label{fig:1pd}
\end{figure}
Fig.~\ref{fig:1pd} shows the diagrammatic perturbation expansion for the electron propagator of a non-interacting $N$-particle system through third order.
Here the directed lines represent the diagonal ``free'' electron propagator elements, reading in energy representation
\begin{equation}
\label{eq:g0er}
G_{pq}^0 (\omega) = \delta_{pq} \left(\frac{\bar n_p}{\omega - \epsilon_p + i\eta} + \frac{n_p}{\omega - \epsilon_p - i\eta}\right)
\end{equation}
and the ``crosses'' stand for the matrix elements $w_{pq}$ of the perturbation part of the hamiltonian~(\ref{eq:Hmb}). There is just one diagram in each order $n$ of the PT expansion, featuring $n$ perturbation crosses and $n+1$ free fermion lines. 
The first-order contribution, for example, reads
\begin{equation}
\label{eq:g1er}
G^{(1)}_{pq}(\omega) = G^0_{p}(\omega) w_{pq} G^0_{q}(\omega)
\end{equation}
In energy representation and matrix form, the diagrammatic PT expansion can be written as
\begin{equation}
\label{eq:PT4G}
 \bs G(\omega) = \bs G^0(\omega) + \bs G^0(\omega)\bs W \bs G^0(\omega)  + \bs G^0(\omega)\bs W \bs G^0(\omega) \bs W \bs G^0(\omega) + \dots 
\end{equation}
which can be cast into a recursion scheme according to 
\begin{equation}
\label{eq:dyson}
 \bs G(\omega) = \bs G^0(\omega) + \bs G^0(\omega)\bs W \bs G(\omega) 
\end{equation}
Obviously, this is just the Dyson equation(see Fetter and Walecka~\cite{Fetter:1971}), where the self-energy part, $\bs \Sigma = \bs W$, is a constant (energy independent) matrix. A graphical representation of Eq.~(\ref{eq:dyson}) is given in Fig.~\ref{fig:dyson}.
\begin{figure}
\begin{equation*}
\vcenterpdf{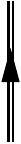} \quad = \quad
\vcenterpdf{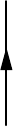} \quad + \quad \vcenterpdf{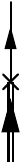} 
\end{equation*}
\caption{Dyson equation in the case of non-interacting particles.}
\label{fig:dyson}
\end{figure}
The Dyson equation can formally be solved, 
\begin{equation}
\label{eq:GDys}
 \bs G(\omega) = \left( \bs G^0(\omega)^{-1} - \bs W \right)^{-1} =   \left( \omega -\bs \epsilon - \bs W \right)^{-1}
\end{equation}
which just reproduces Eq.~(\ref{eq:Gsol}) and, furthermore, the explicit spectral form~(\ref{eq:Gecf}) via the 
solution of the one-particle secular equation~(\ref{eq:seceq}). 
This means that, in the case of non-interacting particles, the diagrammatic PT expansion of the $\bs G(\omega)$ propagator 
is straightforward, the result being fully equivalent to the one-particle secular problem.

As may be noted, the imaginary infinitesimals $\pm i \eta$ disregarded in Eq.~(\ref{eq:GDys})
can be incorporated by using the unabridged expression for $\bs G^0(\omega)$. This allows one to distinguish the ($N\!-\!1$)- and ($N\!+\!1$)-particle solutions according to the location of the respective poles in the upper and lower complex $\omega$-plane.

Diagrammatic perturbation theory also allows for
a direct approach to the ($N\!-\!1$)- or ($N\!-\!1$)-particle parts $\bs G^-(\omega)$ and 
$\bs G^+(\omega)$, respectively. To this end, one has to consider the manifold of 
time-ordered or Goldstone diagrams. To each $n$-th order Feynman diagram in Fig.~1 there are $(n+2)!$ Goldstone diagrams corresponding to the respective ordering of the 2 external vertices (in time representation labelled $t,t'$) 
and $n$ internal vertices (labelled $t_1, \dots , t_n$). Here the diagrams with the order $t > t'$ of the external vertices contribute exclusively to $\bs G^+(\omega)$, while those with the opposite ordering $t < t'$ contribute exclusively to $\bs G^-(\omega)$. 
\begin{figure}
\begin{equation*}
\begin{array}{lll}
\vcenterpdf{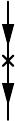} & \quad \quad \vcenterpdf{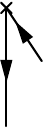} & \quad \vcenterpdf{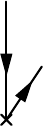}\\
 (a) & \quad \quad (b) & \quad (c) 
\end{array}
\end{equation*}
\caption{Time-orderings of the first-order one-particle diagram contributing to 
$\bs G^-$}.
\label{fig:1pdo1}
\end{figure}
Fig.~\ref{fig:1pdo1} shows the three Goldstone diagrams constituting the first order of the $\bs G^-(\omega)$ part. The algebraic expressions resulting in energy representation for the Goldstone diagrams can readily be written up according to the familiar diagram rules (see Ref.~\cite{Schirm:2018}). For the three first-order diagrams in Fig.~\ref{fig:1pdo1} the resulting expressions read 
\begin{align}
\label{eq:foa}
(a)& \equiv   \frac{1}{(\omega - \epsilon_p - i\eta)} \,w_{pq}\,\frac{1}{(\omega - \epsilon_q - i\eta)}\,\, n_p n_q \\
\label{eq:fob}
(b)& \equiv  \frac{1}{(\omega - \epsilon_p - i\eta)}\,
 \frac{w_{pq}}{(\epsilon_p - \epsilon_q)}\,\, n_p \bar n_q \\
\label{eq:foc}
(c)& \equiv  \frac{1}{(\omega - \epsilon_q - i\eta)}\,
 \frac{w_{pq}}{(\epsilon_q - \epsilon_p)} \,\,n_q \bar n_p
\end{align}

It should be noted that the fragmentation of the compact PT expansion~(\ref{eq:PT4G}) for $\bs G(\omega)$ into separate expansions for $\bs G^-(\omega)$ and $\bs G^+(\omega)$ can also be achieved by using projector integration according to
\begin{equation}
\label{eq:proj}
\bs G^-(\omega) = \frac{1}{2\pi i} \oint \frac{\bs G(\omega')}{\omega -\omega' -i\eta} \,\mathrm{d}\omega'
\end{equation}
For example, the results of Eq.~(\ref{eq:foa}-\ref{eq:foc}) can be derived by inserting the first-order term~(\ref{eq:g1er}) on the right-hand side and using partial fractional decomposition for the mixed ($\pm i \eta$) pole products. 
 
Based on the separate diagrammatic PT expansions, the algebraic-diagrammatic construction (ADC) allows one to derive systematic approximation schemes for $\bs G^-(\omega)$ or, likewise, for $\bs R^I(\omega)$ (and $\bs G^+(\omega)$ or $\bs R^{II}(\omega)$)  through successively higher PT order. This will be addressed in the subsequent section.

\section{ADC procedure for the non-interacting electron propagator part $\bs G^{-}(\omega)$}

\subsection{General concept}

In the following we specifically consider the ADC procedure~\cite{sch98:4734,Schirm:2018} for the ($N\!-\!1$)-part of the electron propagator. 
Adapted to the case of a NIP system, the spectral representation of $\bs{G}^{-}$ could be written in the form 
of Eq.~(\ref{eq:srmat-}),    
\begin{equation}
\bs{G}^{-} (\omega) = \bs x\, (\omega \bs 1 - \bs E_1)^{-1} \bs x^\dagger 
\end{equation}
where $\bs E_1$ is the diagonal matrix of one-particle energies, $e_1, \dots, e_N$,
and the matrix $\bs x$ of spectral amplitudes is given by Eq.~(\ref{eq:X1}). 
The spectral representation results by applying (twice) the resolution of the identity (ROI) in terms of the 
exact energy eigenstates~(\ref{eq:exstats}), $\ket{\Psi_m^{N-1}}, m = 1, \dots, N$, in the defining equations~(\ref{eq:Gmin}). 

In an analogous way,  
the (non-diagonal) ADC form can be established, 
supposing here a ROI in terms of so-called intermediate states, $\ket{\tilde{\Psi}_J}$. These states derive from the ``correlated excited states'' (CES)
$\hat C_J\ket{\Psi_0^N}$, where $\hat C_J$ denote physical excitation operators,
\begin{equation}
\label{eq:hfconfs}
 \{\hat C_J\}  = \{c_k;\, c^\dagger_ac_kc_l,k<l;\dots\}
\end{equation}
for $1h$, $2h$-$1p$, $3h$-$2p$, $\dots$ excitations.  
In the case of non-interacting particles, only the $1h$ intermediate states, 
$\ket{\tilde{\Psi}_k}, k \leq N$,
come into play, as will be shown in Sec.~IV.A, addressing the actual construction of the sofar only postulated 
intermediate states.

Accordingly, one can restrict the ROI to the $1h$ intermediate states,  
\begin{equation}
\label{eq:pROI}
 \hat{\mathbb 1} \sim \sum_{j \leq N} \ket{\tilde{\Psi}_j}\bra{\tilde{\Psi}_j} 
\end{equation}
in the evaluation of Eqs.~(\ref{eq:Gmin}) for $\bs{G}^{-}$. 
For notational convenience and to be consistent 
with the ADC formulation adopted previously in the case of interacting particles, we 
here set out from the transpose of $\bs G^-(\omega)$,
\begin{equation}
\label{eq:Gtilde}
\tilde{\bs G}(\omega) = \bs G^-(\omega)^t 
\end{equation}
that is, $\tilde G_{pq} = G^-_{qp}$. The ADC form, obtained by inserting the ROI~(\ref{eq:pROI}) 
twice on the right-hand side of Eq.~(\ref{eq:Gmin}) (with interchanged indices $p$ and $q$),
is given by 
\begin{equation}
\label{eq:gmadcf}
\tilde{\bs G}(\omega) = \bs{f}^{\dagger} (\omega + \bs M)^{-1} \bs{f}
\end{equation}
Here, $\bs M$ is the ADC secular matrix,
\begin{equation}
\label{eq:issm}
M_{kl} = \dirint{\tilde{\Psi}_k}{\hat H - E^N_0}{\tilde{\Psi}_l}, \;\;k,l \leq N
\end{equation}
and $\bs f$ is the matrix of `effective' transition amplitudes (ETA)
\begin{equation}
\label{eq:istm}
f_{kq} = \dirint{\tilde{\Psi}_k}{c_q}{\Psi^N_0}, \;\; k \leq N
\end{equation}
As will be discussed in the following subsection,
the matrix elements of $\bs M$ and $\bs f$ 
are subject to perturbation expansions, the terms of which 
can successively be derived from the diagrammatic
perturbation expansion for the electron propagator part $\bs G^-(\omega)$.

Once approximate (or exact) expressions for $\bs M$ and $\bs f$ have been established,  
the corresponding energy eigenstates and eigenvalues are obtained by solving  
the ADC secular equations, 
\begin{equation}
\label{eq:adceveq}
  \bs M \bs Y = \bs Y (-\bs E_1),\qquad \bs Y^\dagger \bs Y = \bs 1
\end{equation}
Here $\bs Y$ denotes the matrix of eigenvectors, and $\bs E_1$ 
is the diagonal matrix of the one-particle energies $e_m$,  
or, likewise, the negative detachment energies, $e_m = E^N_0 - E_m^{N-1}$, according to Eq.~(\ref{eq:deten}).

The spectroscopic amplitudes, $x_{pm}$, are obtained  
as the scalar product of the $m$-th eigenvector and the respective column of the matrix 
$\bs f$:
\begin{equation}
\label{eq:spevf}
x_{pm} = \sum_k Y^*_{km} f_{kp}, \; m \leq N
\end{equation}
which can compactly be written as
\begin{equation}
\label{eq:xfY}
\bs x = \bs f^t \bs Y^*
\end{equation} 
The eigenvector components can be seen as the
expansion coefficients
\begin{equation}
\label{eq:adceveqx}
 Y_{km} = \braket{\tilde{\Psi}_k}{\Psi_m^{N\!-\!1}}, \; k, m \leq N
\end{equation}
of the approximate (or exact) energy eigenstates written as linear combinations of
the postulated intermediate states.

\subsection{Explicit ADC procedure through second order}

To formulate the perturbation expansion of the ADC form~(\ref{eq:gmadcf}) 
one has to combine the PT expansions
\begin{align}
  \bs M &= \bs M^{(0)} + \bs M^{(1)} + \bs M^{(2)} + \bs M^{(3)} + \dots\\
\label{eq:ptexp4f}
 \bs f &= \bs f^{(0)} + \bs f^{(1)} + \bs f^{(2)} + \dots
\end{align}
of the ADC matrices $\bs M$ and $\bs f$ with a corresponding expansion of the resolvent type matrix,
$(\omega + \bs M)^{-1}$. To this end we write $\bs M$ in the form 
\begin{equation}
\label{eq:MKC}
\bs M = -(\bs K + \bs C)
\end{equation}
where $\bs K = -\bs M^{(0)}$ is (up to the sign) the zeroth-order part of the secular matrix, and 
$\bs C = -\bs M+ \bs M^{(0)}$ is the remainder, beginning in first order. 
As will be seen, the zeroth-order part is simply given by
the diagonal matrix of orbital energies,
\begin{equation}
\label{eq:kelem}
K_{pq} = \epsilon_p \,\delta_{pq}, \;\; p,q \leq N
\end{equation}

Separating the zeroth-order part of the ADC secular matrix according Eq.~(\ref{eq:MKC}) allows one to expand the resolvent matrix  to
in Eq.~(\ref{eq:gmadcf}) as a geometric series, yielding
\begin{align}
\nonumber
 \tilde{\bs G}(\omega)& = \bs f^\dagger(\omega - \bs K -\bs C)^{-1} \bs f \\
\nonumber 
 &= \bs f^\dagger \,(\omega - \bs K)^{-1} \sum_{\nu = 0}^\infty 
\left(\frac{\bs C}{\omega - \bs K}\right)^\nu \bs f \\
\label{eq:gmadcfps}
&= \bs{f}^{\dagger} (\omega- \bs K )^{-1} \bs{f} + \bs{f}^{\dagger} (\omega- \bs K)^{-1} 
\bs C (\omega- \bs K)^{-1}\bs{f} + \dots
\end{align}
Together with the PT expansions for $\bs C$ (beginning in first order) and $\bs f$, this establishes the formal ADC series for $\tilde{\bs G}(\omega)$, which, in turn, can be compared with diagrammatic PT expansion for $\bs G^-(\omega)$ (or its transpose).   
This comparison allows one to determine the successive terms in the PT expansions of matrix elements of $\bs C$ and $\bs f$. In the following we will perform the ADC procedure for $\bs G^-(\omega)$ through
second order of PT.\\
\\
\textbf{Zeroth order}:\\
The zeroth-order ADC term, 
\begin{equation}
\tilde{\bs G}^{(0)}(\omega) = \bs f^{(0)\dagger} (\omega - \bs K)^{-1} \bs f^{(0)}
\end{equation}
is to be compared with the diagrammatic zeroth-order expression
\begin{equation*}
G_{qp}^{-(0)}(\omega) = \delta_{pq}(\omega - \epsilon_p)^{-1} n_p
\end{equation*}
Obviously, $\bs K$ is consistent with Eq.~(\ref{eq:kelem}), that is,
$K_{pq} = \delta_{pq}\, \epsilon_p, \, n_p = 1$;   
and $\bs f^{(0)}$ is given by
\begin{equation}
 f^{(0)}_{pq} = \delta_{pq}\, n_p
\end{equation}

\noindent
\textbf{First order}:\\
In first order the ADC form reads
\begin{align}
 \nonumber
\tilde{\bs G}^{(1)}(\omega) = \bs f^{(1)\dagger}& (\omega - \bs K)^{-1} \bs f^{(0)} + h.c.\\
+ &\bs f^{(0)\dagger}(\omega - \bs K)^{-1} \bs C^{(1)}(\omega - \bs K)^{-1} \bs f^{(0)}
\end{align}
where the first-order constituents $\bs C^{(1)}$ and $\bs f^{(1)}$ come into play.  
The comparison with the diagrammatic first-order contributions to $G^{-(1)}_{qp}$,  as per 
Eqs.~(\ref{eq:foa}-\ref{eq:foc}), yields the following results:  
\begin{align}
\label{eq:C1pq}
 C^{(1)}_{pq} &= w_{qp}\\
\label{eq:f11}
 f^{(1)}_{pq} &=  \frac{w_{qp}}{\epsilon_p - \epsilon_q}\, n_p \lo n_q
\end{align}
\textbf{Second order}:\\
In second order the ADC expansion takes on the form
\begin{align}
\nonumber
 \tilde{\bs G}^{(2)}(\omega) =
&{\bs f^{(0)}}^\dagger (\omega - \bs K)^{-1} \bs f^{(2)} + h.c. &(A)\\
\nonumber
 +&{\bs f^{(0)}}^\dagger(\omega - \bs K)^{-1} \bs C^{(2)}(\omega - \bs K)^{-1} \bs f^{(0)} &(B)\\
\nonumber
 +&{\bs f^{(1)}}^\dagger (\omega - \bs K)^{-1} \bs f^{(1)}  &(C)\\
\nonumber
 +&{\bs f^{(1)}}^\dagger(\omega - \bs K)^{-1} \bs C^{(1)}(\omega - \bs K)^{-1}\bs f^{(0)} + h.c. &(D)\\
\label{eq:adcexpo2}
+&{\bs f^{(0)}}^\dagger(\omega - \bs K)^{-1} \bs C^{(1)}(\omega - \bs K)^{-1}C^{(1)}
 (\omega - \bs K)^{-1} \bs f^{(0)}
 &(E)
\end{align}
These expressions are to be compared to the 12 second-order diagrams (1)-(12) (for $G^{-(2)}_{qp}$) shown in Fig.~4. 
\begin{figure}
\begin{equation*}
\begin{array}{llllll}
\vcenterpdf{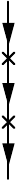} & \quad \quad \vcenterpdf{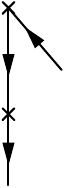} & \quad \vcenterpdf{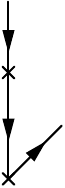} & \quad \vcenterpdf{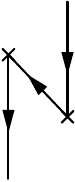} & \quad \vcenterpdf{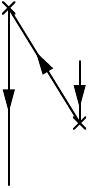} & \quad \vcenterpdf{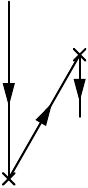}\\
 (1) & \quad \quad (2) & \quad (3) & \quad (4) & \quad (5) & \quad (6) \\
\vcenterpdf{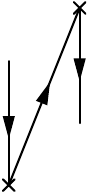} & \quad \quad \vcenterpdf{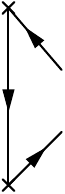} & \quad \vcenterpdf{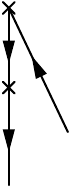} & \quad \vcenterpdf{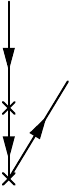} & \quad \vcenterpdf{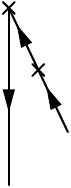} & \quad \vcenterpdf{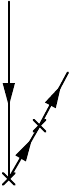}\\
(7) & \quad \quad (8) & \quad (9) & \quad (10) & \quad (11) & \quad (12)
\end{array}
\end{equation*}
\caption{Time-orderings of the second-order one-particle diagram contributing to 
$\bs G^-$.}
\label{fig:1pdo2}
\end{figure}
The terms $(C)$, $(D)$, and $(E)$, featuring the first-order contributions $\bs C^{(1)}$ and $\bs f^{(1)}$ already determined at the previous step, can directly be assigned to corresponding diagrams. Specifically, diagram (1) corresponds to term $(E)$;  
diagrams (2) and (3) correspond to the two parts of term $(D)$; and diagram (8) corresponds to term $(C)$.
The diagrams (9)-(12) match the term $(A)$ and its hermitian conjugate, establishing a 
$hp$-contributions to $f^{(2)}_{pq}$, where $q$ is a particle index, $q > N$.

The remaining 4 diagrams (4)-(7) do not individually match any of the ADC terms, and, moreover, introduce $\omega$-denominators of the type $(\omega + \epsilon_a - \epsilon_k -
\epsilon_l)^{-1}$, apparently suggesting that here $2h$-$1p$ excitations come into play, which would not be compatible with the present $1h$-based ADC scheme. However, when these 4 diagrams are added up, the $2h$-$1p$ denominators drop out and the resulting expressions can can be assigned to the terms $(A)$ and $(B)$, as will be shown in the following. 

Since the diagrams (4)-(7) differ only in their denominator products, they can be combined according to
\begin{equation}
\label{eq:comdiag}
(4) + (5) + (6) + (7)|_{qp} = \sum_{s > N} w_{qs} w_{sp} X_{(pqs)} , \;\; p,q \leq N
\end{equation}
where 
\begin{align}
\nonumber
X_{(pqs)} =& -(\omega - \epsilon_p - \epsilon_q + \epsilon_s)^{-1} 
 \{(\omega-\epsilon_p)^{-1}(\omega-\epsilon_q)^{-1} 
  +(\omega-\epsilon_q)^{-1}(\epsilon_s-\epsilon_q)^{-1}\\
\label{eq:comdiagx}  
&\phantom{xxxxx} + (\omega-\epsilon_p)^{-1}(\epsilon_s-\epsilon_p)^{-1} + (\epsilon_p-\epsilon_s)^{-1}(\epsilon_q-\epsilon_s)^{-1} \}
\end{align}
is the corresponding sum of the denominator products. Straightforward algebra yields
\begin{align*}
X_{(pqs)} &= -(\omega + \epsilon_s - \epsilon_p - \epsilon_q)^{-1}
 \{(\omega-\epsilon_p)^{-1} + (\epsilon_s-\epsilon_q)^{-1}\}\{(\omega-\epsilon_q)^{-1} + (\epsilon_s-\epsilon_p)^{-1}\}\\
 &= -\frac{\omega - \epsilon_p - \epsilon_q + \epsilon_s}{(\omega-\epsilon_p)(\omega-\epsilon_q)(\epsilon_s-\epsilon_p)(\epsilon_s-\epsilon_q)}
 \end{align*} 
Partitioning the numerator according to 
\begin{equation*}
\omega - \epsilon_p - \epsilon_q + \epsilon_s = \half (\omega-\epsilon_p) + \half (\omega-\epsilon_q) + \half (2 \epsilon_s - \epsilon_p - \epsilon_q)
\end{equation*}
shows that $X_{(pqs)}$ splits into three contributions,
\begin{equation*}
X_{(pqs)} = \frac{1}{\omega-\epsilon_p}\frac{1}{\omega-\epsilon_q}\,\frac{(\epsilon_p + \epsilon_q - 2 \epsilon_s)}{2 (\epsilon_s-\epsilon_p)(\epsilon_s-\epsilon_q)} \\
- \left (\frac{1}{\omega-\epsilon_p} +\frac{1}{\omega-\epsilon_q}\right) 
\frac{1}{2(\epsilon_s-\epsilon_p)(\epsilon_s-\epsilon_q)} 
\end{equation*}
which, together with the integrals on the right-hand side of Eq.~(\ref{eq:comdiag}),
now fit the term $(B)$ and the two parts of term $(A)$, respectively. The resulting expressions for the second-order matrix elements of $\bs C$ and $\bs f$ read
\begin{align}
\label{eq:C2pq}
C^{(2)}_{pq} &= \sum_{s > N} w_{qs} w_{sp}\, \frac{\epsilon_p + \epsilon_q - 2 \epsilon_s}{2 (\epsilon_s-\epsilon_p)(\epsilon_s-\epsilon_q)}\, n_p n_q \\
\label{eq:f2pq}
f^{(2)}_{pq} & = -\sum_{s > N} \frac{w_{qs} w_{sp}}{2(\epsilon_s-\epsilon_p)(\epsilon_s-\epsilon_q)}\, n_p n_q 
\end{align}
The expressions for the $hp$ components of 
$\bs f^{(2)}$ derive from diagrams (10) and (12) (or likewise (9) and (11)):
\begin{align}
\label{eq:f2pqx}
f^{(2,1)}_{pq} =  \frac{1}{\epsilon_q -\epsilon_p} \sum_{v \leq N} \frac{w_{vp} w_{qv}}
{\epsilon_v -\epsilon_q}\, n_p \lo n_q \\
\label{eq:f2pqy}
f^{(2,2)}_{pq} =  \frac{1}{\epsilon_q -\epsilon_p}  \sum_{r >N} \frac{w_{rp} w_{qr}}
{\epsilon_r -\epsilon_p}\, n_p \lo n_q 
\end{align}

\begin{figure}
\begin{equation*}
\vcpdf{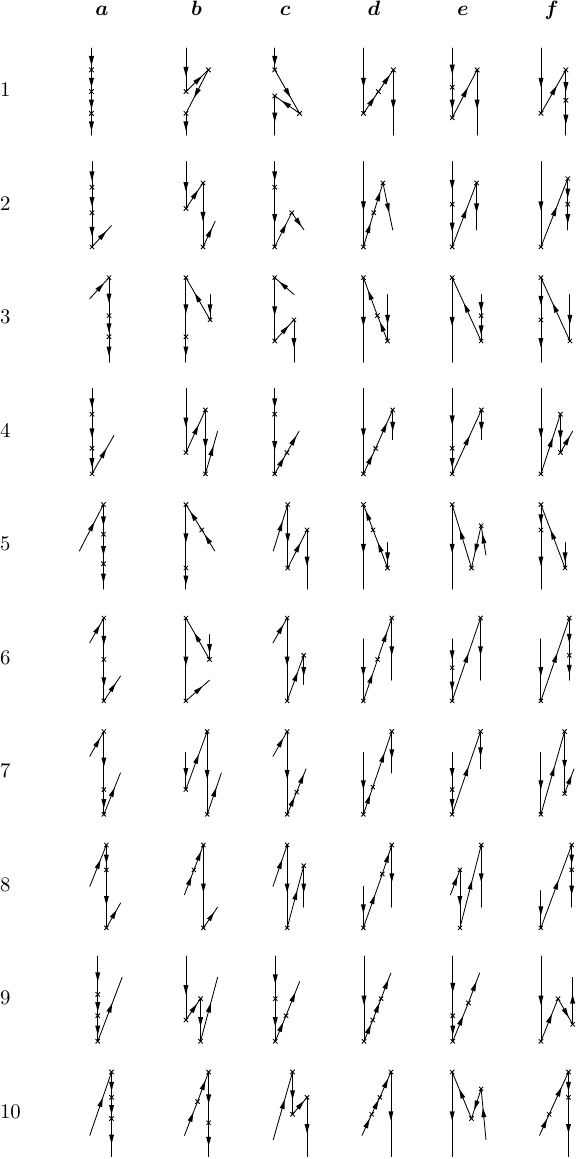}
\end{equation*}
\caption{Time-orderings of the third-order one-particle diagram contributing to $\bs G^-$}
\label{fig:1pdo3}
\end{figure}

In \textbf{third order} there are already 60 diagrams contributing to $\bs G^-(\omega)$ as shown in Fig.~\ref{fig:1pdo3}. Nevertheless, the 
derivation of the ADC terms $\bs C^{(3)}$ and $\bs f^{(3)}$ can still be accomplished without undue 
effort. A brief outline of the derivation and a compilation of the resulting expressions for $C^{(3)}_{pq}$ and $f^{(3)}_{pq}$
are given in App.~A, where also the uniqueness of the ADC procedure is addressed.

Complementary to the ADC procedure, there is a direct construction procedure for the intermediate states and the corresponding ADC secular matrices $\bs M$ and $\bs f$, which will be treated in the following section.

\section{Intermediate-state representation (ISR)}

The ADC procedure for the electron propagator part $\bs G^{-}(\omega)$
considered in the preceding section allows one to construct successively an, in principle, exact 
secular matrix, $\bs M$, interpreted as the representation of the (shifted) hamiltonian, 
$\hat H - E^N_0$, in terms of ``intermediate'' ($N\!-\!1$)-particle states.
In fact, these - so far hypothetical - intermediate states and the consequent representation (ISR) of $\hat H - E^N_0$ 
can explicitly be constructed via a general procedure~\cite{sch91:4647,mer96:2140,Schirm:2018} to be discussed below.
The ISR formulation opens an alternative route to deriving the PT expansions of the elements of the
secular matrices $\bs M$ and $\bs f$. 
Moreover, in the case of non-interacting particles the ISR concept leads to   
explicit closed-form expressions 
for the ADC/ISR secular matrices in terms of eigenvectors and eigenvalues of the 
underlying one-particle secular equations~(\ref{eq:seceq}), which, in turn, establishes a rigorous connection to
QDPT. 

\subsection{General aspects}
 
In the following, we briefly review the general ISR procedure and then consider its specialization to the case of non-interacting particles. The starting point for the construction of the intermediate states 
are the so-called correlated excited (CE) states. For the lowest class of states, that is, the $1h$ states, the CE 
precursors read
\begin{equation}
\label{eq:coexstat}
\ket{\Psi_{k}^{0}}=c_k \ket{\Psi_0^N}, \;\; k= 1, \dots, N
\end{equation}
from which the $1h$ intermediate states are obtained according to
\begin{equation}
\label{eq:isconstr2}
\ket{\tilde{\Psi}_k}=\sum_{l} \ket{\Psi^{0}_l} (\bs S^{-1/2})_{lk}
\end{equation}
as the result of symmetrical orthonormalization of the $1h$ CE states. Here 
\begin{equation}
\label{eq:cesolm}
S_{ij} = \braket{\Psi_{i}^{0}}{\Psi_{j}^{0}} = \dirint{\Psi_0^N}{c_i^\dagger c_j}{\Psi_0^N}, \;\; i,j \leq N 
\end{equation}
are the matrix elements of the CE state overlap matrix $\bs S$, which also can be seen as 
the $h/h$-block of the one-particle density matrix, $\bs S = \bs \rho_{hh}$  

The states of the next higher class, that is, the $2h$-$1p$ states, are constructed by a two-step procedure. First, precursor states are formed by Gram-Schmidt (GS) orthogonalization of the $2h$-$1p$ CE states, $\ket{\Psi^{0}_{aij}}=c_a^\dagger c_i c_j \ket{\Psi_0^N}$, with regard to the already constructed $1h$ intermediate states:  
\begin{equation}
\label{eq:isconstr1}
\ket{\Psi^{\#}_{aij}}=c_a^\dagger c_i c_j \ket{\Psi_0^N}- \sum_{k}\ket{\tilde{\Psi}_k}
\dirint{\tilde{\Psi}_k}{c_a^\dagger c_i c_j}{\Psi_0^{N}}
\end{equation}
In the second step, again symmetrical orthonormalization is applied to the $2h$-$1p$ precursor states, yielding the  
$2h$-$1p$ intermediate states, $\ket{\tilde{\Psi}_{aij}}$.
In an analogous way, the intermediate states of higher excitation classes,
$3h$-$2p$,$4h$-$3p, \dots$, can successively be constructed. 

It should be noted that the GS orthogonalization of a given class of CE states to all states of lower excitation classes 
is of crucial importance. It lies at the core of the so-called canonical order relations~\cite{mer96:2140,Schirm:2018} for the ISR secular matrix elements:
\begin{equation}
\label{eq:cor}
M_{IJ} = \dirint{\tilde{\Psi}_I}{\hat H - E^N_0}{\tilde{\Psi}_J} \sim O(|[I] - [J]|)
\end{equation}
This means that the PT order of the matrix elements $M_{IJ}$ is given by the absolute difference (``distance'') between the respective excitation classes
denoted by $[I]$ and $[J]$. For example, the matrix element $M_{k,abijl}$ for a $1h$ state (class 1) and a $3h$-$2p$ state (class 3) is of the order 2.

\begin{figure}
\begin{equation*}
\vcenterpdf{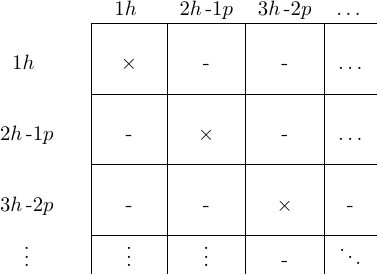}
\end{equation*}
\caption{Diagonal block structure of the full ISR secular matrix $\lo{\bs M}$ for non-interacting particles.}
\label{fig:corfigx}
\end{figure}

\begin{figure}
\begin{equation*}
\vcenterpdf{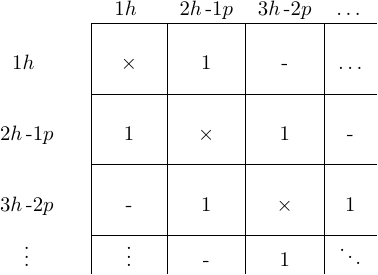}
\end{equation*}
\caption{Order structure of the CI secular matrix $\bs H$ for non-interacting particles.}
\label{fig:ciorfigx}
\end{figure}

\begin{figure}
\begin{equation*}
\vcenterpdf{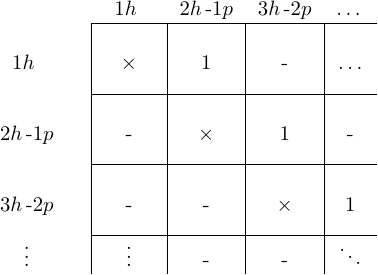}
\end{equation*}
\caption{Order structure of the biorthogonal coupled-cluster (BCC) secular matrix $\bs M^{bcc}$ for non-interacting particles.}
\label{fig:bccorfigx}
\end{figure}

The general ISR construction can easily be applied to the case of non-interacting particles with the Hamiltonian~(\ref{eq:Hmb}), yielding a basis of ($N\!-\!1$)-particle states in terms of the intermediate states, 
$\ket{\tilde{\Psi}_k}; \ket{\tilde{\Psi}_{akl}}, k<l; \dots $ and a corresponding ISR
secular matrix, $\lo{\bs M}$, where the bar is used to distinguish the
full ISR secular matrix from its $1h/1h$-block $\bs M \equiv \lo{\bs M}_{11}$ introduced in Eq.~(\ref{eq:issm}).   
Surpassing the canonical order relations~(\ref{eq:cor}) in the general case, here the secular matrix elements of the ISR secular matrix vanish for states 
of different classes:
\begin{equation}
\label{eq:corx}
\lo{M}_{IJ} = \dirint{\tilde{\Psi}_I}{\hat H - E^N_0}{\tilde{\Psi}_J}  = 0 \;\; \text{if} \;\; [I] \neq [J]
\end{equation}
that is, each excitation class is strictly decoupled from any other excitation class.
The resulting block structure of $\lo{\bs M}$ is depicted in Fig.~\ref{fig:corfigx}. 
The strict decoupling of distinct excitation classes is by no means trivial, as a look at the corresponding CI secular matrix in Fig.~\ref{fig:ciorfigx} shows. 
Here each excitation class features a first-order coupling to the next higher or next lower class. 
A proof of the strict decoupling of the ISR secular blocks in the case of non-interacting particles is given in App.~B, where we also 
consider the biorthogonal coupled-cluster (BCC) representation of a NIP system. The BCC secular matrix, shown in Fig.~\ref{fig:bccorfigx},
can be seen as a hybrid of the ISR case (decoupling) in the lower left part and CI (coupling of adjacent classes) in the upper right part.

The decoupling of the $1h$ intermediate states from those of higher excitations classes means
that the exact $1h$ states, $\ket{\Psi_m^{N-1}} = \tilde c_m \ket{\Psi_0^N},\, m = 1, \dots, N$, are given as 
linear expansions of the $1h$ intermediate states, $\ket{\tilde{\Psi}_j},\, j = 1, \dots, N$, 
which validates the assumptions underlying Eqs.~(\ref{eq:adceveq}, \ref{eq:adceveqx}).  
Moreover, the use of the restricted ROI~(\ref{eq:pROI}) is legitimate as the matrix elements
$\dirint{\tilde{\Psi}_J}{c_p}{\Psi_0^N}$ vanish unless $J$ refers to a $1h$ configuration, $[J] = 1$.
This can be seen by realizing that, according to Eqs.~(\ref{eq:ctilde},\ref{eq:Psi0}), even for $p > N$, the state $c_p \ket{\Psi_0^N}$ is a linear combination 
of exact $1h$ eigenstates, $\ket{\Psi_m^{N-1}},\, m \leq N$, and, thus, also a linear combination of 
$1h$ intermediate states, $\ket{\tilde{\Psi}_j},\, j \leq N$. By construction, the intermediate states of 
higher classes, $[J]>1$, are orthogonal to the $1h$ intermediate states.

Accordingly, we will confine us in the following to the $1h$ part of the secular problem, that is,
the $1h/1h$ block $\bs M$ of the ISR secular matrix according to Eq.~(\ref{eq:issm}), and the transition amplitudes, as 
given by Eq.~(\ref{eq:istm}).

\subsection{Explicit construction of the ISR secular matrix} 

The PT expansions of the intermediate states, 
\begin{equation}
\ket{\tilde \Psi_k} = \ket{\tilde \Psi_k^{(0)}} + \ket{\tilde \Psi_k^{(1)}} + \ket{\tilde \Psi_k^{(2)}} +
\dots
\end{equation}
based on the PT expansion of the $N$-particle ground state 
$\ket{\Psi^N_0}$,
can be translated into PT expansions of the ISR secular matrix elements,
\begin{equation}
\label{eq:isrsmptx}
M_{kl} = M^{(0)}_{kl} + M^{(1)}_{kl} + M^{(2)}_{kl} + \dots
\end{equation}
where the PT expansion of the ground-state energy $E^N_0 = E^{(0)}_0 + E^{(1)}_0 + \dots$
comes into play as well.

In \textbf{zeroth-order} the IR states are given by
\begin{equation}
\ket{\Phi_k} = c_k \ket{\Phi^N_0}
\end{equation}
and the ISR secular matrix elements simply read
\begin{equation}
M^{(0)}_{kl} = \dirint{\Phi_k}{\hat H_0 - E_0^{(0)}}{\Phi_l} = - \epsilon_k \,\delta_{kl}
\end{equation}

In \textbf{first-order} the intermediate states are just given by the corresponding 
first-order CE states,
\begin{equation*}
\ket{\tilde \Psi_k^{(1)}} = c_k \ket{\Psi_0^{(1)}}
\end{equation*}
as the orthonormalization of the CE (or precursor) states does not come into play before second order.
Here, the first-oder ground-state contribution is given by
\begin{equation}
\label{eq:gso1}
\ket{\Psi_0^{(1)}} = \sum t_{ai}^{(1)} c_a^\dagger c_i \ket{\Phi^N_0}
\end{equation}
where 
\begin{equation}
\label{eq:gso1x}
t_{ai}^{(1)} = - \frac{w_{ai}}{\epsilon_a - \epsilon_i} 
\end{equation}
are the first-order $1p$-$1h$ expansion coefficients.
Formally, there are three first-order contributions to the matrix elements $M_{kl}$,
\begin{equation*}
 M^{(1)}_{kl} = \dirint{\Phi_k}{\hat W - E^{(1)}_0}{\Phi_{l}}
 + \dirint{\tilde \Psi_k^{(1)}}{\hat H_0 - E_0^{(0)}}{\Phi_l} + 
\dirint{\Phi_k}{\hat H_0 - E_0^{(0)}}{\tilde \Psi_l^{(1)}} 
\end{equation*}
However, the last two terms, involving the zeroth-order part of the (subtracted) Hamiltonian, vanish, and the  
first term gives
\begin{equation}
M^{(1)}_{kl} = \dirint{\Phi_k}{\hat W - E^{(1)}_0}{\Phi_{l}} = - w_{lk}
\end{equation}
The evaluation of the \textbf{second-order} contributions to the secular matrix elements, being 
somewhat more demanding, is presented in App.~C. The resulting expressions read
\begin{equation}
M^{(2)}_{kl} =  \sum_a w_{ak} w_{la} 
\frac{2 \epsilon_a - \epsilon_k - \epsilon_l}{2(\epsilon_a - \epsilon_k)(\epsilon_a - \epsilon_l)} 
\end{equation}
Noting that $\bs M = - \bs K - \bs C$, the equivalence between the (first and second order) ISR results 
and the original ADC expressions~(\ref{eq:C1pq}) and~(\ref{eq:C2pq}), respectively, is readily established.

In a similar way, PT expansions for the ISR transition amplitudes~(\ref{eq:istm}),
\begin{equation}
\label{eq:isrsmpty}
 f_{kp} = \dirint{\tilde \Psi_k}{c_p}{\Psi^N_0} =
 f^{(0)}_{kp} +  f^{(1)}_{kp} + f^{(2)}_{kp} + \dots
\end{equation}
can be derived. Here, the first- and second-order results are as given by 
Eqs.~(\ref{eq:f11}), (\ref{eq:f2pq}),(\ref{eq:f2pqx}), 
and (\ref{eq:f2pqy}).

\subsection{Closed-form expressions for the ISR secular matrix and transition moments}

The ISR formulation for non-interacting particles outlined above
allows one to establish explicit
closed-form expressions for the secular matrices $\bs M$ and $\bs f$ in terms of the 
eigenvalues and eigenvectors of the underlying one-particle Schr\"{o}dinger equation~(\ref{eq:seceq}).
As these expressions presuppose - at least formally - the full (or partial) solution of the one-particle secular 
equations, they may be designated \textbf{\textit{a posteriori}} expressions. 
Besides enabling some deeper theoretical understanding of the ADC/ISR developments, these \textit{a posteriori} 
closed-form expressions establish a rigorous connection to their QDPT counterparts.

Let us begin with the relations (see Eqs.\ref{eq:ctilde},\ref{eq:Psi0})
\begin{equation}
c_j \ket{\Psi_0^N}= \sum_{k \leq N} X_{jk} \tilde{c}_k \ket{\Psi_0^N}, \;\; j \leq N
\end{equation}
between the exact $1h$ states $\ket{\Psi_k^{N-1}} = \tilde{c}_k \ket{\Psi_0^N}$ and the CE states
$c_j \ket{\Psi_0^N}$, where $X_{jk}$ denote the eigenvector components according to Eq.~(\ref{eq:seceq}).  
These relations can be cast into the compact form, 
\begin{equation}
\label{eq:ceex}
\dul{\Psi}^0 = \dul{\Psi} \bs X_{11}^t
\end{equation}
where $\dul{\Psi}^0$ and $\dul{\Psi}$ denote row 'vectors' defined according to
\begin{align}
&\dul{\Psi}^0 = \left ( c_1 \ket{\Psi_0^N}, \dots, c_N \ket{\Psi_0^N} \right ) \\
&\dul{\Psi} \phantom{x} = \left ( \tilde c_1 \ket{\Psi_0^N}, \dots, \tilde c_N \ket{\Psi_0^N} \right )
\end{align}
and $\bs X_{11}$ is the upper left block of the eigenvector matrix $\bs X$ in the partitioning 
according to Eq.~(\ref{eq:part}).

In a similar way, the $1h$ overlap matrix~(\ref{eq:cesolm}) can be written as
\begin{equation}
\label{eq:1hol}
\bs S = (\dul{\Psi}^{0 \textstyle{\dagger}}, \dul{\Psi}^0 ) 
\end{equation}
where 
\begin{equation}
\dul{\Psi}^{0 \textstyle{\dagger}} =
\left(
\begin{array}{c}
 \bra{\Psi_0^N}c_1^\dagger\\
 \vdots \\
 \bra{\Psi_0^N}c_N^\dagger
\end{array}
\right)
\end{equation}
is the hermitian conjugate of $\dul{\Psi}^0$
and $( , )$ means matrix multiplication (here column times row) plus forming the respective scalar products, that is,
\begin{equation}
(\dul{\Psi}^{0 \textstyle{\dagger}}, \dul{\Psi}^0)_{ij} = \dirint{\Psi_0^N}{c_i^\dagger c_j}{\Psi_0^N}
\end{equation}
Using the relations~(\ref{eq:ceex}) and considering that the exact states are orthonormal, 
Eq.~(\ref{eq:1hol}) can readily be evaluated further, yielding 
\begin{equation}
\label{eq:sxx}
\bs S = \bs X_{11}^*\bs X_{11}^t
\end{equation}
Note that the transpose $\bs S^t$ assumes a less awkward form,
\begin{equation}
\bs S^t = \bs X_{11}\bs X_{11}^\dagger
\end{equation}

Now we may introduce the row 'vector' of $1h$ intermediate states,
\begin{equation}
\dul{\tilde \Psi} = \left (\ket{\tilde \Psi_1}, \dots, \ket{\tilde \Psi_N} \right )
\end{equation}
so that the ISR construction of the intermediate states from the CE states (Eq.~\ref{eq:coexstat})
can compactly be written as
\begin{equation}
\label{eq:is2ce}
\dul{\tilde \Psi} = \dul{\Psi}^0 \,\bs S^{-1/2}
\end{equation}
Using Eq.~(\ref{eq:ceex}) one obtains the relation
\begin{equation}
\label{eq:is2ex}
\dul{\tilde \Psi} = \dul{\Psi}\, \bs X_{11}^t \,\bs S^{-1/2}
\end{equation}
between the intermediate and the exact $1h$ states.
On the other hand, the exact $1h$ states are obtained from the intermediate states according to 
\begin{equation}
\label{eq:ex2is}
\dul{\Psi} = \dul{\tilde \Psi} \,\bs Y
\end{equation}
as the result of the ISR secular equations~(\ref{eq:adceveq},\ref{eq:adceveqx}). 
The comparison of the latter two equations establishes the closed-form expression
\begin{equation}
\label{eq:Ycfr}
\bs Y^\dagger = \bs X_{11}^t\, \bs S^{-1/2}
\end{equation}
for the ISR eigenvector matrix $\bs Y$, where $\bs S$ is given by Eq.~(\ref{eq:sxx}).

Next we turn to the ISR secular matrix $\bs M$ and write the defining equations~(\ref{eq:issm}) 
in the now familiar compact form
\begin{equation}
\bs M =  (\dul{\tilde \Psi}^{\textstyle{\dagger}}, (\hat H - E^N_0)\dul{\tilde \Psi} )
\end{equation}
Using Eq.~(\ref{eq:is2ex}) to replace $\dul{\tilde \Psi}$ with the exact $1h$ states yields
\begin{equation}
\bs M = \bs S^{-1/2} \bs X_{11}^* ( \dul{\Psi}^{\textstyle{\dagger}},(\hat H - E^N_0) \dul{\Psi})\bs X_{11}^t \,\bs S^{-1/2}
\end{equation}
Obviously, the representation of $\hat H - E^N_0$ in terms of the exact $1h$ states leads to the diagonal matrix 
$-\bf E_1$ of the (negative) energies, $-e_1, \dots, -e_N$ (see Eq.~\ref{eq:deten}).
Thus, the final result reads 
\begin{equation}
\label{eq:cf4m}
\bs M = \bs S^{-1/2} \bs X_{11}^* (-\bs E_1) \bs X_{11}^t\, \bs S^{-1/2}
\end{equation}

In a similar way we may deal with the 
effective transition amplitudes. The $hh$-block of $\bs f$ of the elements $f_{kp},\, p\leq N$, 
can be written as
\begin{equation}
\label{eq:f11x}
\bs f_{11} = (\dul{\tilde \Psi}^{\textstyle{\dagger}}, \dul{\Psi}^0  ) 
= \bs S^{-1/2} (\dul{\Psi}^{0 \textstyle{\dagger}},\dul{\Psi}^0 ) = \bs S^{1/2}
\end{equation}
where Eqs.~(\ref{eq:is2ce}) and (\ref{eq:1hol}) have been used in the first and second equation, respectively.
As this result shows, $\bs f_{11}$ is a hermitian matrix. 
The $\bs f_{12}$ block of components $f_{ka},\, a > N$ takes on the form
\begin{equation}
 \bs f_{12} =  (\dul{\tilde \Psi}^{\textstyle{\dagger}}, \dul{\Psi}^{\bs x} ) 
 \end{equation} 
where 
\begin{equation}
\dul{\Psi}^{\bs x} = \left(c_{N+1} \ket{\Psi_0^N}, c_{N+2} \ket{\Psi_0^N}, \dots \right) 
\end{equation}
denotes the row 'vector' of extended CE states, $c_a \ket{\Psi_0^N}, \, a > N$.
Expressing the  intermediate states in terms of the $1h$ CE states via 
Eq.~(\ref{eq:is2ce}) yields   
\begin{equation} 
\label{eq:f12x}
\bs f_{12} = \bs S^{-1/2} (\dul{\Psi}^{0 \textstyle{\dagger}}, \dul{\Psi}^{\bs x} ) = 
\bs S^{-1/2} \bs X_{11}^* \bs X_{21}^t
\end{equation}

So far we have dealt with the ADC/ISR form~(\ref{eq:gmadcf}) for the transpose of $\bs G^-(\omega)$,
\begin{equation*}
\tilde{\bs G}(\omega) = \bs{f}^{\dagger} (\omega + \bs M)^{-1} \bs{f}
\end{equation*}
being better adapted to the treatment of the (interacting or non-interacting) ($N\!-\!1$)-particle system. 
The transition to the proper $\bs G^-(\omega)$ part and, thus, to the $\bs R^I(\omega)$ part of the one-particle resolvent matrix is, of course, straightforward:
\begin{equation}
\label{eq:cfexp}
\bs G^-(\omega) = \bs R^I(\omega) = \bs{f}^t (\omega + \bs M^t)^{-1} \bs{f}^*
\end{equation} 
According to Eq.~(\ref{eq:cf4m}), the transpose of $\bs M$ takes on the form
\begin{equation}
\label{eq:cf4mt}
\bs M^t = - (\bs S^t)^{-1/2} \bs X_{11} \bs E_1 \bs X_{11}^\dagger \, (\bs S^t)^{-1/2}
\end{equation}
where $\bs S^t$ is given by 
\begin{equation}
\label{eq:Stran}
\bs S^t = \bs X_{11}\bs X_{11}^\dagger 
\end{equation}

At this point, it is expedient to make some slight notational adjustments. Let us 
replace $\bs M^t$ and $\bs f^t$ according to 
\begin{align}
\label{eq:snotad1}
\tilde{\bs M} &= - \bs M^t \\
\label{eq:snotad2}
\tilde{\bs f} &= \bs f^t
\end{align}
by the quantities $\tilde{\bs M}$ and $\tilde{\bs f}$ so that Eq.~(\ref{eq:cfexp})
reads
\begin{equation}
\label{eq:cfexpx}
\bs{G}^{-}(\omega) = \bs R^I(\omega) = \tilde{\bs f}(\omega \bs 1 - \tilde{\bs M})^{-1} \tilde{\bs f}^\dagger
\end{equation}
In the new notation, the secular equations~(\ref{eq:adceveq}) take on the form 
\begin{equation}
\label{eq:setilde}
\tilde{\bs M} \tilde{\bs Y} = \tilde{\bs Y} \bs E_1, \;\; \tilde{\bs Y}^\dagger \tilde{\bs Y}= \bs 1
\end{equation}
where the new eigenvector matrix is related to the old one according to $\tilde{\bs Y} = \bs Y^*$.
The relation~(\ref{eq:xfY}) for the spectroscopic amplitudes becomes
\begin{equation}
\label{eq:xftY}
\bs x = \tilde{\bs f} \tilde{\bs Y}
\end{equation}
and the \textit{a posteriori} expression~(\ref{eq:Ycfr}) for $\tilde{\bs Y}$ takes on the form
\begin{equation}
\label{eq:Ycfrx}
\tilde{\bs Y} = (\bs X_{11}\bs X_{11}^\dagger )^{-1/2}\bs X_{11}
\end{equation}

Let us recapitulate: The secular matrix $\tilde{\bs M}$ was introduced as the representation of the NIP hamiltonian~(\ref{eq:Hmb})
in terms of intermediate ($N\!-\!1$)-particle states. The diagonalization of $\tilde{\bs M}$ provides the
energies and spectroscopic amplitudes associated with the ($N\!-\!1$)-particle states of the NIP system, as reflected by the spectral
representation of $\bs{G}^{-}(\omega)$ according to Eq.~(\ref{eq:cfexpx}). 
On the other hand, the latter equation 
also applies to $\bs R^I(\omega)$, that is, the part of the one-particle resolvent matrix associated with the partitioning~(\ref{eq:set12},\ref{eq:set12x}) of the one-particle secular problem. This means that $\tilde{\bs M}$ can just as well be seen 
as the representation of an effective one-particle hamiltonian, $\hat h_{eff}$, in terms of the "model" states
$\phi_k,\, k \leq N$ (Eq.~\ref{eq:h0phi}). Here, the energies obtained by diagonalization of $\tilde{\bs M}$ can directly be interpreted as the energies $e_n,\, n\leq N$ of the first $N$ exact one-particle states $\psi_n$ (Eq.~\ref{eq:hpsi}). 
And Eq.~(\ref{eq:xftY}) provides access to the full eigenvectors, $\ul X_n, \, n \leq N$, in the original representation.
This establishes a definite 
equivalence between the ADC/ISR concept at the NIP level and a corresponding
QDPT scheme at the underlying one-particle level. Moreover, the explicit PT expansions for the matrix elements of 
$\tilde{\bs M}$ and $\tilde{\bs f}$ obtained via the ADC or ISR procedures can directly be transferred to the QDPT approach.

As a rigorous confirmation of this equivalence,  
the \textit{a posteriori} expression for the ADC/ISR matrix $\tilde{\bs M}$, deriving from  
Eqs.~(\ref{eq:cf4mt}), (\ref{eq:Stran}), and (\ref{eq:snotad1}), 
\begin{equation}
\label{eq:cf4mtxx}
\tilde{\bs M}  = (\bs X_{11}\bs X_{11}^\dagger)^{-1/2} \bs X_{11} \bs E_1 \bs X_{11}^\dagger \, (\bs X_{11}\bs X_{11}^\dagger)^{-1/2}
\end{equation}
can be obtained in a completely independent way within the QDPT context, 
being here associated with the
secular matrix of the hermitian effective QDPT hamiltonian (see Eq.~\ref{eq:cfheff}).
This will be addressed in the ensuing Sec.~V.  

The equivalence of NIP-ISR and QDPT comprises the matrix $\tilde{\bs f}$ too.
Combining the separate expressions~(\ref{eq:f11x},\ref{eq:f12x})    
within a single matrix, yields 
\begin{equation}
\label{eq:cf4ft}
\tilde{\bs f} =
\left ( \begin{array}{c} 
\bs{f}^t_{11}\\
\bs{f}^t_{12}
\end{array}
\right)
= 
\left ( \begin{array}{c} 
(\bs X_{11}\bs X_{11}^\dagger)^{1/2}\\
\bs X_{21} \bs X_{11}^\dagger (\bs X_{11}\bs X_{11}^\dagger)^{-1/2}
\end{array}
\right)
\end{equation}
This is exactly the matrix of "effective eigenvector amplitudes"(EEA) displayed in Eq.~(\ref{eq:cffeff}), which in the QDPT formulation allows one 
to generate the full eigenvectors of the "target states" via an equation analogous to Eq.~(\ref{eq:xfY}).
It should be noted that orthonormalization of the partial eigenvector matrix $\bs x$ defined by Eq.~(\ref{eq:X1}) is reflected by a corresponding relation 
\begin{equation}
\tilde{\bs f}^\dagger \tilde{\bs f} = \bs 1
\end{equation} 
for the ETA matrix $\tilde{\bs f}$. This can be verified by using here the \textit{a posteriori}   
expression~(\ref{eq:cf4ft}) together with the orthonormalization of the original eigenvectors.

With regard to the uniqueness of the effective quantities it is instructive to inspect the 
$hh$ block of Eq.~(\ref{eq:xftY}): 
\begin{equation}
\label{eq:XfY}
\bs X_{11} = \tilde{\bs f}_{11} \tilde{\bs Y}
\end{equation}
Together with the explicit expressions for $\tilde{\bs f}_{11}$ and $\tilde{\bs Y}$ according to Eq.~(\ref{eq:cf4ft}) and 
Eq.~(\ref{eq:Ycfrx}), this equation represents the \textbf{\textit{polar decomposition}} of the (non-singular) matrix 
$\bs{\mathsf{X}}_{11}$ into the product
of the hermitian matrix $\tilde{\bs f}_{11}$ and the unitary matrix $\tilde{\bs Y}$. The polar decomposition is unique. This means that both $\tilde{\bs f}_{11}$ and $\tilde{\bs Y}$  uniquely determined by the $\bs X_{11}$ block  
of the eigenvector matrix $\bs X$, if $\tilde{\bs f}_{11}$ is required to be hermitian. This, in turn, implies that 
$\bs X_{11}$ and $\bs E_{1}$ uniquely determine the effective secular matrix, 
$\tilde{\bs M} = \tilde{\bs Y} \bs E_1 \tilde{\bs Y}^\dagger$; the $\tilde{\bs f}_{21}$ block is uniquely determined 
by $\bs X_{11}$ and $\bs X_{21}$.

The ISR construction of the secular matrices $\tilde{\bs M}$ and $\tilde{\bs f}$ and their \textit{a posteriori} expressions~(\ref{eq:cf4mtxx},\ref{eq:cf4ft}) presented here presupposes, strictly speaking, the particle concept, that is, they were derived assuming a one-particle system (and the concomitant NIP system) as introduced in Sec. II.A. 
However, the derivation and the results can readily be generalized as will be discussed 
in the next section.\\
\\

It should be mentioned that,
in an essentially analogous way, the ADC and ISR concepts can be applied to the ($N\!+\!1$)-particle part $\bs G^+(\omega)$. A few key points are as follows. 
Eq.~(\ref{eq:srmat+}) may be complemented according to  
\begin{equation}
\label{eq:srmat++}
 \bs R^{II}(\omega)= \bs G^+(\omega) = \bs y (\omega \bs 1 - \bs E_2)^{-1} \bs y^\dagger =
 \lo{\bs f}(\omega \bs 1 - \bs M^+)^{-1} \lo{\bs f}^\dagger
\end{equation}
with the ADC form for $\bs G^+(\omega)$ (note small notational changes with respect to Eq.~\ref{eq:gmadcf}).
Here, the ADC 
secular matrix, $\bs M^+$, is given as the representation of the (shifted) hamiltonian 
in terms of intermediate ($N\!+\!1$)-particle states $\ket{\tilde{\Psi}_a}, a > N$,
\begin{equation}
\label{eq:issm+}
M^+_{ab} = \dirint{\tilde{\Psi}_a}{\hat H - E^N_0}{\tilde{\Psi}_b}, \;\;a,b > N
\end{equation}
The matrix elements of $\lo{\bs f}$ are given by
\begin{equation}
\label{eq:istm+}
\lo f_{qa} = \dirint{\Psi^N_0}{c_q}{\tilde{\Psi}_a}, \;\; a > N
\end{equation}
PT expansions for $\bs M^+$ and $\lo{\bs f}$ can be established either by performing the ADC procedure for $\bs G^+(\omega)$, in analogy to Chapter~III and App.~A, or, alternatively, by adapting the 
ISR concept of Chap.~IV and App.~C to the case of ($N\!+\!1$)-particle states.
In analogy to Eq.~(\ref{eq:adceveq}), the ADC/ISR secular equations are given by
\begin{equation}
\label{eq:adceveq+}
  \bs M^+ \lo{\bs Y} = \lo{\bs Y} \bs E_2,\qquad \lo{\bs Y}^\dagger \lo{\bs Y} = \bs 1
\end{equation}
where $\bs E_2$ is the diagonal matrix of the one-particle energies $e_m, \, m > N$ 
(Eq. ~\ref{eq:hpsi}), or the negative particle attachment energies, $e_m = E^{N+1}_m - E^N_0$.
The original eigenvector matrix $\bs y$ can be recovered from the ADC/ISR eigenvector matrix $\lo{\bs Y}$ according to
\begin{equation}
\label{eq:yfY}
\bs y = \lo{\bs f}\, \lo{\bs Y}
\end{equation} 
Closed-form expressions for $\bs M^+$,
\begin{equation}
\label{eq:cf4mty}
\bs M^+  = (\bs X_{22}\bs X_{22}^\dagger)^{-1/2} \bs X_{22} \bs E_2 \bs X_{22}^\dagger \, (\bs X_{22}\bs X_{2}^\dagger)^{-1/2}
\end{equation}
 and $\lo{\bs f}$,
\begin{align}
\lo{\bs f}_{22} =& (\bs X_{22}\bs X_{22}^\dagger)^{1/2}\\
\lo{\bs f}_{12} =& \bs X_{12} \bs X_{22}^\dagger (\bs X_{22}\bs X_{22}^\dagger)^{-1/2}
\end{align}
can be obtained in a similar way as above.

\section{General QDPT and the NIP-ADC/ISR approach}

So far, the line of argumentation was based on the particle concept, supposing a system 
of non-interacting (yet fermionic) particles subject to an external potential.  
Here, an equivalence could be established between the NIP-ADC/ISR procedures for ($N\!-\!1$)-particle states and 
a corresponding QDPT scheme for the underlying one-particle system. The particle premise seems to restrict
the applicability of the NIP-ADC/ISR formulation of QDPT, that is, to systems being of one-particle type. 
However, as will be addressed in the following, the results of the NIP-ADC/ISR approach can readily be 
generalized to arbitrary quantum systems.


\subsection{Beyond the particle concept}

Like PT for the ground state, or the determination of eigenvalues and eigenfunctions, QDPT is essentially 
a mathematical-computational approach that can be applied regardless of the particular physics under consideration.  
Accordingly, we will now go beyond the one-particle system and consider a completely general physical system at the basic level. Here, no further specifications are needed, such as to the number of degrees of freedom, or the permutation symmetry in a possible composite system. The only requirement to be satisfied is that a PT-type partitioning of the 
hamiltonian can be established,
\begin{equation}
\label{eq:Hgen}
 \hat{\mathcal{H}}= \hat{\mathcal{H}}_0 + \hat{\mathcal{W}}
\end{equation}
where 
$\hat{\mathcal{H}}_0$ and $\hat{\mathcal{W}}$ are the unperturbed and perturbation part, respectively.
Note that here and in the following we use slightly modified notations (and/or typography) to distinguish the 
generalized expressions from those used within the preceding particle-based formulation.   
As before, we assume that the eigenvalue problem of $\hat{\mathcal{H}}_0$,  
\begin{equation}
\label{eq:evpupgen}
\hat{\mathcal{H}}_0 \Phi_\nu = \mathcal E_\nu \Phi_\nu, \; \nu = 1,2, \dots
\end{equation} 
has been solved, and the (energetically ordered) eigenfunctions $\Phi_\nu,\, \nu = 1,2, \dots$, also referred to as unperturbed states,
form an orthonormal basis set,
\begin{equation*}
\braket{\Phi_\mu}{\Phi_\nu} = \delta_{\mu \nu}
\end{equation*}
In the corresponding representation, 
the Schr\"{o}dinger equation for the full hamiltonian,
\begin{equation}
\label{eq:sefh}
\hat{\mathcal{H}} \Psi_n = E_n \Psi_n, \; n = 1,2, \dots
\end{equation} 
takes on the form
\begin{equation}
\label{eq:seceqx}
\bs{\mathcal{H}} \bs{\mathsf{X}} = \bs{\mathsf{X}} \bs{\mathsf{E}}, \; \bs{\mathsf{X}}^\dagger \bs{\mathsf{X}} = \bs 1
\end{equation}
Here, $\bs{\mathcal{H}}$ is the matrix representation of $\hat{\mathcal{H}}$ with respect to the unperturbed states, 
\begin{equation}
\label{eq:Hmunu}
\mathcal{H}_{\mu \nu} = \dirint{\Phi_\mu}{\hat{\mathcal{H}}}{\Phi_\nu} =
\mathcal E_\mu \delta_{\mu \nu} + W_{\mu \nu}
\end{equation}
$\bs{\mathsf{E}}$ denotes the diagonal matrix of eigenvalues, and
$\bs{\mathsf{X}}$ is the eigenvector matrix of elements 
\begin{equation}
\mathsf{X}_{\mu n} = \braket{\Phi_\mu}{\Psi_n}
\end{equation}
The expansions of the eigenfunctions take on the form
\begin{equation}
\Psi_n = \sum_\mu \mathsf X_{\mu n} \Phi_\mu
\end{equation}

In analogy to Eq.~(\ref{eq:rop}), a resolvent operator can be defined according to
\begin{equation}
\hat{\mathcal{R}}(\omega) = (\omega - \hat{\mathcal{H}})^{-1}
\end{equation}
which in matrix representation can be written as
\begin{equation}
\label{eq:ropmatx}
\bs{\mathsf{R}}(\omega) = (\omega \bs 1 - \bs{\mathcal{H}} )^{-1} = \bs{\mathsf{X}} 
(\omega \bs 1 - \bs{\mathsf{E}})^{-1} \bs{\mathsf{X}}^\dagger
\end{equation}

As in the particle case considered in Sec.II.A, a partitioning of the Hilbert space
is established by the unperturbed hamiltonian $\hat{\mathcal{H}}_0$ and a choice of the integer $N \geq 1$. 
According to 
\begin{align}
\nonumber
(\bs 1)& \equiv \{\Phi_\nu, \, \nu \leq N \}\\
\label{eq:set12gen} 
(\bs 2)& \equiv \{\Phi_\nu, \, \nu  > N \}
\end{align}
the unperturbed eigenstates are divided into two groups $(\bs 1)$ and $(\bs 2)$,
where group $(\bs 1)$ contains the $N$ energetically lowest states and group $(\bs 2)$
all other states. The states of group $(\bs 1)$, referred to as ``model states'', span the ``model space''.
The orthogonal complement of the model space is spanned by the states of group $(\bs 2)$.
An analogous partitioning is established with regard to the eigenstates of the full 
hamiltonian $\hat{\mathcal{H}}$. Here the two groups, given according to 
\begin{align}
\nonumber
(\bs 1)& \equiv \{\Psi_n, \, n \leq N \}\\
\label{eq:set12genx} 
(\bs 2)& \equiv \{\Psi_n, \, n > N \}
\end{align}
establish the ``target space'', being the counterpart of the model space, and the orthogonal complement
of the target space, respectively.

It should be clear that the practical usefulness of perturbation theory based on the partitioning schemes ($\hat{\mathcal{H}}_0, N$) and
($\hat{\mathcal{H}}, N$) depends on whether there is a sufficiently large separation of the model and non-model state energies, $\Delta \sim \mathcal E_{N+1} - \mathcal E_N$, compared with their mutual coupling strenghts, 
$\Gamma \sim |W_{\mu \nu|},\, \mu \leq N, \nu > N$.
Concerning the target states, it has to be assumed that 
there is a "PT genealogy" with respect to the model states, that is, the 
span of the model states evolves into the span of the target states when the interaction part of the hamiltonian is
ramped up via a scaling factor from 0 to 1 . Problems do arise if there are "intruder" states, that is, final states  
among the $N$ energetically lowest states lacking a PT descent from the model space~\cite{eva87:4930,Helgaker:2000}.

As in Eq.~(\ref{eq:part}), the partitioning of the states
entails a corresponding partitioning of the
eigenvector matrix $\bs{\mathsf{X}}$,  
\begin{equation}
\label{eq:genXpart}
\bs{\mathsf{X}} = 
\left(
\begin{array}{cc}
\bs{\mathsf{X}}_{11} & \bs{\mathsf{X}}_{12} \\
 \bs{\mathsf{X}}_{21}& \bs{\mathsf{X}}_{22}
\end{array}
\right)
\end{equation}
Here the collective indices $ 1, 2$ in the first position refer to the model and complementary model states,
respectively, while the second index distinguishes target and complementary target states.

Again, the resolvent matrix~(\ref{eq:ropmatx}) can be split into two parts,
\begin{equation}
\label{eq:ropmat12}
\bs{\mathsf{R}}(\omega) = \bs{\mathsf{R}}^I(\omega) + \bs{\mathsf{R}}^{II}(\omega)
\end{equation}
where the first part
\begin{equation}
\label{eq:srmatgen}
\bs{\mathsf{R}}^I(\omega) = \bs{\mathsf{x}} (\omega \bs 1 - \bs{\mathsf{E}}_1)^{-1} \bs{\mathsf{x}}^\dagger 
\end{equation}
is associated with the physical information about the target states.
Here, $\bs{\mathsf{x}}$ is the matrix formed by the first $N$ eigenvectors,
\begin{equation}
\label{eq:X1sf} 
\bs{\mathsf{x}}  = 
\left(
\begin{array}{c}
\bs{\mathsf{X}}_{11}\\
\bs{\mathsf{X}}_{21}  
\end{array}
\right)
\end{equation}
and $\bs{\mathsf{E}}_1$ is the diagonal matrix of the corresponding energies, $E_\nu, \, \nu \leq N$.
An analogous expression applies to $\bs{\mathsf{R}}^{II}(\omega)$ 
(see Eqs.~\ref{eq:X1},\ref{eq:srmat+}).

A unitary transformation takes us from the diagonal spectral form~(\ref{eq:srmatgen}) to the non-diagonal
form
\begin{equation}
\label{eq:cfexpxx}
 \bs{\mathsf{R}}^I(\omega) = \bs{\mathsf{x}} \tilde{\bs{\mathsf{Y}}}^\dagger \tilde{\bs{\mathsf{Y}}}(\omega \bs 1 - \bs{\mathsf{E}}_1)^{-1} \tilde{\bs{\mathsf{Y}}}^\dagger \tilde{\bs{\mathsf{Y}}}
  \bs{\mathsf{x}}^\dagger =
 \tilde{\bs{\mathsf{f}}}(\omega \bs 1 - \tilde{\bs{\mathsf{M}}})^{-1} 
 \tilde{\bs{\mathsf{f}}}^\dagger
\end{equation}
Here $\tilde{\bs{\mathsf{M}}}$ is a secular matrix which can be seen as the representation of an effective
hamiltonian $\hat{\mathcal{H}}_{eff}$ in terms of the $N$ model states, while $\tilde{\bs{\mathsf{f}}}$
denotes a matrix of effective eigenvector amplitudes. Obviously, the unitary transformation matrix, $\tilde{\bs{\mathsf{Y}}}$ ,
is identical with the eigenvector matrix of $\tilde{\bs{\mathsf{M}}}$. 
The associated 
secular equations (see Eq.~\ref{eq:setilde}), 
\begin{equation}
\tilde{\bs{\mathsf{M}}} \tilde{\bs{\mathsf{Y}}} = \tilde{\bs{\mathsf{Y}}}\bs{\mathsf{E}}_1, \;\; 
\tilde{\bs{\mathsf{Y}}}^\dagger \tilde{\bs{\mathsf{Y}}}= \bs 1
\end{equation}
reproduce (or provide) the target state energies,
and the relation (see Eq.~\ref{eq:XfY})
\begin{equation}
\label{eq:xxfY}
\bs{\mathsf{x}} = \tilde{\bs{\mathsf{f}}} \tilde{\bs{\mathsf{Y}}}
\end{equation}
allows one to generate the original target state eigenvectors from the eigenvectors of $\tilde{\bs{\mathsf{M}}}$.

But how can $\tilde{\bs{\mathsf{M}}}$ and $\tilde{\bs{\mathsf{f}}}$, or likewise $\tilde{\bs{\mathsf{Y}}}$, actually be determined, as we are no longer operating within the framework of the particle concept? 
In principle, one can resort to the general \textit{a posteriori} relations~(\ref{eq:cf4mtxx},\ref{eq:cf4ft}),  
adapted to the eigenvectors and eigenvalues of the general hamiltonian~(\ref{eq:Hgen}) according to the secular equations~(\ref{eq:seceqx}). This leads to the following expressions for $\tilde{\bs{\mathsf{M}}}$ and 
$\tilde{\bs{\mathsf{f}}}$:
\begin{align}
\label{eq:cf4mtxxx}
\tilde{\bs{\mathsf{M}}}  =& (\bs{\mathsf{X}}_{11} \bs{\mathsf{X}}_{11}^\dagger)^{-1/2} \bs{\mathsf{X}}_{11} 
\bs{\mathsf{E}}_1 \bs{\mathsf{X}}_{11}^\dagger \, (\bs{\mathsf{X}}_{11}\bs{\mathsf{X}}_{11}^\dagger)^{-1/2}
\\
\label{eq:cf4ftx}
\tilde{\bs{\mathsf{f}}} =&
\left ( \begin{array}{c} 
(\bs{\mathsf{X}}_{11}\bs{\mathsf{X}}_{11}^\dagger)^{1/2}\\
\bs{\mathsf{X}}_{21} \bs{\mathsf{X}}_{11}^\dagger (\bs{\mathsf{X}}_{11}\bs{\mathsf{X}}_{11}^\dagger)^{-1/2}
\end{array}
\right)
\end{align}
The associated expression for $\tilde{\bs{\mathsf{Y}}}$, reading
\begin{equation}
\label{eq:Ycfry}
\tilde{\bs{\mathsf{Y}}}= (\bs{\mathsf{X}}_{11}\bs{\mathsf{X}}_{11}^\dagger )^{-1/2}\bs{\mathsf{X}}_{11}
\end{equation}
is obtained from the adaption of Eq.~(\ref{eq:Ycfrx}). 

While the latter three equations might simply be postulated as analogues to 
Eqs.~(\ref{eq:cf4mtxx}), (\ref{eq:cf4ft}), and (\ref{eq:Ycfrx}), respectively, 
there is a rigorous derivation 
within the traditional formulation of QDPT (see Eqs.~\ref{eq:cfheff},\ref{eq:cffeff},\ref{eq:evmz}) as will be discussed in the following two sections.
The important implication here is that the PT expansions for the secular matrices derived within the NIP-ADC/ISR
approach can - \textit{mutatis mutandis} - directly be transferred to $\tilde{\bs{\mathsf{M}}}$ and $\tilde{\bs{\mathsf{f}}}$. This is the consequence of the fact that the PT expansions of the eigenvectors and eigenvalues 
of the one-particle hamiltonian~(\ref{eq:H}) and those of the general hamiltonian~(\ref{eq:Hgen}) are formally equivalent. 
In practice, this means that one just has to replace the unperturbed energies $\epsilon_p$ of $\hat h_0$ with the 
unperturbed energies $\mathcal E_\nu$ of $\hat{\mathcal{H}}_0$, and the interaction integrals $w_{pq}$ of $\hat w$ with 
the integrals $W_{\mu \nu}$ of the operator $\hat{\mathcal{W}}$ (Eq.~\ref{eq:Hmunu}).

There is yet an alternative direct way of dealing with general physical systems, namely by 
a formal generalization of the particle concept itself. Here we take the system underlying the
hamiltonian $\hat{\mathcal{H}}$ as constituting a "hyper-particle" (HP), and conceive an ensemble (super-system)
consisting of many such non-interacting hyper-particles supposed to behave as non-interacting fermions. 
(It should be noted that there is no "exchange interaction" for non-interacting particles.) 
 
Let us have a look at some specifics. For an ensemble of $N$ hyper-particles, the 
hamiltonian is given by the analogue to Eq.~(\ref{eq:nparth}),  
\begin{equation}
\hat{\bs{H}}= \sum_{j=1}^N \hat{\mathcal{H}}(j)
\end{equation}
where $\hat{\mathcal{H}}(j)$ is the operator~(\ref{eq:Hgen}) acting on the HP subsystem $j$.
The Slater determinant 
\begin{equation}
\ket{\bs \Phi_0^N} = |\Phi_1 \Phi_2 \dots \Phi_N|
\end{equation}
of the energetically lowest eigenstates of Eq.~(\ref{eq:evpupgen}), $\Phi_\mu, \, \mu = 1, \dots N$,
constitutes the "unperturbed" ground state of the system of $N$ hyper-particles. 
In a similar way, the exact HP ground state
is given by the Slater determinant
\begin{equation}
\ket{\bs \Psi_0^N} = |\Psi_1 \Psi_2 \dots \Psi_N|
\end{equation}
in terms of the energy eigenstates, $\Psi_n,\, n =1, \dots, N$, of the hamiltonian $\hat{\mathcal{H}}$,
where
\begin{equation}
E^N_0 = \sum_{n = 1}^N E_n
\end{equation}
is the corresponding energy.

Moreover, the formalism of second quantization can be extended to the case of hyper-particles  
(and the corresponding HP Fock space), where the field operators
$\mathsf{c}^\dagger_\mu$ and $\mathsf{c}_\mu$   
denote the creation and destruction operator, respectively, for a hyper-particle in  
the unperturbed state $\Phi_\mu$. Analogous to Eq.~(\ref{eq:Hmb}), this allows one to write the HP-ensemble hamiltonian independently of the respective
number of hyper-particles. 
In addition, one may introduce field operators $\tilde{\mathsf{c}}^\dagger_n, \tilde{\mathsf{c}}_n$ based on the eigenfunctions $\Psi_n$ of $\hat{\mathcal{H}}$. 
Analogous to Eq.~(\ref{eq:ex1h}), the exact "$1h$" states 
of the ($N\!-\!1$)-HP system can be written as
\begin{equation}
\ket{\bs \Psi^{N-1}_n} = \tilde{\mathsf{c}}_n \ket{\bs \Psi_0^N}, \;\, n \leq N  
\end{equation}
Eq.~(\ref{eq:ctilde}), establishing the relation between the two kinds of destruction operators,
applies in the form
\begin{equation}
\mathsf{c}_\mu = \sum_n \mathsf{X}_{\mu n} \tilde{\mathsf{c}}_n
\end{equation}
where of course the eigenvector components are those of Eq.~(\ref{eq:seceqx}).
In analogy to Eqs.~(\ref{eq:coexstat},\ref{eq:isconstr2},\ref{eq:cesolm}), HP intermediate states, $\ket{\tilde{\bs \Psi}_\mu}, \mu \leq N$,   
can be formed based on the HP CE states 
\begin{equation}
\ket{\bs{\Psi}^0_\mu} = \mathsf{c}_\mu \ket{\bs \Psi_0^N}, \;\, \mu \leq N  
\end{equation}
In fact, the entire formal apparatus outlined in the preceding sections, comprising
the electron propagator, its diagrammatic PT, the ADC procedure, and the ISR concept, can be transferred 
essentially one-to-one to the super-system of non-interacting hyper-particles. 
One may go through all the previous derivations and translate them into the hyper-particle language. 
This leads straightforwardly to the expressions~(\ref{eq:cf4mtxxx},\ref{eq:cf4ftx}), and, moreover,
to the direct NIP-ADC/ISR PT expansions for the QDPT matrices $\tilde{\bs{\mathsf{M}}}$ and $\tilde{\bs{\mathsf{f}}}$.

\subsection{Traditional formulation of quasi-degenerate perturbation theory}

As already pointed out, the traditional formulation of QDPT 
(see, e.g., Refs.\cite{Lindgren:1982,dur87:321})
begins with the partitioning
schemes~(\ref{eq:set12gen},\ref{eq:set12genx}) of the eigenstates of  $\hat{\mathcal{H}}_0$ and 
$\hat{\mathcal{H}}$, respectively. The unperturbed states of group $(\bs 1)$, 
are referred to as model states, spanning the "model subspace" of the Hilbert space.
One may define a model space projection operator according to 
\begin{equation}
\hat P_0 = \sum_{\nu = 1}^N \ket{\Phi_\nu} \bra{\Phi_\nu}
\end{equation}
and the complementary projector
\begin{equation}
\hat Q_0 = \hat 1 - \hat P_0
\end{equation}
The counterpart to the model space is the "target space", spanned by the
exact states of group $(\bs 1)$, that is, the $N$ energetically lowest states of the 
full hamiltonian $\hat{\mathcal{H}}$. As mentioned above, there should be a PT genealogy
relating the target space to the model space. 
In analogy to $\hat P_0$ and $\hat Q_0$, target space projectors are defined according to 
\begin{equation}
\hat P = \sum_{n = 1}^N \ket{\Psi_n} \bra{\Psi_n}
\end{equation}
and
\begin{equation}
\hat Q = \hat 1 - \hat P
\end{equation}
As a starting point for the further procedure, one introduces the states
\begin{equation}
\label{eq:pts}
\Psi^0_n = \hat P_0\, \Psi_n , \;\; n = 1, \dots ,N
\end{equation}
obtained by projecting the target states onto the model space. 
Obviously, the projected states are not orthonormal, so they may be supplemented 
with the set of biorthogonal states, $\lo \Psi^0_n, \, n = 1, \dots ,N$, 
satisfying the relations
\begin{equation}
\braket{\lo \Psi^0_n}{\Psi^0_m} = \delta_{n m}
\end{equation}

A central term is the \textbf{wave operator} $\hat \Omega$, which acts onto the model space 
and generates the exact target states according to
\begin{equation}
\label{eq:wop}
\Psi_n = \hat \Omega \Psi^0_n, \;\;n = 1, \dots ,N
\end{equation}
The action onto the complementary model space is trivial, namely,
$\hat \Omega \Psi^0_n = 0, \, n > N$.
That such an operator exists is seen by the explicit construction  
\begin{equation}
\label{eq:wopx}
\hat \Omega = \sum_{n = 1}^N \ket{\Psi_n} \bra{\lo \Psi^0_n}
\end{equation}
Using the explicit form, the following properties of the wave operator can easily be established: 
\begin{align}
\label{eq:wopp1}
&\hat \Omega \hat P_0 = \hat \Omega, \;\hat P_0 \hat \Omega = \hat P_0
 \\
\label{eq:wopp2} 
&\hat P \hat \Omega = \hat \Omega, \; \hat \Omega \hat P = \hat P \\
\label{eq:wopp3}
&\hat \Omega^2 = \hat \Omega 
\end{align}

To derive an \textbf{effective hamiltonian} for the model space, one
applies the projector $\hat P_0$ to both sides of the Schr\"{o}dinger equation~(\ref{eq:sefh}), 
\begin{equation}
\hat P_0 \hat{\mathcal{H}} \Psi_n = E_n \Psi^0_n, \;\;n = 1,\dots, N 
\end{equation}
Using Eq.~(\ref{eq:wop}) on the left-hand side one obtains
\begin{equation}
\label{eq:}
\hat P_0 \hat{\mathcal{H}} \hat \Omega \Psi^0_n =  E_n \Psi^0_n,  \;\;  n = 1,\dots,N
\end{equation}
This represents a subspace Schr\"{o}dinger equation for the exact energies $E_n, n \leq N$, where the subspace 
is the model space, featuring an effective hamiltonian, 
\begin{equation}
\label{eq:effh}
\hat{\mathcal{H}}^{nh}_{eff} = \hat P_0 \hat{\mathcal{H}} \hat \Omega 
\end{equation}
The desired energies are obtained by diagonalizing $\hat{\mathcal{H}}^{nh}_{eff}$ in the model space. It should be noted, though, that  
$\hat{\mathcal{H}}^{nh}_{eff}$ is not hermitian, as indicated by the superscript $nh$.

Alternatively, as suggested by writing the Schr\"{o}dinger equation~(\ref{eq:sefh}) in the form
\begin{equation*}
\hat{\mathcal{H}} \hat \Omega \Psi^0_n =  E_n \hat \Omega \Psi^0_n
\end{equation*}
one may consider the diagonalization of the original 
hamiltonian $\hat{\mathcal{H}}$ in the space spanned by the states  $\hat \Omega \Phi_\mu,\, \mu = 1, \dots, N$.
The latter states are not orthonormal,  
and one may proceed to (symmetrically) orthonormalized states, $\tilde \Phi_\mu$, according to
\begin{equation}
\label{eq:owos}
\tilde \Phi_\mu = \hat \Omega (\hat \Omega^\dagger \hat \Omega)^{-1/2}\Phi_\mu, \,\, \mu \leq N
\end{equation}
The operator
\begin{equation}
\label{eq:uwop}
\hat{\Omega}_S
= \hat \Omega (\hat \Omega^\dagger \hat \Omega)^{-1/2}
\end{equation}
is referred to as the symmetrized wave operator (SWO). 
As should be noted, the operator $\hat \Omega^\dagger \Omega$ is hermitian and positive definite,
so that the square root operator $(\hat \Omega^\dagger \Omega)^{-1/2}$ is well defined. 
The orthonormality of the SWO states, $\hat \Omega_S \Phi_\mu$, is easily verified:
\begin{equation}
\braket{\tilde \Phi_\mu}{\tilde \Phi_\nu} = 
\dirint{\Phi_\mu}{(\hat \Omega^\dagger \hat \Omega)^{-1/2} \hat \Omega^\dagger \hat \Omega (\hat \Omega^\dagger \hat \Omega)^{-1/2}}{\Phi_\nu} = \delta_{\mu\nu}
\end{equation}  
The symmetrized states $\tilde \Phi_\mu$ can be obtained as well by the usual orthonormalization procedure, 
\begin{equation}
\tilde \Phi_\mu = \sum_\kappa \hat \Omega \Phi_\kappa (\bs \sigma^{-1/2})_{\kappa \mu}
\end{equation}
where $\bs \sigma$ is the overlap matrix of elements
\begin{equation}
\sigma_{\mu \nu} = \dirint{\Phi_\mu}{\hat \Omega^\dagger \hat \Omega}{\Phi_\nu}, \;\; \mu,\nu \leq N
\end{equation}  
The matrix $\tilde{\bs{\mathcal H}}$ of the elements
\begin{equation}
\label{eq:sworep}
\tilde{\mathcal H}_{\mu \nu} = \dirint{\tilde \Phi_\mu}{\hat{\mathcal{H}}}{\tilde \Phi_\nu} =
\dirint{\Phi_\mu}{\hat \Omega_S^\dagger \,\hat{\mathcal{H}} \,\hat \Omega_S}{\Phi_\nu}, \;\; \mu,\nu \leq N 
\end{equation} 
can either be seen as a representation of the original hamiltonian $\hat{\mathcal{H}}$ in terms of the 
symmetrized wave operator (SWO) states, $\tilde \Phi_\mu$,
or likewise as the representation of a hermitian effective hamiltonian 
in terms of the model states, ${\Phi_\mu},\, \mu = 1,\dots, N$, where the hamiltonian is
given by
\begin{equation}
\label{eq:heffh}
\hat{\mathcal{H}}^{h}_{eff} =  \hat \Omega_S^\dagger \,\hat{\mathcal{H}}\,\hat \Omega_S  
\end{equation}
In either interpretation one obtains the common algebraic secular equations,
\begin{equation}
\tilde{\bs{\mathcal H}} \bs Z = \bs Z \bs{\mathsf{E}}_1, \;\; \bs Z^\dagger \bs Z = \bs 1   
\end{equation}   
where $\bs{\mathsf{E}}_1$ is the diagonal matrix of the $N$ target state energies, and 
$\bs Z$ is the corresponding eigenvector matrix. 

In agreement with the former interpretation of the secular problem, the target states can be expanded according to 
\begin{equation}
\label{eq:eeax}
\Psi_n = \sum_{\mu =1}^N Z_{\mu n} \tilde \Phi_\mu,\,\, n = 1, \dots, N
\end{equation}
in terms of the wave operator states. The components of the original eigenvectors (Eq.~\ref{eq:seceqx}) 
are obtained according to
\begin{equation}
\label{eq:eeay}
\mathsf X_{\kappa n} = \braket{\Phi_\kappa}{\Psi_n} = \sum_{\mu=1}^N Z_{\mu n} \braket{\Phi_\kappa}{\tilde \Phi_\mu},
\;\; n \leq N; \kappa = 1,2, \dots
\end{equation}
Introducing the matrix $\bs{\mathsf{f}}$ of "effective eigenvector amplitudes", or, likewise, SWO matrix elements,  
\begin{equation}
\label{eq:eeam}
\mathsf{f}_{\kappa \mu} = \braket{\Phi_\kappa}{\tilde \Phi_\mu}
= \dirint{\Phi_\kappa}{\hat \Omega_S}{\Phi_\mu},\,\,\kappa = 1,2,\dots ; \mu \leq N
\end{equation} 
the relations~(\ref{eq:eeay}) can compactly be written as
\begin{equation}
\label{eq:eeaz}
\bs{\mathsf{x}}  = \bs{\mathsf{f}} \bs Z
\end{equation}
where $\bs{\mathsf{x}}$ is the matrix~(\ref{eq:X1sf}) formed by the first $N$ eigenvectors. \\
\\

A perturbation theory (PT) approach to the wave operator $\hat \Omega$ and the non-hermitian effective hamiltonian $\hat{\mathcal{H}}^{nh}_{eff}$
can be established by means of the Bloch equation~\cite{blo58:329,lin74:2441},
\begin{equation}
\label{eq:bloch}
[\hat \Omega,\hat{\mathcal{H}}_0] = \hat{\mathcal{W}}\hat \Omega -\hat \Omega \hat{\mathcal{W}}\hat \Omega
\end{equation}
This equation can be derived by applying the projector $\hat P_0$ to the Schr\"{o}dinger equation~(\ref{eq:sefh}),
which gives
\begin{equation*}
(E_n - \hat{\mathcal{H}}_0)\hat P_0 \Psi_n = \hat P_0 \hat{\mathcal{W}} \Psi_n, \; n \leq N
\end{equation*}
Now applying $\hat \Omega$ on both sides, and using $\Psi_n = \hat \Omega \Psi^0_n$
and $E_n \Psi_n = \hat{\mathcal{H}}\hat \Omega \Psi^0_n$, 
where $\Psi^0_n = \hat P_0\Psi_n$, leads to
\begin{equation*}
[\hat \Omega,\hat{\mathcal{H}}_0] \Psi^0_n = (\hat{\mathcal{W}}\hat \Omega -\hat \Omega \hat{\mathcal{W}}\hat \Omega)
\Psi^0_n
\end{equation*}
The latter equation holds for all states $\Psi^0_n, \, n = 1, \dots, N$, that is, for the entire model space,
which implies the operator identity
\begin{equation*}
[\hat \Omega,\hat{\mathcal{H}}_0] \hat P_0 =(\hat{\mathcal{W}}\hat \Omega - 
\hat \Omega\hat{\mathcal{W}}\hat \Omega)\hat P_0
\end{equation*}
Here, the $\hat P_0$ operator can be incorporated within the wave operator, which then leads to the form of Eq.~(\ref{eq:bloch}). 
In App.~D the explicit construction of PT expansions via the Bloch equation is reviewed. It should be noted that there is no Bloch-type equation for the symmetrized wave operator 
$\hat \Omega_S$.

\subsection{\textit{A posteriori} closed-form expressions}

Analogous to Sec.~IV.C, closed-form expressions can be established
for the wave operator, the effective hamiltonians, and all other constituents of  
the QDPT formulation based on the solution of the secular equations~(\ref{eq:seceqx}). 
These very instructive \textit{a posteriori} expressions shall be considered in the following.

Following the notations used in Sec.~VI.C,
we introduce a row 'vector' of target states according to 
\begin{equation}
\dul{\Psi} = \left (\Psi_1, \dots, \Psi_N \right )
\end{equation}
The row of target states is represented by the matrix of the first $N$ eigenvectors,
$\ul X_\nu, \, \nu \leq N$,
\begin{equation}
\label{eq:tsrep}
\dul{\Psi} \equiv  (\dul{\Phi}^{x \textstyle{\dagger}}, \dul{\Psi} ) =
\left ( \begin{array}{c} 
\bs{\mathsf{X}}_{11}\\
\bs{\mathsf{X}}_{21}
\end{array}
\right)
\end{equation}
Here, the basis set expansion is written in the compact form 
$(\dul{\Phi}^{x \textstyle{\dagger}}, \dul{\Psi} )$,
where $\dul{\Phi}^x$ 
is the row vector of all unperturbed states,
\begin{equation}
\dul{\Phi}^x = \left (\Phi_1, \Phi_2, \dots  \right )
\end{equation}
and the operation $(,)$ combines matrix multiplication and forming scalar products, as in Eq.~(\ref{eq:1hol}):
\begin{equation}
(\dul{\Phi}^{x \textstyle{\dagger}}, \dul{\Psi} )_{\mu\nu} = \braket{\Phi_\mu}{\Psi_\nu} = \mathsf X_{\mu\nu}
\end{equation}
The corresponding representation of the row of model states 
\begin{equation}
\dul{\Phi} = \left (\Phi_1, \dots, \Phi_N \right )
\end{equation}
simply becomes
\begin{equation}
\label{eq:msrep}
\dul{\Phi} \equiv 
\left ( \begin{array}{c}
\bs 1\\ \bs 0 
\end{array}
\right)
\end{equation}
For the projected target states, introduced by Eq.~(\ref{eq:pts}), the explicit representation 
takes on the form 
\begin{equation}
\label{eq:epts}
\dul{\Psi}^0 = 
\left (\Psi^0_1, \dots, \Psi^0_N\right ) \equiv 
\left ( \begin{array}{c}
\bs{\mathsf{X}}_{11} \\ \bs 0 
\end{array}
\right)
\end{equation}
Their biorthogonal counterparts,
\begin{equation}
\dul{\lo \Psi}^0 =
\left (\lo\Psi^0_1, \dots, \lo\Psi^0_N\right ) 
\end{equation}
are obtained according to 
\begin{equation}
\dul{\lo \Psi}^0 = \dul{\Psi}^0 \bs{\mathsf{S}}^{-1} 
\end{equation}
where the overlap matrix $\bs{\mathsf{S}}$ is given by
\begin{equation}
\bs{\mathsf{S}} = (\dul{\Psi}^{0 \textstyle{\dagger}}, \dul{\Psi}^0 ) = \bs{\mathsf{X}}_{11}^\dagger \bs{\mathsf{X}}_{11} 
\end{equation}
Accordingly, the explicit representation of $\dul{\lo \Psi}^0$ is given by
\begin{equation}
\label{eq:bpts}
\dul{\lo \Psi}^0 
\equiv 
\left ( 
\begin{array}{c}
\bs{\mathsf{X}}_{11} (\bs{\mathsf{X}}_{11}^\dagger \bs{\mathsf{X}}_{11})^{-1}  \\ \bs 0 
\end{array}
\right) = 
\left ( \begin{array}{c}
(\bs{\mathsf{X}}_{11}^\dagger)^{-1}  \\ \bs 0 
\end{array}
\right)
\end{equation}

Now we may turn to the various operators. Let us note that here the matrix representations refer to the unperturbed states.  
The matrix representations of the model space projection operators read
\begin{equation}
\label{eq:p0q0}
\bs P_0 = 
\left(
\begin{array}{cc}
\bs 1 & \bs 0\\
\bs 0 & \bs 0
\end{array}
\right), \;\; \; 
\bs Q_0 = \bs 1 - \bs P_0 = 
\left(
\begin{array}{cc}
\bs 0 & \bs 0\\
\bs 0 & \bs 1
\end{array}
\right)
\end{equation}
The target state projector is given by 
\begin{equation}
\label{eq:tsp}
\bs P = \left ( 
\begin{array}{c}
\bs{\mathsf{X}}_{11} \\
\bs{\mathsf{X}}_{21} 
\end{array}
\right) 
\left ( 
\begin{array}{cc}
\bs{\mathsf{X}}_{11}^\dagger &
\bs{\mathsf{X}}_{21}^\dagger 
\end{array}
\right) 
=
\bs{\mathsf{X}} \bs P_0 \bs{\mathsf{X}}^\dagger = 
\left(
\begin{array}{cc}
\bs{\mathsf{X}}_{11}\bs{\mathsf{X}}_{11}^\dagger & \bs{\mathsf{X}}_{11}\bs{\mathsf{X}}_{21}^\dagger \\
 \bs{\mathsf{X}}_{21}\bs{\mathsf{X}}_{11}^\dagger  & \bs{\mathsf{X}}_{21}\bs{\mathsf{X}}_{21}^\dagger
\end{array}
\right)
\end{equation}
In a similar way, $\bs Q = \bs{\mathsf{X}} \bs Q_0 \bs{\mathsf{X}}^\dagger$ can be evaluated.

Using the representations of the target states~(\ref{eq:tsrep}) and the 
biorthogonal states~(\ref{eq:bpts}), the representation of the wave operator $\hat \Omega$ can easily be established, e.g. via Eq.~(\ref{eq:wopx}), as follows:
\begin{equation}
\label{eq:rwopx}
\bs \Omega = \left ( 
\begin{array}{c}
\bs{\mathsf{X}}_{11} \\
\bs{\mathsf{X}}_{21} 
\end{array}
\right) 
\left ( 
\begin{array}{cc}
\bs{\mathsf{X}}_{11}^{-1} & \bs 0
\end{array}
\right) =
\left ( 
\begin{array}{cc}
\bs 1 & \bs 0\\
\bs{\mathsf{X}}_{21} \bs{\mathsf{X}}_{11}^{-1} & \bs 0
\end{array}
\right)
\end{equation}
Proceeding with the product form for $\bs \Omega$, 
the representation of the operator product $\hat \Omega^\dagger \hat \Omega$ can readily be 
evaluated, yielding
\begin{equation}
\bs \Omega^\dagger \bs \Omega = 
\left ( 
\begin{array}{cc}
(\bs{\mathsf{X}}_{11}\bs{\mathsf{X}}_{11}^\dagger)^{-1} & \bs 0\\
\bs 0 & \bs 0
\end{array}
\right)
\end{equation}
where the orthonormality of the target state eigenvectors,
$\bs{\mathsf{X}}_{11}^\dagger \bs{\mathsf{X}}_{11} 
+ \bs{\mathsf{X}}_{21}^\dagger \bs{\mathsf{X}}_{21} = \bs 1$, has been used.
This allows us to write $(\hat \Omega^\dagger \hat \Omega)^{-1/2}$ as
\begin{equation}
(\bs \Omega^\dagger \bs \Omega)^{-1/2} =
 \left ( 
\begin{array}{cc}
(\bs{\mathsf{X}}_{11}\bs{\mathsf{X}}_{11}^\dagger)^{1/2} & \bs 0\\
\bs 0 & \bs 0
\end{array}
\right)
\end{equation}
The symmetrisized wave operator $\hat \Omega_S = \hat \Omega(\hat \Omega^\dagger \hat \Omega)^{-1/2}$ becomes 
\begin{equation}
\label{eq:wops}
\bs \Omega_S = 
\bs \Omega (\bs \Omega^\dagger \bs \Omega)^{-1/2} =
\left ( 
\begin{array}{cc}
(\bs{\mathsf{X}}_{11}\bs{\mathsf{X}}_{11}^\dagger)^{1/2} & \bs 0\\
\bs{\mathsf{X}}_{21}\bs{\mathsf{X}}_{11}^{-1}(\bs{\mathsf{X}}_{11}\bs{\mathsf{X}}_{11}^\dagger)^{1/2}  & \bs 0
\end{array}
\right)
\end{equation}
Noting that $\bs{\mathsf{X}}_{11}^{-1} (\bs{\mathsf{X}}_{11}\bs{\mathsf{X}}_{11}^\dagger)^{1/2} = \bs{\mathsf{X}}_{11}^\dagger (\bs{\mathsf{X}}_{11}\bs{\mathsf{X}}_{11}^\dagger)^{-1/2}$
the latter expression can be written in the more expedient product form
\begin{equation}
\label{eq:wopspf}
\bs \Omega_S  = 
\left ( 
\begin{array}{cc}
\bs{\mathsf{X}}_{11} & \bs 0\\
\bs{\mathsf{X}}_{21} & \bs 0
\end{array}
\right)
\left ( 
\begin{array}{cc}
\bs{\mathsf{X}}_{11}^\dagger (\bs{\mathsf{X}}_{11}\bs{\mathsf{X}}_{11}^\dagger)^{-1/2}& \bs 0\\
 \bs 0& \bs 0
\end{array}
\right)
\end{equation}

Combining the product form of Eq.~(\ref{eq:rwopx}) and the partitioning of the matrix representation~(\ref{eq:Hmunu}) of the hamiltonian $\hat{\mathcal{H}}$,
\begin{equation}
\bs{\mathcal{H}} =
\left ( 
\begin{array}{cc}
\bs{\mathsf{H}}_{11} & \bs{\mathsf{H}}_{12}\\
\bs{\mathsf{H}}_{21} & \bs{\mathsf{H}}_{22}
\end{array}
\right) 
\end{equation}
one obtains the closed-form expression for the non-hermitian effective hamiltonian~(\ref{eq:effh}) according to
\begin{equation}
\label{eq:nheffh}
\bs{\mathcal{H}}^{nh}_{eff} = \bs P_0 \bs{\mathcal{H}} \bs \Omega = 
\left ( 
\begin{array}{cc}
\bs{\mathsf{H}}_{11} & \bs{\mathsf{H}}_{12}\\
\bs 0 & \bs 0
\end{array}
\right) 
\left ( 
\begin{array}{c}
\bs{\mathsf{X}}_{11} \\
\bs{\mathsf{X}}_{21} 
\end{array}
\right) 
\left ( 
\begin{array}{cc}
\bs{\mathsf{X}}_{11}^{-1} & \bs 0
\end{array}
\right)
= 
\left ( 
\begin{array}{cc}
\tilde{\bs{\mathcal{H}}}' & \bs 0\\
\bs 0 & \bs 0
\end{array}
\right) 
\end{equation}
where, according to 
the partial secular equations, $\bs{\mathsf{H}}_{11} \bs{\mathsf{X}}_{11} +  \bs{\mathsf{H}}_{12}\bs{\mathsf{X}}_{21} = \bs{\mathsf{X}}_{11} \bs{\mathsf{E}}_1$,
the $11$-block is given by
\begin{equation}
\label{eq:nheffh11}
\tilde{\bs{\mathcal{H}}}' =
\bs{\mathsf{X}}_{11} \bs{\mathsf{E}}_1 \bs{\mathsf{X}}_{11}^{-1}
\end{equation} 
The three other blocks vanish.

The same matrix structure results for the representation of the hermitian effective hamiltonian~(\ref{eq:heffh}),   
\begin{equation}
\bs{\mathcal{H}}^h_{eff} = \bs{\Omega}_S^\dagger \bs{\mathcal{H}}
\bs{\Omega}_S =
\left ( 
\begin{array}{cc}
\tilde{\bs{\mathcal{H}}} & \bs 0\\
\bs 0 & \bs 0
\end{array}
\right) 
\end{equation}
Here the non-vanishing $11$-block of $\bs{\mathcal{H}}^h_{eff}$ is given by
\begin{equation}
\label{eq:cfheff}
 \tilde{\bs{\mathcal{H}}} =
(\bs{\mathsf{X}}_{11}\bs{\mathsf{X}}_{11}^\dagger)^{-1/2}\bs{\mathsf{X}}_{11} \bs{\mathsf{E}}_1 \bs{\mathsf{X}}_{11}^\dagger(\bs{\mathsf{X}}_{11}\bs{\mathsf{X}}_{11}^\dagger)^{-1/2}
\end{equation}
In deriving this expression use has been made of the product form~(\ref{eq:wopspf}) of the symmetrized wave operator  
and the partial secular (and orthonormality) equations
\begin{equation}
\left ( 
\begin{array}{cc} 
\bs{\mathsf{X}}_{11}^\dagger & \bs{\mathsf{X}}_{21}^\dagger 
\end{array}
\right ) 
\left ( 
\begin{array}{cc}
\bs{\mathsf{H}}_{11} & \bs{\mathsf{H}}_{12}\\
\bs{\mathsf{H}}_{21} & \bs{\mathsf{H}}_{22}
\end{array}
\right) 
\left ( 
\begin{array}{c}
\bs{\mathsf{X}}_{11} \\
\bs{\mathsf{X}}_{21} 
\end{array}
\right) 
= \bs{\mathsf{E}}_1
\end{equation}
The result of Eq.~(\ref{eq:cfheff}) is identical with that of Eq.~(\ref{eq:cf4mtxxx}), obtained by the generalized NIP-ADC/ISR formulation. This allows us to state the identities 
\begin{align}
\label{eq:heqm}
\tilde{\bs{\mathcal{H}}} =& \tilde{\bs{\mathsf{M}}} \\
\bs{\mathsf{Z}} =& \tilde{\bs{\mathsf{Y}}}
\end{align}
of the respective effective hamiltonians and their eigenvector matrices.

It remains to inspect the \textit{a posteriori} expression for the EEA matrix $\bs{\mathsf{f}}$. 
According to Eq.~(\ref{eq:eeam}), $\bs{\mathsf{f}}$ is just the representation of the symmetrized wave operator of $\hat \Omega (\hat \Omega^\dagger \hat \Omega)^{-1/2}$ with respect to the unperturbed 
states ${\Phi_\nu}$, where the second (row) index is restricted to model states (group $(\bs 1)$). Accordingly,  
$\bs{\mathsf{f}}$ is given by the non-vanishing
$11$- and $21$-blocks of $\bs \Omega (\bs \Omega^\dagger \bs \Omega)^{-1/2}$ as specified in Eq.~(\ref{eq:wops}):
\begin{equation}
\label{eq:cffeff}
\bs{\mathsf{f}} = 
\left (
\begin{array}{c}
\bs{\mathsf{f}}_{11}\\
\bs{\mathsf{f}}_{21}
\end{array}
\right )
=
\left ( 
\begin{array}{c}
(\bs{\mathsf{X}}_{11}\bs{\mathsf{X}}_{11}^\dagger)^{1/2} \\
\bs{\mathsf{X}}_{21}\bs{\mathsf{X}}_{11}^\dagger (\bs{\mathsf{X}}_{11}\bs{\mathsf{X}}_{11}^\dagger)^{-1/2}
\end{array}
\right)
\end{equation}
This reproduces the NIP-ADC/ISR result~(\ref{eq:cf4ftx}), which establishes the
further identification 
\begin{equation}
\label{eq:id4f}
\bs{\mathsf{f}} = \tilde{\bs{\mathsf{f}}}
\end{equation}
Moreover, using the result for $\bs{\mathsf{f}}_{11}$ in the $11$-block of Eq.~(\ref{eq:eeaz}), 
\begin{equation}
\label{eq:xfZ}
\bs{\mathsf{X}}_{11} = \bs{\mathsf{f}}_{11} \bs{\mathsf{Z}}
\end{equation}
yields the \textit{a posteriori} expression 
\begin{equation}
\label{eq:evmz}
\bs{\mathsf{Z}} = (\bs{\mathsf{f}}_{11})^{-1} \bs{\mathsf{X}}_{11} = (\bs{\mathsf{X}}_{11}\bs{\mathsf{X}}_{11}^\dagger)^{-1/2} \bs{\mathsf{X}}_{11}
\end{equation}
for the eigenvector matrix $\bs{\mathsf{Z}}$, 
which is consistent with Eq.~(\ref{eq:Ycfry}).
As in the prior discussion of Eq.~(\ref{eq:XfY}) in Sec.~III.C, Eq.~(\ref{eq:xfZ}) 
can be seen as the \textit{polar decomposition} of the matrix $\bs{\mathsf{X}}_{11}$
into the product of the hermitian matrix, $\bs{\mathsf{f}}_{11} = (\bs{\mathsf{X}}_{11}\bs{\mathsf{X}}_{11}^\dagger)^{1/2}$, and a unitary matrix, 
$\bs{\mathsf{Z}} = (\bs{\mathsf{X}}_{11}\bs{\mathsf{X}}_{11}^\dagger)^{-1/2} \bs{\mathsf{X}}_{11}$. 
More generally, Eq.~(\ref{eq:xfZ}) 
implies that, if $\bs{\mathsf{f}}_{11}$
is hermitian, $\tilde{\bs{\mathcal{H}}}$ and $\bs{\mathsf{f}}$ are uniquely defined by the eigenvector matrix blocks $\bs{\mathsf{X}}_{11}$ and $\bs{\mathsf{X}}_{21}$. 

To summarize the essential findings: The 
expressions~(\ref{eq:cfheff}), (\ref{eq:cffeff}), and (\ref{eq:evmz}), derived within the traditional 
QDPT approach are identical with the generalized NIP-ADC/ISR results of Eqs.~(\ref{eq:cf4mtxxx}),(\ref{eq:cf4ftx}), and 
(\ref{eq:Ycfry}), respectively. 
The equivalence between the QDPT and NIP expressions means that the 
PT expansions generated by the NIP-ADC/ISR procedures for 
$\tilde{\bs{\mathsf{M}}}$ and $\tilde{\bs{\mathsf{f}}}$ can directly be transferred to their QDPT counterparts. The corresponding PT expansions through third-order are listed in App.~E.

Let us recall the explicit forms of the identifications~(\ref{eq:heqm}) and (\ref{eq:id4f}), reading (in hyper-particle notation)  
\begin{align}
\nonumber
\tilde{\mathcal H}_{kl} &= \dirint{\tilde \Phi_k}{\hat{\mathcal{H}}}{\tilde \Phi_l} \;\;\; = \\
\label{eq:refl1}
\tilde{ \mathsf M}_{kl} &= \dirint{\tilde{\bs \Psi}_l}{E^N_0 - \hat{\bs H}}{\tilde{\bs \Psi}_k}, \;\;\; k,l \leq N
\end{align} 
for the secular matrix elements, and
\begin{align}
\nonumber
\mathsf f_{pk} &= \braket{\Phi_p}{\tilde \Phi_k}  \;\;\;= \\
\label{eq:refl2}
\tilde f_{pk} &= \dirint{\tilde{\bs \Psi}_k}{\mathsf{c}_p}{\bs{\Psi}^N_0},
\;\;\; k \leq N,\,\,  p = 1,2,\dots  
\end{align}
for the effective eigenvector amplitudes. 
These equations juxtapose the one-particle (or hyper-particle) QDPT description (respective first line) and the many-(hyper)particle NIP description (respective second line).
One may note that both descriptions feature a set of "intermediate states", that is, the SWO states, $\tilde \Phi_k = \hat \Omega_S \Phi_k$, at the one-particle level and the intermediate $(N\!-\!1)$-particle states, $\ket{\tilde{\bs \Psi}_k}$, at the NIP level. So what is afforded at the NIP level beyond the one-particle level? It is the well-established machinery of PT expansions available here. For the SWO states of the one-particle level, the PT expansion is cumbersome since there is no direct Bloch-type PT expansion for the symmetrized wave operator $\hat \Omega_S$. At the NIP level, by contrast, one can resort to the expedient and versatile ADC and ISR procedures presented in Secs. III and IV.

\subsection{QDPT extensions} 

The QDPT formulation outlined above aimed at the treatment of the target states group $(\bs1)$ via an 
effective hamiltonian acting on the model states. A completely analogous formulation can be 
devised for the treatment of the complementary target states by simply reversing the roles of the states of group $(\bs1)$ (model and target states) and group $(\bs2)$ (complementary model and target states). 
For example, one may define the analogue to the symmetrized wave operator (Eq.~\ref{eq:uwop}) according to 
\begin{equation}
\label{eq:uwopp}
\hat{\Omega}'_S
= \hat \Omega' (\hat \Omega^{'\dagger} \hat \Omega')^{-1/2}
\end{equation}
where, in analogy to Eq.~(\ref{eq:wops}), the (\textit{a posteriori}) matrix representation takes on the form  
\begin{equation}
\label{eq:wopsp}
\bs \Omega'_S =
\left ( 
\begin{array}{cc}
\bs 0 & \bs{\mathsf{X}}_{12}\bs{\mathsf{X}}_{22}^{\dagger}(\bs{\mathsf{X}}_{22}\bs{\mathsf{X}}_{22}^\dagger)^{-1/2} \\
\bs 0 & (\bs{\mathsf{X}}_{22}\bs{\mathsf{X}}_{22}^\dagger)^{1/2}
\end{array}
\right)
\end{equation}
The effective hamiltonian for the complementary states is obtained according to
\begin{equation}
\hat{\mathcal{H}}_{eff}' = \hat \Omega_S^{'\dagger} \hat{\mathcal{H}} \hat \Omega_S'
\end{equation}
which in terms of the explicit \textit{a posteriori} expressions reads
\begin{equation}
\bs{\mathcal{H}}_{eff}' = \bs{\Omega}_S^{'\dagger} \bs{\mathcal{H}}
\bs{\Omega}_S' =
\left ( 
\begin{array}{cc}
\bs 0 & \bs 0\\
\bs 0 & \tilde{\bs{\mathcal{H}}}^{II}
\end{array}
\right) 
\end{equation}
where the non-vanishing $22$-block, $\tilde{\bs{\mathcal{H}}}^{II}$, is given by 
\begin{equation}
\label{eq:cfheff2}
 \tilde{\bs{\mathcal{H}}}^{II} =
(\bs{\mathsf{X}}_{22}\bs{\mathsf{X}}_{22}^\dagger)^{-1/2}\bs{\mathsf{X}}_{22} \bs{\mathsf{E}}_2 \bs{\mathsf{X}}_{22}^\dagger(\bs{\mathsf{X}}_{22}\bs{\mathsf{X}}_{22}^\dagger)^{-1/2} 
\end{equation}

Rather than dealing with two separate cases 
a joint treatment can be devised by adopting a unitary transformation of the original hamiltonian, 
\begin{equation}
\label{eq:hefftot}
\hat{\mathcal{H}}^{bd} =  \hat U^\dagger \,\hat{\mathcal{H}}\,\hat U  
\end{equation}
resulting in a block-diagonal hamiltonian $\hat{\mathcal{H}}^{bd}$. 
In the present case, the unitary operator $\hat U$ combines $\hat \Omega_S$ and $\hat \Omega'_S$, which, in matrix representation, is given by 
\begin{equation}
\label{eq:utrafot}
\bs{U} = \bs{\Omega}_S + \bs{\Omega}_S' =
\left ( 
\begin{array}{cc}
(\bs{\mathsf{X}}_{11}\bs{\mathsf{X}}_{11}^\dagger)^{1/2} & \bs{\mathsf{X}}_{12}\bs{\mathsf{X}}_{22}^{\dagger}
(\bs{\mathsf{X}}_{22}\bs{\mathsf{X}}_{22}^\dagger)^{-1/2} \\
\bs{\mathsf{X}}_{21}\bs{\mathsf{X}}_{11}^{\dagger}(\bs{\mathsf{X}}_{11}\bs{\mathsf{X}}_{11}^\dagger)^{-1/2}  & 
(\bs{\mathsf{X}}_{22}\bs{\mathsf{X}}_{22}^\dagger)^{1/2} 
\end{array}
\right)
\end{equation}
It should be mentioned that a simple direct route to this \textit{a posteriori} expression was proposed in Ref.~\cite{ced89:2427}.
Again, $\bs U$ can be written in product form according to
\begin{equation}
\bs{U} = \bs{\mathsf{X}}
\left ( 
\begin{array}{cc}
\bs{\mathsf{X}}_{11}^\dagger (\bs{\mathsf{X}}_{11}\bs{\mathsf{X}}_{11}^\dagger)^{-1/2}& \bs 0\\
 \bs 0& \bs{\mathsf{X}}_{22}^\dagger (\bs{\mathsf{X}}_{22}\bs{\mathsf{X}}_{22}^\dagger)^{-1/2}
\end{array}
\right)
\end{equation}
This makes the unitarity of $\bs U$ apparent, and the transformation~(\ref{eq:hefftot}) can easily be performed. The resulting 
 matrix representation of $\mathcal{H}^{bd}$ reads
\begin{equation}
\label{eq:bdiag}
\bs{\mathcal{H}}^{bd} = \bs{U}^\dagger \bs{\mathcal{H}} \bs{U}
 =
\left ( 
\begin{array}{cc}
\tilde{\bs{\mathcal{H}}}^I & \bs 0\\
\bs 0 & \tilde{\bs{\mathcal{H}}}^{II}
\end{array}
\right) 
\end{equation}
where the $11$-block,
$\tilde{\bs{\mathcal{H}}}^I$, and the $22$-block, $\tilde{\bs{\mathcal{H}}}^{II}$,  are given by Eqs.~(\ref{eq:cfheff}) and (\ref{eq:cfheff2}), respectively.
This particular unitary transformation, featuring hermitian diagonal blocks $\bs U_{11}$ and  $\bs U_{22}$, has been referred to as the canonical Van Vleck 
transformation~\cite{sha80:5711} or des Cloiseaux transformation~\cite{dur83:3184}.

In accordance with Eq.~(\ref{eq:bdiag}), the resolvent matrix (Eq.~\ref{eq:ropmatx}) can be written as
\begin{equation}
\bs{\mathsf{R}}(\omega) = (\omega \bs 1 - \bs{\mathcal{H}} )^{-1} 
= \bs U (\omega - \bs{\mathcal{H}}^{bd})^{-1} \bs U^\dagger
\end{equation}
which naturally decomposes into two parts (Eq.~\ref{eq:ropmat12}),
\begin{equation}
\bs{\mathsf{R}}(\omega) = \bs{\mathsf{R}}^I(\omega) + \bs{\mathsf{R}}^{II}(\omega)
\end{equation}
with $\bs{\mathsf{R}}^I(\omega)$ given by
\begin{equation}
\bs{\mathsf{R}}^I(\omega) = \left (
\begin{array}{c} \bs U_{11} \\ \bs U_{21} \end{array} \right )
\left ( 
\begin{array}{cc}
(\omega -\tilde{\bs{\mathcal{H}}}^I)^{-1} & \bs 0\\
\bs 0 & \bs 0
\end{array}
\right) 
\left (\begin{array}{cc} \bs U_{11}^\dagger & \bs U_{21}^\dagger \end{array} \right)
\end{equation}
and $\bs{\mathsf{R}}^{II}(\omega)$ by a corresponding expression. 
Here, the $\bs U_{11}, \bs U_{21}, \bs U_{12}$, and $\bs U_{22}$ matrix blocks are as given in 
Eq.~(\ref{eq:utrafot}). Note that here the diagonal blocks $\bs U_{11}$ and $\bs U_{22}$ are hermitian.

Analogous to Eq.~(\ref{eq:xfZ}), the comparison of $\bs{\mathsf{R}}^I(\omega)$ and 
$\bs{\mathsf{R}}^{II}(\omega)$ with the corresponding spectral representations according to 
Eq.~(\ref{eq:srmatgen}) establishes the identities
re are the relations
\begin{align}
\nonumber
\bs{\mathsf{X}}_{11} & = \bs U_{11} \tilde{\bs{\mathsf{Y}}}^I \\
\label{eq:XUY}
\bs{\mathsf{X}}_{22} & = \bs U_{22} \tilde{\bs{\mathsf{Y}}}^{II}
\end{align}
where $\tilde{\bs{\mathsf{Y}}}^I$ and $\tilde{\bs{\mathsf{Y}}}^{II}$ are the eigenvector matrices of $\tilde{\bs{\mathcal{H}}}^I$ and $\tilde{\bs{\mathcal{H}}}^{II}$, respectively. As discussed above, these equations
represent unique polar decompositions of the original eigenvector blocks 
$\bs{\mathsf{X}}_{11}$ and 
$\bs{\mathsf{X}}_{22}$, respectively, if the $\bs U_{11}$ and $\bs U_{22}$ diagonal blocks are hermitian. The hermiticity of the diagonal blocks of $\bs U$ further implies that also the non-diagonal blocks are uniquely determined by the blocks of $\bs{\mathsf{X}}$, and 
$\bs U$ is uniquely given in the form of Eq.~(\ref{eq:utrafot}). 
 
Expediently, the unitary transformation operator can be written in the form
\begin{equation*}
\hat U = e^{\hat S}
\end{equation*}
where $\hat S = - \hat S^\dagger$ is an anti-hermitian operator~\cite{Kemble:1937,pri63:710}.
To make $\hat U$ unique, one may require that the diagonal blocks $\bs S_{11}$ and 
$\bs S_{22}$ in the representation $\bs S$ of $\hat S$ vanish~\cite{sha80:5711}. 
Obviously, the latter restriction on $\hat S$ implies the hermiticity of the $\bs U_{11}$ and $\bs U_{22}$ diagonal blocks (note that also the reverse is valid), yielding the canonical 
Van Vleck (or des Cloiseaux) transformation.

More generally, block-diagonalization can be achieved by means of a similarity transformation,
\begin{equation}
\label{eq:sthbd}
\hat{\mathcal{H}}^{bd} =  \hat U^{-1} \,\hat{\mathcal{H}}\,\hat U  
\end{equation}
where $\hat U$ does not need to be unitary. This allows a rather wide scope for the choice of  
$\hat U$. Durand, for example, has presented a whole range of such decoupling operators 
in the form~\cite{dur83:3184} 
\begin{equation}
\label{eq:Ugamma}
\hat U_\gamma  = \hat P (\hat P_0 \hat P \hat P_0 )^{-\gamma} + \hat Q (\hat Q_0 \hat Q \hat Q_0 )^{-\gamma}
\end{equation}
where $\gamma$ is a non-negative number. The representations of the operators $\hat U_\gamma$
can be evaluated by using the expressions~(\ref{eq:tsp}) for the projector operators 
$\hat P$ and $\hat Q$,
\begin{equation}
\bs P = \left ( 
\begin{array}{c}
\bs{\mathsf{X}}_{11} \\
\bs{\mathsf{X}}_{21} 
\end{array}
\right) 
\left ( 
\begin{array}{cc}
\bs{\mathsf{X}}_{11}^\dagger &
\bs{\mathsf{X}}_{21}^\dagger 
\end{array}
\right), 
\;\;\;
\bs Q = \left ( 
\begin{array}{c}
\bs{\mathsf{X}}_{12} \\
\bs{\mathsf{X}}_{22} 
\end{array}
\right) 
\left ( 
\begin{array}{cc}
\bs{\mathsf{X}}_{12}^\dagger &
\bs{\mathsf{X}}_{22}^\dagger 
\end{array}
\right)
\end{equation}
Hereby, we obtain
\begin{equation}
\hat P_0 \hat P \hat P_0  =
\left ( 
\begin{array}{cc}
\bs{\mathsf{X}}_{11}\bs{\mathsf{X}}_{11}^\dagger & \bs 0\\
\bs 0 & \bs 0
\end{array}
\right) 
\end{equation}
and
\begin{equation}
\hat P (\hat P_0 \hat P \hat P_0 )^{-\gamma}
= 
\left ( 
\begin{array}{c}
\bs{\mathsf{X}}_{11} \\
\bs{\mathsf{X}}_{21} 
\end{array}
\right) 
\left ( 
\begin{array}{cc}
\bs{\mathsf{X}}_{11}^\dagger (\bs{\mathsf{X}}_{11}\bs{\mathsf{X}}_{11}^\dagger)^{-\gamma}&
\bs 0
\end{array}
\right) 
\end{equation}
The corresponding result for the second term in Eq.~(\ref{eq:Ugamma}) reads 
\begin{equation}
\hat Q (\hat Q_0 \hat Q \hat Q_0 )^{-\gamma} 
=
\left ( 
\begin{array}{c}
\bs{\mathsf{X}}_{12} \\
\bs{\mathsf{X}}_{22} 
\end{array}
\right) 
\left ( 
\begin{array}{cc}
\bs 0 & 
\bs{\mathsf{X}}_{22}^\dagger (\bs{\mathsf{X}}_{22}\bs{\mathsf{X}}_{22}^\dagger)^{-\gamma}
\end{array}
\right)
\end{equation}
The latter two expressions can be combined 
to give the representation of $\hat U_\gamma$: 
\begin{equation}
\bs U_\gamma = 
\left(
\begin{array}{cc}
\bs{\mathsf{X}}_{11} & \bs{\mathsf{X}}_{12} \\
 \bs{\mathsf{X}}_{21}& \bs{\mathsf{X}}_{22}
\end{array}
\right)
\left(
\begin{array}{cc}
\bs{\mathsf{X}}_{11}^\dagger (\bs{\mathsf{X}}_{11}\bs{\mathsf{X}}_{11}^\dagger)^{-\gamma} & \bs 0 \\
 \bs 0 & \bs{\mathsf{X}}_{22}^\dagger (\bs{\mathsf{X}}_{22}\bs{\mathsf{X}}_{22}^\dagger)^{-\gamma}
\end{array}
\right)
\end{equation}
The inverse of this matrix can easily be formed, yielding
\begin{equation}
\bs U_\gamma^{-1}
= 
\left(
\begin{array}{cc}
(\bs{\mathsf{X}}_{11}\bs{\mathsf{X}}_{11}^\dagger)^{\gamma -1} \bs{\mathsf{X}}_{11} & \bs 0 \\
 \bs 0 &  (\bs{\mathsf{X}}_{22}\bs{\mathsf{X}}_{22}^\dagger)^{\gamma -1}\bs{\mathsf{X}}_{22}
\end{array}
\right)
\left(
\begin{array}{cc}
\bs{\mathsf{X}}_{11}^\dagger & \bs{\mathsf{X}}_{21}^\dagger \\
 \bs{\mathsf{X}}_{12}^\dagger & \bs{\mathsf{X}}_{22}^\dagger
\end{array}
\right)
\end{equation}
This expression can be identified with the representation of 
Durand's projector formula,
\begin{equation}
\bs U_\gamma^{-1} = (\hat P_0 \hat P \hat P_0 )^{\gamma -1}\hat P  +  (\hat Q_0 \hat Q \hat Q_0 )^{\gamma -1} \hat Q
\end{equation}
Now the block-diagonalization according to Eq.~(\ref{eq:sthbd})
expression for the blocks of $H^{bd}$, specifying the diagonal blocks according to
\begin{align}
\tilde{\bs{\mathcal{H}}}_\gamma^{I} &=
(\bs{\mathsf{X}}_{11}\bs{\mathsf{X}}_{11}^\dagger)^{\gamma -1}\bs{\mathsf{X}}_{11} \bs{\mathsf{E}}_1 \bs{\mathsf{X}}_{11}^\dagger(\bs{\mathsf{X}}_{11}\bs{\mathsf{X}}_{11}^\dagger)^{-\gamma} \\ 
\tilde{\bs{\mathcal{H}}}_\gamma^{II} &=
(\bs{\mathsf{X}}_{22}\bs{\mathsf{X}}_{22}^\dagger)^{\gamma -1}\bs{\mathsf{X}}_{22} \bs{\mathsf{E}}_2 \bs{\mathsf{X}}_{22}^\dagger(\bs{\mathsf{X}}_{22}\bs{\mathsf{X}}_{22}^\dagger)^{-\gamma} 
\end{align}

Let us note that the manifold of similarity transformations $U_\gamma$ comprises, for $\gamma = 1/2$, the canonical Van Vleck (or des Cloizeaux) approach associated with Eqs.~(\ref{eq:utrafot},\ref{eq:bdiag}). For $\gamma = 1$ one obtains the non-hermitian effective hamiltonian variant according to Eqs.~(\ref{eq:rwopx},\ref{eq:nheffh}). A hermitan conjugate version of the latter results for $\gamma = 0$.

\section{Concluding remarks}

The NIP-ADC/ISR approach to QDPT presented in this article is based on 
the recognition that physical information of a one-particle system turns up in specific ways in a corresponding many-particle system formed by $N$ non-interacting fermionic "replicas" of the original particle, where the number $N$ can be chosen arbitrarily. For example, the first $N$ energy eigenvalues of the one-particle system are obtained as particle-detachment energies of that NIP system. Moreover, the components of the corresponding eigenvectors can be identified with the spectroscopic amplitudes of the detachment process.
The electron propagator, $\bs G(\omega)$, for the NIP system becomes the resolvent matrix of the one-particle system, $\bs R(\omega)$, and its particle detachment part, $\bs G^-(\omega)$, is identical with the part 
$\bs R^I(\omega)$ of the resolvent matrix associated with the model and target space partitionings in the QDPT formulation of the one-particle problem. This, in turn establishes the identity of the representation of
the QDPT effective hamiltonian, in its distinguished hermitian form, 
and the representation of the NIP hamiltonian in terms of $N$
intermediate $(N-1)$-particle states. 
Similarly, the representation of the symmetrized wave operator, $\bs \Omega_S$, can be obtained as the matrix of effective transition amplitudes in the NIP-ADC form of $\bs G^-(\omega)$. 
The NIP concept is not confined to actual particles but can readily be transferred to an arbitrary quantum system, which is then supposed to represent a "hyper-particle", from which a system of non-interacting hyper-particles can be formed.  

The essential equivalence of the one-particle/hyper-particle problem and the many-particle problem of non-interacting particles/hyper-particles means that the ADC and ISR methods, originally devised for the treatment of real many-body systems, can be 
adapted to the NIP case and in this form establish an alternative formulation of QDPT for general quantum systems.
Thus, the NIP-ADC procedure, based on the diagrammatic PT expansion of 
$\bs G^-(\omega)$, has been used to derive the PT expansions   
of the matrix elements of the hermitian effective hamiltonian and the symmetrized wave operator through third order. The results for the effective hamiltonian, collected in App.~E, agree with the expressions derived within the traditional QDPT formulations. This applies as well for the symmetrized wave operator or effective eigenvector amplitudes, at least through second order, as third-order expressions have apparently not been published previously. 

Alternatively to NIP-ADC, the PT expansions can also be obtained via the NIP-ISR procedure, which here, for illustration, was performed up to second order. The essential benefit of the NIP-ISR approach, however, is that it allows the derivation of exact closed-form expressions for the ISR secular matrix and effective transition amplitudes in terms of the first $N$ eigenvectors and eigenvalues of the one-particle/hyper-particle hamiltonian, being accordingly referred to as a \textit{posteriori expressions}. The same expressions can also be derived within the traditional QDPT formulation, which provides an independent confirmation of the validity of the NIP-ADC/ISR approach to QDPT.

The NIP-ADC/ISR formulation has also allowed us to review the question of uniqueness of the effective QDPT hamiltonians. The requirement of hermiticity of the effective hamiltonian is not sufficient to this end. Uniqueness is obtained by requiring that the 11-block 
$\bf{\mathsf f}_{11}$ of the matrix of effective eigenvector amplitudes is hermitian. This results in a polar decomposition of the 11-block
$\bf{\mathsf X}_{11}$ of the original eigenvector matrix, which, in turn, uniquely determines the hermitian effective hamiltonian and, together with  the 21-block $\bf{\mathsf X}_{21}$, the full $\bf{\mathsf f}$ matrix.  

We note in passing that the present development comprises the ordinary Rayleigh-Schr\"{o}dinger perturbation theory (RSPT) as the special case $N = 1$ and also the case of degenerate model states, that is, degenerate perturbation theory (DPT).

In view of the ever-increasing computer power, the savings in computing time and memory requirements afforded by QDPT may no longer be of paramount importance. Still, 
a brief evaluation of the scaling properties of the QDPT computational schemes may be of interest.
Let $m$ and $n$ denote the dimensions of the model space and the complementary space, respectively, where usually $n$ will be much larger than $m$, $m \ll n$. To be specific, we assume that all $m$ states associated with the model space are to be determined. Furthermore, it is supposed that the original secular matrix is essentially a full matrix
of non-zero elements.
Solving the original secular problem for the $m$ lowest eigenstates (regular mode) scales as $m (m+n)^2 = m^3 + 2m^2 n + m n^2 \sim m n^2$.
In the QDPT schemes the main computational cost arises in the computation of the effective secular matrix, whereas the subsequent diagonalization, scaling as $m^3$, is only a minor contribution.
As the inspection of the second-order expressions shows, the QDPT(2) scales as $m^2 n$, which compares favourably with the scaling of the  
regular mode.
At the third-order level, the first term in Eq.~(\ref{eq:m3kl}) seems to introduce an 
$m^2 n^2$ scaling in the computation of the effective secular matrix. 
However, by forming suitable intermediates the cost can be reduced, resulting in an overall scaling according to $m n^2 + m^2 n \sim m n^2$.  
This means the computational cost of QDPT(3) matches that of the regular mode. At fourth and higher order the QDPT treatment will be more costly than the regular one.

The third-order step of the NIP-ADC procedure, demonstrated in App.~A, was already somewhat involved. At the fourth-order stage one already has to deal with a set of 360 diagrams for $\bs G^-(\omega)$, 
so that the ADC procedure becomes a major challenge even though many of those diagrams are only repetitive. The NIP-ISR derivation, being essentially based on the PT expansion of the $N$-particle
 ground state, becomes cumbersome already at third order. Nonetheless, the ISR approach seems to be more suitable than the diagrammatic one for an
 automated generation of PT expressions, as was demonstrated recently in the development of the fourth-order computational scheme for the polarization propagator 
~\cite{lei22:184101}. So if an extension to fourth-order PT expansions were ever envisaged, automated ISR would be the method of choice.

\appendix
\renewcommand{\theequation}{A.\arabic{equation}}
\setcounter{equation}{0}
\section*{Appendix A: NIP-ADC procedure in third order}

In third order the ADC expansion for the NIP electron propagator part 
$\bs G^-$ or rather its transpose takes on the form
\begin{align}
\label{eq:adcexpo3}
 \tilde{\bs G}^{(3)}(\omega) =&\\
\nonumber 
&{\bs f^{(0)}}^\dagger (\omega - \bs K)^{-1} \bs f^{(3)} +  h.c.\;\; + &(A)\\
\nonumber
&{\bs f^{(0)}}^\dagger(\omega - \bs K)^{-1} \bs C^{(3)}(\omega - \bs K)^{-1} \bs f^{(0)}
\;\; + &(B)\\
\nonumber
 &{\bs f^{(1)}}^\dagger (\omega - \bs K)^{-1} \bs f^{(2)}  + h.c. \;\; +&(C)\\
\nonumber
 &{\bs f^{(1)}}^\dagger(\omega - \bs K)^{-1} \bs C^{(2)}(\omega - \bs K)^{-1}\bs f^{(0)} + h.c. \;\; +&(D)\\  
\nonumber
 &{\bs f^{(0)}}^\dagger(\omega - \bs K)^{-1} \bs C^{(1)}(\omega - \bs K)^{-1}\bs f^{(2)} + h.c.\;\; + &(E)\\
 \nonumber 
 &{\bs f^{(0)}}^\dagger(\omega - \bs K)^{-1} \bs C^{(1)}(\omega - \bs K)^{-1}\bs C^{(2)}(\omega - \bs K)^{-1}\bs f^{(0)} + h.c. \;\; +&(F)\\  
\nonumber
 &{\bs f^{(1)}}^\dagger(\omega - \bs K)^{-1} \bs C^{(1)}(\omega - \bs K)^{-1}\bs f^{(1)}
 \;\; +  &(G)\\
\nonumber
 &{\bs f^{(0)}}^\dagger(\omega - \bs K)^{-1} \bs C^{(1)}(\omega - \bs K)^{-1}\bs C^{(1)}(\omega - \bs K)^{-1}\bs f^{(1)} + h.c. \;\; +&(H)\\   
\nonumber 
&{\bs f^{(0)}}^\dagger(\omega - \bs K)^{-1} \bs C^{(1)}(\omega - \bs K)^{-1}\bs C^{(1)}(\omega - \bs K)^{-1}
 \bs C^{(1)}(\omega - \bs K)^{-1}\bs f^{(0)}  &(I)   
\end{align}
The new quantities to be determined at the third-order level are $\bs C^{(3)}$ and $\bs f^{(3)}$ in the terms $(B)$ and $(A)$, respectively. The remaining terms $(C)$-$(I)$ are repetitive, as they involve only contributions to $\bs C$ and $\bs f$ already determined in the preceding ADC steps.

The third-order ADC form is to be compared with the third-order diagrams 
for $\bs G^-$ shown in Fig.~\ref{fig:1pdo3}. Here the 60 time-ordered (Goldstone) diagrams are divided into 10 groups of 6 diagrams each. The groups reflect the 10 possibilities of arranging the two outer vertices $t,t'$ at the available positions $1,2, \dots, 5$ under the constraint $t' > t$. The 6 diagrams $(a,b, \dots, f)$ within each group correspond to the 6 permutations of the three inner vertices (crosses).

A part of the diagrams can directly be assigned to corresponding terms in the ADC form. The terms $(G)$, $(H)$, and $(I)$, containing only contributions up to first order in $\bs C$ and $\bs f$, can readily be retrieved in the diagrams $a1$ ($I$), $a2, a3$ ($H$), and $a6$ ($G$).
The $\bs f^{(2)}$ contributions arising in the terms $(E)$ and $(C)$ can be distinguished as being of $hh$-type (Eq. \ref{eq:f2pq}) or $hp$-type (Eqs. \ref{eq:f2pqx},\ref{eq:f2pqy}). The corresponding $hp$ parts of $(E)$ and $(C)$ can be retrieved in the diagrams $a4,a5,c4,b5$ ($E$) and $a7,a8,c7,b8$ ($C$). Moreover, the 6 diagrams $a9, \dots, f9$ and their hermitian conjugate counterparts $a10, \dots, f10$ can directly be assigned to the term 
$(A)$, which allows for the determination of corresponding 6 $hp$ contributions to $\bs f^{(3)}$.

The remaining 36 diagrams feature one or two $2h$-$1p$ denominators, which means that they cannot individually be assigned to the ADC terms but only upon forming suitable linear combinations. To this end one may distinguish 5 different diagram types $t1, \dots, t6$ according to their respective integral product structure. For example, the 8 diagrams $d1, \dots, d8$, constituting type $t1$, have the same integral product structure, $w_{rp} w_{sr} w_{qs} n_p n_q \lo n_r \lo n_s$, or, in terms of the directions of subsequent fermion lines 'down-up-up-down` $[duud]$. The complete list of the 5 types and their constituting diagrams is as follows:
\begin{align*}
t1 \equiv [duud] &\equiv \{d1, \dots, d8\}\\
t2 \equiv [ddud] &\equiv \{c1,c2, e1, \dots, e4,e6,e7\}\\
t3 \equiv [dudd] &\equiv \{b1,b3,f1,f2,f3,f5,f6,f8\}\\
t4 \equiv [udud] &\equiv \{c3,c5,c6,c8,e5,e8\}\\
t5 \equiv [dudu] &\equiv \{b2,b4,b6,b7,f4,f7\}
\end{align*} 
Note that $t3$ and $t5$ can be seen as the hermitian conjugate counterparts to $t2$ and $t4$, respectively. 
The diagram types $t1-t3$ comprise 8 diagrams each, whereas only 6 diagrams belong $t4$ and $t5$, respectively. The latter sets could each be augmented by the two diagrams $b9,f9$ and $c10,e10$, respectively, which have already been addressed.\\   
\\
\noindent
\textbf{$\bs{t1}$ diagrams}\\
Let us first inspect the diagrams of type $t1$. Here the 4 diagrams with 
two $2h$-$1p$ denominators can be combined as follows:
\begin{equation}
\label{eq:t1}
(d1 + d2 + d3 + d6)|_{pq} = \sum_{r,s > N}  w_{rp} w_{sr} w_{qs}  X_{(pqrs)},  \;\; p,q \leq N
\end{equation}
where 
\begin{align*}
X_{(pqrs)} =& (\omega - \epsilon_p - \epsilon_q + \epsilon_r )^{-1} 
(\omega - \epsilon_p - \epsilon_q + \epsilon_s)^{-1}\\
 &\{(\omega-\epsilon_p)^{-1}(\omega-\epsilon_q)^{-1} 
  +(\omega-\epsilon_p)^{-1}(\epsilon_r-\epsilon_p)^{-1} \\
  &\quad +(\omega-\epsilon_q)^{-1}(\epsilon_s-\epsilon_q)^{-1}
   + (\epsilon_r-\epsilon_p)^{-1}(\epsilon_s-\epsilon_q)^{-1} \}
\end{align*}
The 4 terms in the bracket can be written as the product
\begin{align*}
\{\dots \} =& ( \frac{1}{\omega - \epsilon_p} +  \frac{1}{\epsilon_s - \epsilon_q} )( \frac{1}{\omega - \epsilon_q} +  \frac{1}{\epsilon_r - \epsilon_p} )\\
= & \frac{\omega - \epsilon_p - \epsilon_q + \epsilon_s }{(\omega - \epsilon_p) (\epsilon_s-\epsilon_q)}\,\, 
\frac{\omega - \epsilon_p - \epsilon_q + \epsilon_r }{(\omega - \epsilon_q) (\epsilon_r-\epsilon_p)}
\end{align*}
so that $X_{(pqrs)}$ takes on the form 
\begin{equation*}
X_{(pqrs)} = \frac{1}{\omega - \epsilon_p} \frac{1}{\omega - \epsilon_q} \frac{1}{\epsilon_r - \epsilon_p} \frac{1}{\epsilon_s - \epsilon_q}
\end{equation*} 
where the $2h$-$1p$ denominators have cancelled out. 
Obviously, the sum $(d1 + d2 + d3 + d6)$ matches the term $(B)$ of the third-order ADC form~(\ref{eq:adcexpo3}), which allows us to specify a corresponding contribution,
\begin{equation}
\label{eq:Cpqt1}
C_{pq}^{(3,t1)} = \sum_{r,s > N}  \frac{w_{rp} w_{sr} w_{qs}}{(\epsilon_r - \epsilon_p) (\epsilon_s - \epsilon_q)}
\end{equation}
to the third-order ADC secular matrix.

In a similar way, we may combine the two $t1$ diagrams $d4$ and $d7$, having one $2h$-$1p$ denominator to the expression 
\begin{equation}
(d4 + d7)|_{pq} = \frac{1}{\omega - \epsilon_p} \sum_{r,s > N}  \frac{w_{rp} w_{sr} w_{qs}}
{(\epsilon_s-\epsilon_p)(\epsilon_r-\epsilon_p)(\epsilon_s-\epsilon_q)}, \;\; p,q \leq N 
\end{equation}
which can be assigned to the ADC term $(A)$. The corresponding contribution to $\bs f$
reads
\begin{equation}
\label{eq:t1f}
f_{pq}^{(3,t1)} = \sum_{r,s > N}  \frac{w_{rp} w_{sr} w_{qs}}{(\epsilon_r - \epsilon_p) (\epsilon_s - \epsilon_p) (\epsilon_s - \epsilon_q)}n_p n_q
\end{equation}
The remaining two diagrams $d5$ and $d8$ of type $t1$ are redundant as they just reproduce the \textit{h.c.} contribution in the ADC term $(A)$.\\

\noindent
$\bs{t2}$ \textbf{and} $\bs{t3}$ \textbf{diagrams}\\   
The $t2$ and $t3$ diagrams reproduce the ADC terms $(E)$ ($hh$ part) and $(F)$ and, moreover, establish contributions to $(B)$ and $(A)$. Let us specifically inspect the $t2$ diagrams.
Combining again the diagrams with each two $2h$-$1p$ denominators,
\begin{equation}
\label{eq:t2}
(e1 +e2 + e3 + e6)|_{pq} = - \sum_{v \leq N,s > N}  w_{vp} w_{sv} w_{qs}  X_{(pqvs)},  \;\; p,q \leq N
\end{equation} 
where 
\begin{align*}
X_{(pqvs)} =& (\omega - \epsilon_p - \epsilon_q + \epsilon_s )^{-1} 
(\omega - \epsilon_q - \epsilon_v + \epsilon_s)^{-1}\\
 &\{(\omega-\epsilon_p)^{-1}(\omega-\epsilon_q)^{-1} 
  +(\omega-\epsilon_p)^{-1}(\epsilon_s-\epsilon_v)^{-1} \\
  &\quad +(\omega-\epsilon_q)^{-1}(\epsilon_s-\epsilon_q)^{-1}
   + (\epsilon_s-\epsilon_v)^{-1}(\epsilon_s-\epsilon_q)^{-1} \}
\end{align*} 
yields the expression
\begin{equation}
 \label{eq:t2x}
(e1 +e2 + e3 + e6)|_{pq} = \frac{1}{\omega -\epsilon_p}
\sum_{v \leq N,s > N}  \frac{w_{vp} w_{sv} w_{qs}}{(\epsilon_q - \epsilon_s)(\epsilon_s-\epsilon_v)}\,\, \frac{1}{\omega -\epsilon_q}
\end{equation}
which matches the ADC term $(B)$. However, as will be seen below, only half of this expression 
contributes to $(B)$ as one half is spent in reproducing the ADC terms $(E)$ and $(F)$. 
The two diagrams with one $2h$-$1p$-denominators combine to a $(A)$-type contribution
\begin{equation}
\label{eq:t2fx}
(e4 + e7)|_{pq} = \frac{1}{\omega - \epsilon_p} f_{pq}^{(3,t2)}, \;\; p,q \leq N  
\end{equation}
where
\begin{equation}
\label{eq:t2f}
f_{pq}^{(3,t2)} = \sum_{v \leq N,s > N}  \frac{ w_{vp} w_{sv} w_{qs}}{(\epsilon_p-\epsilon_s)(\epsilon_q-\epsilon_s)(\epsilon_v-\epsilon_s)} n_p n_q
\end{equation}
The remaining two $t2$ diagrams feature three $\omega$-dependent denominators. They can be combined according to
\begin{equation}
\label{eq:t2xx}
(c1 + c2)|_{pq} =  \frac{1}{\omega - \epsilon_p} \sum_{v \leq N,s > N} \frac{1}{\omega - \epsilon_v} \frac{ w_{vp} w_{sv} w_{qs}}{(\epsilon_v-\epsilon_s)} \frac{1}{\omega - \epsilon_q}, \;\; p,q \leq N
\end{equation}
This is a further contribution to the reproduction of the ADC terms $(E)$ and $(F)$, which will be addressed in the following. 

According to Eq.~(\ref{eq:adcexpo3}), the $hh$ matrix elements of $(E)$ and $(F)$ read
\begin{align*}
(E)|_{pq} =& (\omega - \epsilon_p)^{-1} \sum_{v\leq N}  C_{pv}^{(1)} (\omega - \epsilon_v)^{-1} f_{vq}^{(2)}, \; p,q \leq N \;\;\; + \; h.c. \\
(F)|_{pq} =& (\omega - \epsilon_p)^{-1} \sum_{v\leq N} C_{pv}^{(1)} (\omega - \epsilon_v)^{-1} C_{vq}^{(2)} (\omega - \epsilon_q)^{-1}, \; p,q \leq N \;\; \;+ \; h.c.
\end{align*}
Inserting here the second-order results~(\ref{eq:C2pq},\ref{eq:f2pq}) for $C_{vq}^{(2)}$ and $f_{vq}^{(2)}$, respectively, and $C_{pv}^{(1)}= w_{vp}$ yields the following expression for 
the sum of the $(E)$ and $(F)$ matrix elements: 
\begin{equation*}
(E + F)|_{pq} = (\omega - \epsilon_p)^{-1} 
\sum_{v\leq N, s>N} \frac{1}{\omega - \epsilon_v} \frac{w_{vp}w_{sv}w_{qs}}{(\epsilon_s-\epsilon_v)(\epsilon_s-\epsilon_q)} \left ( - \half + \frac{\epsilon_v/2 + \epsilon_q /2 - \epsilon_s}{\omega -\epsilon_q} \right ) + h.c.
\end{equation*}
The round bracket on the right-hand side can be rewritten as
\begin{equation*}
\left (\frac{\epsilon_v/2 + \epsilon_q /2 - \epsilon_s}
{\omega -\epsilon_q} - \half \right )
=  \frac{1}{\omega -\epsilon_q}\{(\epsilon_q -\epsilon_s) -\half (\omega - \epsilon_v) \}
\end{equation*}
so that $(F) + (E)$ becomes the sum of two terms,
\begin{align}
\nonumber
(E + F)|_{pq} =&\frac{1}{\omega - \epsilon_p} 
\sum_{v\leq N, s>N}\frac{1}{\omega - \epsilon_v} \frac{w_{vp}w_{sv}w_{qs}}{\epsilon_v - \epsilon_s}  \frac{1}{\omega - \epsilon_q} \\
\label{eq:FEpq}
&+ \half \frac{1}{\omega - \epsilon_p}  \sum_{v\leq N, s>N} \frac{w_{vp}w_{sv}w_{qs}}{(\epsilon_q - \epsilon_s)(\epsilon_s - \epsilon_v)} \,\,\frac{1}{\omega - \epsilon_q} + h.c. 
\end{align}
As the comparison with Eqs.~(\ref{eq:t2x},\ref{eq:t2xx}) shows, the first contribution is identical with the sum $(c1 + c2)|_{pq}$ of the diagrams $c1$ and $c2$, whereas the second contribution differs from the sum $(e1 + e2 + e3 + e6)|_{pq}$ by the factor $1/2$. This means that one half of $(e1 + e2 + e3 + e6)$ enters the reproduction of $(F) + (E)$, while one half contributes to the ADC term $(B)$:
\begin{equation*}
\half (e1 + e2 + e3 + e6)|_{pq} \rightarrow (B)|_{pq}
\end{equation*}
This establishes the contribution
\begin{equation}
\label{eq:Cpqt2}
C_{pq}^{(3,t2)} = \half \sum_{v\leq N, s>N} \frac{w_{vp}w_{sv}w_{qs}}{(\epsilon_q - \epsilon_s) (\epsilon_s - \epsilon_v)} 
\end{equation}
to the third-order ADC secular matrix $\bs C^{(3)}$.

The $t3$ diagrams play an analogous role for the \textit{h.c.} parts of 
$(E)$ and $(F)$. The four diagrams $f1$-$f3$ and $f6$ contribute in equal parts to
$(F) + (E)$ (\textit{h.c.} parts) and $(B)$, furnishing the contribution 
\begin{equation}
\label{eq:Cpqt3}
C_{pq}^{(3,t3)} = \half \sum_{v\leq N, s>N} \frac{w_{sp}w_{vs}w_{qv}}{(\epsilon_p - \epsilon_s) (\epsilon_s - \epsilon_v)} 
\end{equation}
to $\bs C^{(3)}$.
The latter contribution is the hermitian conjugate of the $t2$ contribution~(\ref{eq:Cpqt2}) so that the overall contribution is hermitian. 

In analogy to Eq.~(\ref{eq:FEpq}), the two diagrams $b1$ and $b3$ enter the reproduction of  the \textit{h.c.} parts of $(F) + (E)$.   
Finally, the diagrams $f5$ and $f6$, being analogues to 
$e4$ and $e7$, respectively, contribute to $(A)$, specifically to the \textit{h.c.} part, reconfirming the result~(\ref{eq:t2fx}) for a third-order contribution to $\bs f^{(3)}$.\\ 

\noindent
$\bs{t4}$ \textbf{and} $\bs{t5}$ \textbf{diagrams}\\ 
It remains to inspect the $t4$ and $t5$ diagrams, which reproduce the ADC terms $(C)$ and $(D)$ and, moreover, contribute to $(A)$, thereby establishing a $hp$ constituent to $\bs f^{(3)}$.

Let us first consider the ADC terms $(C)$ and $(D)$. 
According to Eq.~(\ref{eq:adcexpo3}), the respective matrix (of $ph$ type) elements read
\begin{align*}
(C)|_{pq} =&  \sum_{v \leq N}  (f^{(1)\dagger})_{pv} (\omega - \epsilon_v)^{-1} f_{vq}^{(2)}, \;\; p> N, q \leq N, \;\;\; + \; h.c.\\
(D)|_{pq} =&  \sum_{v\leq N} (f^{(1)\dagger})_{pv} (\omega - \epsilon_v)^{-1} C_{vq}^{(2)} (\omega - \epsilon_q)^{-1}, \;\; p> N, q \leq N, \;\;\; + \; h.c.
\end{align*}
Note that index restrictions $p \leq N, q > N$ apply to the $h.c.$ parts.

Using now the first-order expression~(\ref{eq:f11}) and the second-order results~(\ref{eq:C2pq},\ref{eq:f2pq}) for $C_{vq}^{(2)}$ and $f_{vq}^{(2)}$, respectively,
the $(C)$ and $(D)$ matrix elements can be combined 
 
\begin{equation*}
(C + D)|_{pq} = 
\sum_{v\leq N, s>N} \frac{1}{\omega - \epsilon_v} \frac{w_{vp}w_{sv}w_{qs}}
{(\epsilon_v - \epsilon_p)(\epsilon_s-\epsilon_v)(\epsilon_s-\epsilon_q)} \left ( - \half + \frac{\epsilon_v/2 + \epsilon_q /2 - \epsilon_s}{\omega -\epsilon_q} \right ) + h.c.
\end{equation*} 
The term in round brackets can be rewritten according to
\begin{equation*}
\left ( \dots \right) = \frac{1}{\omega -\epsilon_q}\, \{-\half (\omega -\epsilon_v) + 
(\epsilon_q - \epsilon_s)\} 
\end{equation*}
so that $(C + D)|_{pq}$ takes on the form
\begin{equation*}
(C + D)|_{pq} = X_{pq} + Y_{pq} + h.c.
\end{equation*}
where
\begin{align}
\label{eq:CDI}
X_{pq} =& \half \sum_{v\leq N, s>N} \frac{w_{vp}w_{sv}w_{qs}}{(\epsilon_p - \epsilon_v)(\epsilon_s-\epsilon_v)(\epsilon_s-\epsilon_q)}\,\, \frac{1}{\omega - \epsilon_q} \\
Y_{pq} =& \sum_{v\leq N, s>N} \frac{1}{\omega - \epsilon_v} \frac{w_{vp}w_{sv}w_{qs}}
{(\epsilon_p - \epsilon_v)(\epsilon_s-\epsilon_v)}\,\, \frac{1}{\omega - \epsilon_q}
\end{align} 
These expressions can be compared with the 6 $\,t4$ diagrams. 
The diagrams $c5$, $c8$, $e5$, and $e8$, featuring both a $2h$-$1p$ and an $\omega$-independent $2p$-$2h$ denominator, can be combined such that these denominators cancel out.
Skipping the essentially obvious algebra, one arrives at the following result: 
\begin{equation}
(c5 +c8 + e5 + e8)|_{pq} = 2\,X_{pq}  
\end{equation}
The sum of the two remaining diagrams $c3$ and $c6$ can directly be identified with
$Y_{pq}$, 
\begin{equation}
(c3 +c6)|_{pq} =  Y_{pq}
\end{equation}
In an analogous way, the 6 $t5$ diagrams $b2$, $b4$, $b6$, $b7$, $f4$, and $f7$  relate to the $h.c.$ parts of $(C)$ and $(D)$.
Altogether, the $t4$ and $t5$ diagrams reproduce the $(C)$ and $(D)$ terms, leaving surpluses of $1/2 (c5 +c8 + e5 + e8)$ and $1/2 (b4 + b7 + f4 + f7)$, respectively, to be accomodated in $(A)$. This results in the following  
$hp$ contribution to $\bs f^{(3)}$:
\begin{equation}
\label{eq:t45f}
f_{pq}^{(3,t4/5)} = \half \sum_{v \leq N,s > N}  \frac{ w_{sp} w_{vs} w_{qv}}{(\epsilon_s-\epsilon_p)(\epsilon_q-\epsilon_v)(\epsilon_s-\epsilon_v)} \,n_p \lo n_q
\end{equation} 

Herewith the third-order ADC procedure for the NIP propagator part $\bs G^-$ is completed. As we have seen, the 60 diagrams of Fig.~\ref{fig:1pdo3} reproduce the ADC terms $(C)$ - $(I)$, consisting of lower-order ingredients, and specify the terms $(A)$ and $(B)$ from which the third-order contributions to $\bs C$ and $\bs f$ are derived. However, there is a general uniqueness issue concerning the distribution between certain $(A)$ and $(B)$ terms, which shall be briefly  addressed in the following.\\
\\
\noindent
\textbf{Uniqueness of the ADC procedure}\\
At a given order $n > 1$, the allocation of contributions to the ADC terms $(A)$ and $(B)$ 
(in obvious generalization of the third-order expressions in 
Eq.~\ref{eq:adcexpo3})
is not uniquely determined. The $hh$ part of $\bs f^{(n)}$ is not necessarily hermitian, and eventual anti-hermitian contributions can be shifted to the non-diagonal matrix elements of $\bs C^{(n)}$ (and vice versa). 
Let us consider the general $n$-th order expression for $(A)$,
\begin{equation}
(A) \equiv {\bs f^{(0)}}^\dagger (\omega - \bs K)^{-1} \bs f^{(n)} + {\bs f^{(n)}}^\dagger (\omega - \bs K)^{-1} \bs f^{(0)}
\end{equation}
and here, in particular, a $hh$ matrix element, $(A)|_{pq}, p \neq q$ and $p,q \leq N$. 
The matrix element $f_{pq}^{(n)}$, which here comes into play, 
can be decomposed 
into a hermitian and a (possibly non-vanishing) anti-hermitian part,
\begin{equation*}
f_{pq}^{(n)} = h_{pq} +  h'_{pq}
\end{equation*}
where
\begin{equation*}
h_{pq} = \half( f_{pq}^{(n)} +f_{qp}^{(n)*}), \;h'_{pq} = \half ( f_{pq}^{(n)} - f_{qp}^{(n)*})
\end{equation*}
The anti-hermitian part, $h'_{pq}$, generates a contribution to $(A)|_{pq}$, which is of the form
\begin{equation*}
 Z_{pq} = \frac{1}{\omega -\epsilon_p} h'_{pq} + (h'_{qp})^* \frac{1}{\omega -\epsilon_q}  
\end{equation*}
Using that  $(h'_{qp})^* = - h'_{pq}$ this becomes  
\begin{equation*}   
 Z_{pq} = \left (\frac{1}{\omega -\epsilon_p} -\frac{1}{\omega -\epsilon_q} \right ) h'_{pq}
  = \frac{1}{\omega -\epsilon_p}\frac{1}{\omega -\epsilon_q} (\epsilon_p - \epsilon_q)
    h'_{qp}
\end{equation*}
As the right-hand side of the last equation shows, $Z_{pq}$ can now be seen as a part of 
the matrix element $(B)|_{pq}$ of the ADC term $(B)$, establishing here a contribution 
\begin{equation}
(\epsilon_p - \epsilon_q)  h'_{qp} \rightarrow C_{pq}^{(n)}
\end{equation}
to the (non-diagonal) matrix element $C_{pq}^{(n)}$ of the secular matrix. 

In such a way, any anti-hermitian contributions can be removed from the $hh$ part of $\bs f^{(n)}$and transferred to $\bs C^{(n)}$. 
Implementing hermiticity in the $hh$ part of $\bs f^{(n)}$ at successive orders $n$ removes any ambiguity in the  
ADC procedure, resulting in uniquely-defined expressions for $\bs f$ and $\bs C$.

The second-order ADC procedure presented in Sec.~III gave directly a hermitian result for the
$hh$ part of $\bs f^{(2)}$, as shown by Eq.~(\ref{eq:f2pq}). This is not the case in the
third-order results derived above, 
comprising three contributions~(\ref{eq:Cpqt1},\ref{eq:Cpqt2},\ref{eq:Cpqt3}) 
to $\bs C^{(3)}$ and two contributions~(\ref{eq:t1f},\ref{eq:t2f}) to the $hh$ part of 
$\bs f^{(3)}$. Obviously, the $\bs f$ expressions are not hermitian, so that here the  procedure effecting a hermitian $hh$ block of $\bs f^{(3)}$ applies. 
Of course, this does not
disqualify the use of the original results for computational purposes. They constitute a fully legitimate third-order scheme, being in fact simpler than the 'unique' scheme. The two variants differ at third order in the non-diagonal matrix elements of $\bs C^{(3)}$, which means that computational discrepancies of the respective energy eigenvalues are of fourth order (as the PT expansion of the non-diagonal matrix elements begins in first order). 
\\
\\
\noindent
\textbf{Unique third-order results}\\
The reformulation of the original third-order expressions can readily be performed according to the operations 
\begin{align*}
C_{pq}^{(3)}\; &\leftarrow \;C_{pq}^{(3)} + (\epsilon_p - \epsilon_q) \half ( f_{pq}^{(3)} - f_{qp}^{(3)*})\\
f_{pq}^{(3)}\; &\leftarrow \;\half ( f_{pq}^{(3)} + f_{qp}^{(3)*})
\end{align*}
where in $f_{pq}^{(3)}$ both the $t1$ part~(\ref{eq:t1f}) and the $t2/t3$ part ~(\ref{eq:t2f}) have to be taken into account.
This results in the following three contributions to $\bs C^{(3)}$:
\begin{align}
\label{eq:Cpq31}
C_{pq}^{(3,1)} =& \half \sum_{r,s > N}  w_{rp} w_{sr} w_{qs}\left (\frac{1}{(\epsilon_p - \epsilon_r) (\epsilon_p - \epsilon_s)} + \frac{1}{(\epsilon_q - \epsilon_r) (\epsilon_q - \epsilon_s)}\right )\\
\label{eq:Cpq32}
C_{pq}^{(3,2)} =& \half \sum_{v\leq N, s>N} \frac{w_{vp}w_{sv}w_{qs}}{(\epsilon_p - \epsilon_s) (\epsilon_s - \epsilon_v)}\\ 
\label{eq:Cpq33}
C_{pq}^{(3,3)} =& \half \sum_{v\leq N, s>N} \frac{w_{sp}w_{vs}w_{qv}}{(\epsilon_q - \epsilon_s) (\epsilon_s - \epsilon_v)} 
\end{align}
Note that the new expressions $C_{pq}^{(3,2)}$ and $C_{pq}^{(3,2)}$ differ from their original
counterparts~(\ref{eq:Cpqt2}), (\ref{eq:Cpqt3}) only via the inconspicuous exchange of the 
denominators $(\epsilon_p - \epsilon_s)^{-1}$ and $(\epsilon_q - \epsilon_s)^{-1}$.

The new $hh$ part of  $\bs f^{(3)}$ is constituted by the contributions 
\begin{align}
\label{eq:f31}
f_{pq}^{(3,1)} =& - \half \sum_{r,s > N}  \frac{w_{rp} w_{sr} w_{qs}}{(\epsilon_p - \epsilon_r) (\epsilon_q - \epsilon_s)} \left (\frac{1}{\epsilon_p - \epsilon_s} + 
\frac{1}{\epsilon_q - \epsilon_r} \right )\, n_p n_q\\
\label{eq:f32}
f_{pq}^{(3,2)} =& \half \sum_{v\leq N, s>N} \frac{w_{vp}w_{sv}w_{qs}}{(\epsilon_p - \epsilon_s) (\epsilon_q - \epsilon_s)(\epsilon_v - \epsilon_s)} \, n_p n_q\\
\label{eq:f33}
f_{pq}^{(3,3)} =& \half \sum_{v\leq N, s>N} \frac{w_{sp}w_{vs}w_{qv}}
{(\epsilon_p - \epsilon_s) (\epsilon_q - \epsilon_s)(\epsilon_v - \epsilon_s)} \,n_p n_q
\end{align}
Note that the original $t2/t3$ contribution~(\ref{eq:t2f}) has now been replaced by two 
hermitian conjugate parts $f_{pq}^{(3,2)}$ and $f_{pq}^{(3,3)}$.

Finally, there are altogether 7 contributions to the $hp$ part of $\bs f^{(3)}$. One $hp$ contribution is given by Eq.~(\ref{eq:t45f}). Further six contributions can directly be derived from the diagrams $a9$ - $f9$ (or their \textit{h.c.} counterparts $a10$ -$f10$). The corresponding explicit expressions are listed in App. E.

\appendix
\renewcommand{\theequation}{B.\arabic{equation}}
\setcounter{equation}{0}
\section*{Appendix B: Biorthogonal coupled-cluster approach to non-interacting particles}

The purpose of this appendix is to give a proof for the block-diagonal structure of the NIP-ADC/ISR secular matrix. Since this proof makes explicit use of the biorthogonal coupled-cluster (BCC) representation of the ($N\!-\!1$)-particle excitations in a NIP system, we begin with a look at the BCC approach to non-interacting particles, which is of interest in itself.

In the BCC formulation (see e.g. Refs.~\cite{Helgaker:2000,Shavitt:2009}), the ground state of a system of $N$ non-interacting
particles (NIP) is given by  
\begin{equation}
\label{eq:bccgs}
\ket{\Psi^N_0} = e^{\hat T}\ket{\Phi^N_0} = ( 1 + \hat T + \half \hat T^2 + \sixth \hat T^3 + \dots)
\ket{\Phi^N_0}
\end{equation}
where the $\hat T$ operator is restricted to single excitations,
\begin{equation}
\hat T \equiv \hat T_1 = \sum_{a,k} t_{ak} c_a^\dagger c_k, \; \; a > N, k \leq N 
\end{equation}
The amplitudes $t_{ak}$ can be determined by the NIP-CC equations
\begin{align}
\nonumber
0  =& \dirint{\Phi_{ak}}{e^{-\hat T_1} (\hat H_0 + \hat W) e^{\hat T_1}}{\Phi^N_0} \\
\label{eq:ccampl}
   =& \dirint{\Phi_{ak}}{(1 - \hat T_1)(\hat H_0 + \hat W) (1 + \hat T_1 + \half \hat T_1^2)}{\Phi^N_0}
\end{align}
where $\ket{\Phi_{ak}} = c_a^\dagger c_k \ket{\Phi^N_0}$.
For obvious reasons, the left and right exponential expansions terminate after the linear and quadratic term, respectively. 
The ground state energy is given by
\begin{equation}
\label{eq:ccgse}
E^N_0 = \dirint{\Phi^N_0}{(\hat H_0 + \hat W) e^{\hat T_1}}{\Phi^N_0} =
\sum_k (\epsilon_k + w_{kk}) + \sum_{b,l}
t_{bl}\,w_{lb} 
\end{equation} 
Closed-form expressions for the NIP-CC amplitudes have been derived by Thouless~\cite{tho60:225}. In matrix form, they
read  
\begin{equation}
\label{eq:toft}
\bs t = \bs X_{21} \bs X_{11}^{-1}
\end{equation}
where $\bs t$ denotes the matrix of the amplitudes, $t_{ak}, a > N, k \leq N$, 
and $\bs X_{11}$ and $\bs X_{21}$ are blocks of the eigenvector matrix $\bs X$ of the 
secular Eq.~(\ref{eq:seceq}); the partitioning scheme is as specified by Eq.~(\ref{eq:part}). 
While the original proof of Eq.~(\ref{eq:toft}) is constructive, that is, independent of the 
CC amplitude equations~(\ref{eq:ccampl}), it can be verified that the Thouless expression  
solves the latter equations. Using the Thouless expression in Eq.~(\ref{eq:ccgse}) reproduces the direct ground state energy result of Eq.~(\ref{eq:mbgsen}).

The biorthogonal CC approach to general excitations, here in the ($N\!-\!1$)-particle system, is based on the 
non-hermitian representation of the 
(shifted) hamiltonian $\hat{H}-E^N_{0}$, 
\begin{equation}
\label{eq:ccsm}
M^{cc}_{IJ}=\bra{\lo{\Phi}_{I}}\hat{H}-E^N_{0}\ket{\Psi_{J}^{0}}
= \dirint{\Phi^N_{0}}{\hat{C}_{I}^{\dagger} e^{-\hat{T}_1}
[\hat{H},\hat{C}_{J}] e^{\hat{T}_1}}{\Phi^N_{0}}.
\end{equation}
in terms of two distinct sets of ($N\!-\!1$)-particle states.
There is a right expansion manifold $\{R\}$ formed by
the correlated excited states,
\begin{equation}
\label{eq:ccstates}
\ket{\Psi^0_{J}}=
\hat{C}_{J} \ket{\Psi^{cc}_0}=
\hat{C}_{J} e^{\hat{T}_1}\ket{\Phi^N_{0}} =  e^{\hat{T}_1}\hat{C}_{J} \ket{\Phi^N_{0}}
\end{equation}
where $\hat C_J$ denote physical excitation operators of the manifold of $1h$, $2h$-$1p$, $\dots$, excitations,
\begin{equation}
\label{eq:bccmf}
\{\hat C_J\} = \{c_k; c_a^\dagger c_k c_l, k<l; \dots\}
\end{equation}
The left expansion manifold $\{L\}$ is formed by the associated biorthogonal states, 
\begin{equation}
\label{eq:bostates}
\bra{\lo{\Phi}_{I}}=
\bra{\Phi^N_{0}}\hat{C}_{I}^{\dagger} e^{-\hat{T}_1}
\end{equation}

The block structure of the NIP-BCC secular matrix $\bs M^{cc}$ is as shown in Fig.~\ref{fig:bccorfigx}, that is, 
featuring a CI structure in the upper right part, but being block-diagonal in the lower left part. 
The CI-type structure in the upper right part can be stated as
\begin{equation}
\label{eq:bstruct1}
M^{cc}_{IJ} = \begin{cases} \dirint{\Phi_I}{\hat W}{\Phi_J}, \; [J] = [I] +1 \\
0, \; [J] > [I] +1
\end{cases}
\end{equation}
This is a trivial consequence of the states in the right and left expansion manifolds. 
Obviously, a CE state of class $[I]$ can be written according to
\begin{equation}
\label{eq:rset}
\ket{\Psi^0_I} = 
\ket{\Phi_I} +\sum_{K,\, [K]>[I]}   z^{(I)}_K \ket{\Phi_K}   
\end{equation}
as a linear combination of $\ket{\Phi_I}$ and CI configurations of \emph{higher} excitation classes,
$[K]>[I]$. 
By contrast, the CI expansion of a state of the $\{L\}$ set reads 
\begin{equation}
\label{eq:lset}
\bra{\lo{\Phi}_I} = 
\bra{\Phi_I} +\sum_{K,\, [K]<[I]} \lo z^{(I)}_K \bra{\Phi_K}   
\end{equation}
that is, a linear combination of $\bra{\Phi_I}$ and CI excitations of \emph{lower} classes, 
$[K]=1,\dots,[I]-1$ 
As $\hat W$ is a one-particle operator, the matrix element~(\ref{eq:ccsm}) vanishes if $[J] > [I] + 1$, and 
becomes the CI result for $[J] = [I] + 1$.

The remarkable finding is the block-diagonal structure of the lower left part,
\begin{equation}
\label{eq:eq:bstruct2}
M^{cc}_{IJ} = 0 \;\;\text{for}\;\; [I] > [J]
\end{equation}
which will be proven in the following.
First we consider Eq.~(\ref{eq:ccsm}) in the form
\begin{equation}
M^{cc}_{IJ}= \dirint{\Phi^N_{0}}{\hat{C}_{I}^{\dagger} e^{-\hat{T}_1}
\hat{H}  e^{\hat{T}_1} \hat{C}_{J}}{\Phi^N_{0}}, \; I\neq J
\end{equation}
adapted to non-diagonal elements of the BCC secular matrix, 
where the commutator on the right-hand side of Eq.~(\ref{eq:ccsm}) is redundant.
The Baker-Campbell-Hausdorff (BCH) expansions for the similarity-transformed hamiltonian or its constituents 
$\hat H_0$ and $\hat W$,
\begin{align*}
e^{-\hat{T}_1} \hat{H}_0 e^{\hat{T}_1} =& \hat{H}_0 + [\hat{H}_0, \hat T_1]\\
e^{-\hat{T}_1} \hat W e^{\hat{T}_1} =& \hat W + [\hat W, \hat T_1] + 
\half [[\hat W, \hat T_1], \hat T_1]
\end{align*}
terminate after the single and double commutator, respectively. ($\hat W$ has at most two 'unphysical' fermion operators, and each commutator with $\hat T_1$ eliminates one unphysical operator.) Moreover, this shows that $e^{-\hat{T}_1} \hat{H} e^{\hat{T}_1}$ is an operator of rank 1. Let us note that the rank of a fermion operator product denotes the number of creation operators; the rank of a general operator is given by the maximal rank of its operator product constituents. Both $\hat H_0$ and $\hat W$ are of rank 1, and the commutator with the  operator $\hat T_1$, being of rank 1, does not increase the rank.
Obviously, a matrix element of the type  
\begin{equation*}
\dirint{\Phi^N_{0}}{\hat{C}_{I}^{\dagger}\hat O(1) \hat{C}_{J}}{\Phi^N_{0}} 
\end{equation*}
whith $\hat O(1)$ being an operator of rank 1, vanishes if $[I] > [J] + 1$. Note that 
the rank of an ($N\!-\!1$)-particle excitation operator $\hat C_K$ is given by $[K]-1$.

So it remains to show the assertion for $[I] = [J] + 1$.
Let us here focus on the case 
where $I = aij$ is a $2h$-$1p$ excitation (class 2) and  $J = k$ is a $1h$ excitation 
(class 1) and return to the more general commutator expression of Eq.~(\ref{eq:ccsm}), 
\begin{equation}
\label{eq:maijk}
M^{cc}_{aij, k} = \dirint{\Phi^N_{0}}{c_i^\dagger c_j^{\dagger} c_a e^{-\hat{T}_1}
[\hat H, c_k] e^{\hat{T}_1}}{\Phi^N_{0}}
\end{equation}
The commutator $\hat K_k = [\hat H, c_k]$ can readily be evaluated to give 
\begin{equation}
\hat K_k = [\hat H_0 + \hat W, c_k] = - \epsilon_k c_k - \sum w_{kq} c_q
\end{equation}
and the ensuing similarity transformation becomes
\begin{equation}
e^{-\hat{T}_1} \hat K_k e^{\hat{T}_1} = \hat K_k + [\hat K_k,\hat{T}_1]
\end{equation}
This is a sum over $1h$ operators (being of rank 0), so that the matrix element (Eq.~\ref{eq:maijk}) with a $2h$-$1p$ excitation $c_a^\dagger c_i c_j$ (being of rank 1) vanishes.
The latter argumentation can easily be generalized to arbitrary excitations $I,J$ with $[I] = [J] + 1$. As should be noted, the latter proof can of course also be done without resorting to the commutator, which, however, requires explicit use of the CC amplitude equations.

\subsection*{Proof of the block-diagonal structure of the NIP-ISR secular matrix}
The full ISR secular matrix $\lo{\bs M}$ introduced in Sec.~IV.A is block-diagonal, that is,
\begin{equation}
\label{eq:corxx}
\lo M_{IJ} = \dirint{\tilde{\Psi}_{I}}{\hat H}{\tilde{\Psi}_{J}} = 0 \; \text{if} \; 
[I] \neq [J]
\end{equation}
The proof of this property given in the following will build on the block-diagonal structure established 
above for the lower left part of the NIP-BCC secular matrix $\bs M^{cc}$. 
Here, we follow the essentially analogous proof for the canonical order relations 
in the general ISR secular matrices~\cite{mer96:2140}.

The BCC expansion manifolds, given by the CE states~(\ref{eq:ccstates}) and their biorthogonal counterparts~(\ref{eq:bostates}), allow one to define a 
biorthogonal resolution of the identity (ROI) according to
\begin{equation}
\hat{\mathbb 1} = \sum_K \ket{\Psi_{K}^{0}}\bra{\lo{\Phi}_{K}}
\end{equation}
Assuming $[I] > [J]$ and inserting the biorthogonal ROI twice on the right-hand side of Eq.~(\ref{eq:corxx}) yields the following expressions 
\begin{equation}
\label{eq:mroi}
\lo M_{IJ}
= \sum_{K,L} \braket{\tilde{\Psi}_{I}}{\Psi_{K}^{0}}\dirint{\lo{\Phi}_{K}}{\hat H}{\Psi_{L}^{0}}\braket{\lo{\Phi}_L}{\tilde{\Psi}_{J}}, \; [I] > [J]
\end{equation}
which need to be further analyzed.
By construction, the intermediate state $\ket {\tilde{\Psi}_{I}}$ is orthogonal to the CE states 
$\ket{\Psi_{K}^{0}}$ of lower classes, $[K] < [I]$. 
Therefore the summation over $K$ is restricted to $[K] \geq [I]$. 
An opposite restriction, namely, $[J] \geq [L]$ applies to the summation over $L$. 
To see this, consider the 
matrix element 
\begin{equation}
 \braket{\lo{\Phi}_L}{\tilde{\Psi}_{J}} = \dirint{\Phi^N_0}{\hat C^\dagger_L e^{-\hat T}}{\tilde{\Psi}_J} 
\end{equation} 
and recall that by construction $\ket{\tilde{\Psi}_J}$ is of the form 
\begin{equation} 
\ket{\tilde{\Psi}_J} = \sum_{R, [R] \leq [J]} \tilde{z}^{(J)}_R \hat C_{R} \ket{\Psi^N_0} 
\end{equation} 
where the excitation classes of the operators $\hat C_{R}$ are restricted to $[R] \leq [J]$. 
As a consequence, 
\begin{equation} 
\braket{\lo{\Phi}_L}{\tilde{\Psi}_{J}} =   \sum_{R, [R] \leq [J]} \tilde{z}^{(J)}_R\, \dirint{\Phi_0}{\hat C^\dagger_L  \hat C_{R} }{\Phi^N_0} = 0, \;\;\text{if} \;[L] > [J]
\end{equation} 
Here $e^{-\hat T}$ has been commuted to the right, yielding $e^{-\hat T} \ket{\Psi^N_0} = \ket{\Phi^N_0}$. 
Altogether, the index restrictions in the summations on the right-hand side of 
Eq.~(\ref{eq:mroi}) are $[K] \geq [I] > [J] \geq [L]$, and, in particular, $[K] > [L]$, so that the BCC matrix elements in the sum~(\ref{eq:mroi}), 
\begin{equation}
M^{cc}_{KL} = \dirint{\lo{\Phi}_{K}}{\hat H}{\Psi_{L}^{0}}   
\end{equation}
vanish according to Eq.~(\ref{eq:eq:bstruct2}). This means that $\lo M_{IJ} = 0$ for
$[I] > [J]$. Since  $\lo{\bs M}$ is hermitian, this applies also to $[I] < [J]$, that is,
$\lo M_{IJ} = 0$ if $[I] \neq [J]$.

\subsection*{A look at unitary coupled-cluster (UCC) for NIP systems}
For comparison, let us briefly consider how the unitary coupled-cluster (UCC) method~\cite{Muk:1989} behaves in the case of a NIP system. (For a review of the general UCC approach to electron detachment and further references the reader is also referred to a recent paper by Hodecker et al.~\cite{hod22:074104}).  

In the UCC formulation, the $N$-particle ground state is given by  
\begin{equation}
\label{uccent}
\ket{\Psi_0^N} = e^{\hat \sigma}\ket{\Phi_0^N} 
\end{equation}
where the anti-hermitian operator in the exponential, 
\begin{equation}
\label{eq:siop}
\hat \sigma \equiv  \hat \sigma_1 = \hat S_1 - \hat S_1^\dagger 
\end{equation}
is constituted by single excitations, 
\begin{equation}
\hat S_1 = \sum s_{ak} c_a^\dagger c_k, \; \; a > N, k \leq N
\end{equation}
Note that in the UCC form the ground state is normalized to 1, whereas intermediate normalization applies to the BCC expression~(\ref{eq:bccgs}), 
which, after all, is the only difference.

In contrast to the BCC method, the UCC amplitude equations, reading 
\begin{equation}
\label{ucc1} 
0 = \dirint{\Phi_{ak}}{e^{-\hat \sigma}(\hat H_0 + \hat W)e^{\hat \sigma}}{\Phi_0^N}
\end{equation}
do not terminate at a finite power in the BCH expansion of $e^{-\hat \sigma} \hat H e^{\hat \sigma}$. In fact, the UCC amplitude equations exhibit a remarkable complexity, as can be seen, for instance,
by performing a PT analysis of the NIP-UCC amplitudes through third order.

The UCC treatment of ($N\!-\!1$)-particle excitations is based on 
on $1h, 2h$-$1p, \dots$ states being of the form   
\begin{equation} 
\ket{\vtilde{\Psi}_J} = e^{\hat \sigma}\hat C_J \ket{\Phi_0^N}
\end{equation}
where $\hat C_J$ are excitation operators from the manifold~(\ref{eq:bccmf}). Obviously, these states are orthonormal, so that they can directly be used 
to establish a hermitian representation $\bs M^{ucc}$ of the (shifted) hamiltonian (UCC secular matrix). In the NIP case, 
$\bs M^{ucc}$ is block-diagonal. Moreover, the $1h$ UCC states are identical with the corresponding intermediate states (\ref{eq:isconstr2})
\begin{equation}
\label{eq:uccisr}
\ket{\vtilde{\Psi}_k} = \ket{\tilde{\Psi}_k}, \;\; k \leq N
\end{equation}
A proof and analysis of these findings is given in a separate paper~\cite{tbp1}, where also an alternative proof for the block-diagonality of the ISR secular matrix $\lo{\bs M}$ is presented.

\appendix
\renewcommand{\theequation}{C.\arabic{equation}}
\setcounter{equation}{0}
\section*{Appendix C: NIP-ISR procedure in second order}
In the following we address the evaluation of 
the \textbf{second-order} contributions to the ISR secular matrix elements,
\begin{equation}
\label{eq:mkl1}
M_{kl} = \dirint{\tilde \Psi_k}{\hat H - E^N_0}{\tilde \Psi_{l}}
\end{equation}
where the $1h$ intermediate states are given by
\begin{equation}
\label{eq:psik}
\ket{\tilde \Psi_k} = \sum_i c_{i}\ket{\Psi_0^N} (\bs S^{-1/2})_{ik}
\end{equation}
and $\bs S $ is the overlap matrix of the CE states $\ket{\Psi_k^0} = c_k\ket{\Psi_0^N}$, as specified in Eq.~(\ref{eq:cesolm}).

All what is needed to this end, is the first-order expansion of the ground state,
$\ket{\Psi_0^N} = \ket{\Phi_0^N} + \ket{\Psi_0^{(1)}} + O(2)$.
We recall that the first-oder ground-state contribution reads 
\begin{equation}
\label{eq:gsao1}
\ket{\Psi_0^{(1)}} = \sum t_{ai}^{(1)} c_a^\dagger c_i \ket{\Phi_0}
\end{equation}
where the first-order amplitudes are given by
\begin{equation}
\label{eq:gsao1x}
t_{ai}^{(1)} = - \frac{w_{ai}}{\epsilon_a - \epsilon_i} 
\end{equation}

Let us first consider the PT expansion of $\bs S^{-1/2}$. Since 
$\bs S$ is a hermitian and positive definite matrix, the matrix $\bs S^{1/2}$
is well-defined. Moreover, the PT expansion of $\bs S$ has the structure
\begin{equation*}
\bs S = \bs 1 +  \bs S^{(2)} + O(3)
\end{equation*}
since the first-order contributions vanish,
\begin{equation*}
S^{(1)}_{kl} = \dirint{\Psi_0^{(1)}}{c_k^\dagger c_l}{\Phi_0^N} + \dirint{\Phi_0^N}{c_k^\dagger c_l}{\Psi_0^{(1)}} =
\delta_{kl}( \braket{\Psi_0^{(1)}}{\Phi_0^N} + \braket{\Phi^N_0}{\Psi_0^{(1)}}) = 0
\end{equation*}
This, in turn, entails the PT expansion 
\begin{equation}
\label{eq:smhalf}
\bs S^{-1/2} = \bs 1 - \half  \bs S^{(2)} + O(3)
\end{equation}
for the inverse of $\bs S^{1/2}$. 

In second-order there are three contributions to $\bs S$, 
\begin{equation}
S_{kl}^{(2)} = \dirint{\Psi^{(1)}_0}{c_k^\dagger c_l}{\Psi^{(1)}_0}
+ \dirint{\Psi^{(2)}_0}{c_k^\dagger c_l}{\Phi_0^N} + \dirint{\Phi_0^N}{c_k^\dagger c_l}{\Psi^{(2)}_0}
\end{equation}
The latter two, involving $\ket{\Psi^{(2)}_0}$, are seen to vanish, as, for example, 
\begin{equation}
\dirint{\Phi_0^N}{c_k^\dagger c_l}{\Psi^{(2)}_0} = \delta_{kl}\braket{\Phi_0^N}{\Psi_0^{(2)}} = 0 
\end{equation}
which is a consequence of the intermediate normalization supposed for $\ket{\Psi_0^N}$.
The remaining second-order contribution can readily be evaluated (where it is helpful to distinguish 
the cases $k<l$ and $k=l$):
\begin{equation}
\label{eq:so2}
 S_{kl}^{(2)} =  \dirint{\Psi^{(1)}_0}{c_k^\dagger c_l}{\Psi^{(1)}_0}
= \delta_{kl} I^{(2)}_0 - \sum_{a} t_{ak}^{(1)}t^{(1)*}_{al}
\end{equation}
Here $I^{(2)}_0$ is the second-order contribution to the normalization integral $\braket{\Psi^N_0}{\Psi^N_0}$,
\begin{equation*}
I^{(2)}_0 = \sum_{a,k}|t_{ak}^{(1)}|^2
\end{equation*} 
To deal with the secular matrix elements, we write Eq.~(\ref{eq:mkl1}) in a more explicit form,
\begin{align}
\nonumber
M_{kl} =& \sum_{k',l'} 
\bs (\bs S^{-1/2})_{kk'} \dirint{\Psi_0^N}{c_{k'}^\dagger (\hat H - E^N_0) c_{l'} }{\Psi_0^N}(\bs S^{-1/2})_{l'l} \\
\nonumber
=& \dirint{\Psi_0^N}{c_{k}^\dagger (\hat H - E^N_0) c_{l} }{\Psi_0^N} 
- \half \sum_{l'} \dirint{\Psi_0^N}{c_{k}^\dagger (\hat H - E^N_0) c_{l'} }{\Psi_0^N}S^{(2)}_{l'l}\\
\label{eq:mklexp}
&-\half \sum_{k'} \dirint{\Psi_0^N}{c_{k'}^\dagger (\hat H - E^N_0) c_{l} }{\Psi_0^N}S^{(2)}_{kk'} + O(3)
\end{align}
and specify the second-order contributions $M_{kl}^{(2)}$ arising here. 
Obviously, in the last two terms the second-order 
matrix elements $S^{(2)}_{ij}$ need to be combined with zeroth-order contributions of the
ground-state expectation values. The resulting second-order contribution to $M_{kl}$,
reads
\begin{equation}
(C1) \equiv  \half (\epsilon_k + \epsilon_l) S^{(2)}_{kl}
\end{equation}
with $S^{(2)}_{kl}$ as given by Eq.~(\ref{eq:so2}).
There are three more second-order contributions stemming from
the first term on the right-hand side of Eq.~(\ref{eq:mklexp}):
\begin{align*}
(C2) \equiv & \dirint{\Psi^{(1)}_0}{c_{k}^\dagger (\hat H_0 - E^{(0)}_0) c_{l}}{\Psi^{(1)}_0}\\
(C3) \equiv & \dirint{\Psi^{(1)}_0}{c_{k}^\dagger \hat H_I c_{l}}{\Phi^N_0} + 
\dirint{\Phi^N_0}{c_{k}^\dagger \hat H_I c_{l}}{\Psi^{(1)}_0}\\
(C4) \equiv & - E^{(2)}_0 \delta_{kl}
\end{align*}
Note that contributions involving $\ket{\Psi^{(2)}_0}$, such as
$\dirint{\Psi^{(2)}_0}{c_{k}^\dagger (\hat H_0 - E^{(0)}_0) c_{l}}{\Phi_0^N} = \delta_{kl} \epsilon_k
 \braket{\Psi^{(2)}_0}{\Phi_0^N}$, vanish (again supposing intermediate normalization).  
Moreover, there are no contributions involving $E^{(1)}_0$, since here the
accompanying matrix elements vanish, as, for example, in
$\dirint{\Psi^{(1)}_0}{c_{k}^\dagger c_{l}}{\Phi_0^N} = \delta_{kl}\braket{\Psi^{(1)}_0}{\Phi_0^N} = 0$.

In evaluating ($C2$) and ($C3$), it is advantageous to treat the
cases $k = l$ and $k\neq l$ separately. 
Supposing $k<l$, one obtains the following expressions
\begin{align*}
(C2) =&  -\sum_{a} t_{ak}^{(1)}t^{(1)*}_{al} (\epsilon_a - \epsilon_k -\epsilon_l)\\
(C3) =& + \sum_{a} (t_{ak}^{(1)}w_{al}^* + w_{ak}t_{al}^{(1)*}) 
\end{align*}
Now the three non-vanishing second-order contributions ($C1$), ($C2$), and ($C3$), 
differing only in their respective orbital energy factors,
can be combined, yielding 
\begin{equation}
\label{eq:mklo2}
M^{(2)}_{kl} =  \sum_{a} w_{ak} w_{al}^* \frac{(\epsilon_a - \half \epsilon_k -\half \epsilon_l)}
{(\epsilon_a - \epsilon_k)(\epsilon_a - \epsilon_l)}  
\end{equation}
It is interesting to note that $(C1)+(C2) = -\half (C3)$ or $(C1)+(C2)+(C3) = \half (C3)$.
The expression~(\ref{eq:mklo2}), derived for $k<l$, applies also to $k>l$ and, in particular, to $k = l$, where, of course, the diagonal ($C4$) contribution has to be taken into account.

In a similar way the explicit ISR(2) expressions for
\begin{equation}
\label{eq:ftilde}
 \tilde f_{kp} = \dirint{\tilde \Psi_k}{c_p}{\Psi_0^N}\\
\end{equation}
can be evaluated. Note that in these matrix elements $\ket{\Psi_0^N}$ has to be normalized to 1.

\appendix
\renewcommand{\theequation}{D.\arabic{equation}}
\setcounter{equation}{0}
\section*{Appendix D: Perturbation expansions for the wave operator and non-hermitian effective hamiltonian}
The following is a brief review of the derivation of PT expansions for the wave operator and the associated effective hamiltonian via the Bloch equation, as presented for example in Refs.~\cite{kva74:605,lin74:2441,svr87:625}. It should be noted that there is also an alternative approach to the Bloch equation based on an iteration procedure~\cite{dur83:3184}.

The wave operator $\hat \Omega$, as given by Eq.~\ref{eq:bloch}, fulfils the Bloch equation~\cite{blo58:329,lin74:2441}
\begin{equation}
\label{eq:blochx}
[\hat \Omega,\hat{\mathcal{H}}_0] = \hat{\mathcal{W}}\hat \Omega -\hat \Omega \hat{\mathcal{W}}\hat \Omega
\end{equation}
which allows one to establish a perturbation expansion for $\hat \Omega$ based on the 
partitioning~(\ref{eq:Hgen}) of the underlying hamiltonian.  
It should be noted that there is a more general form of the Bloch equation~\cite{dur83:3184},
\begin{equation}
\label{eq:gbloch}
\hat{\mathcal{H}}\hat \Omega = \hat \Omega \hat{\mathcal{H}}\hat \Omega
\end{equation}
which does not take recourse to the PT partitioning of the hamiltonian. The PT 
variant~(\ref{eq:blochx}) can easily be retrieved from the general equation upon resorting to the partitioned hamiltonian. Both forms of the Bloch equation can also be verified by employing the \textit{a posteriori} expression of Eq.~(\ref{eq:rwopx}) for $\hat \Omega$.
 
Supposing the formal PT expansion 
\begin{equation}
\hat \Omega = \hat \Omega^{(0)} + \hat \Omega^{(1)} + \hat \Omega^{(2)} + \dots
\end{equation}
for the wave operator, where $\hat \Omega^{(0)} = \hat P_0$, and 
using this expansion in the Bloch equation results in a recursion relation~\cite{lin74:2441} for the successive terms $\hat \Omega^{(m)}$ in the PT expansion:
\begin{equation}
\label{eq:woprec}
[\hat \Omega^{(m)}, \hat{\mathcal{H}}_0] = \hat{\mathcal{W}} \hat \Omega^{(m-1)} -
\sum_{\kappa = 0}^{m-1} \hat \Omega^{(m- \kappa - 1)} \hat{\mathcal{W}}\hat \Omega^{(\kappa)},
\; m = 1,2,\dots 
\end{equation}
A particularity of this recursion equation is that the commutator on the left-hand side can only be evaluated in application to a distinct model state, 
\begin{equation*}
[\hat \Omega^{(m)}, \hat{\mathcal{H}}_0] \Phi_\nu = (\mathcal E_\nu - \hat{\mathcal{H}}_0)
\hat \Omega^{(m)}\Phi_\nu 
\end{equation*}
This means that the recursion does not provide the operator terms $\hat \Omega^{(n)}, n= 1,2,\dots$ themselves but rather the states $\hat \Omega^{(n)} \Phi_\nu, \nu \leq N$.

Let us consider the explicit construction in first and second order. 
In first order Eq.~(\ref{eq:woprec}) becomes 
\begin{equation}
[\hat \Omega^{(1)}, \hat{\mathcal{H}}_0] = \hat{\mathcal{W}} \hat P_0 -\hat P_0 \hat{\mathcal{W}}\hat P_0 = \hat Q_0 \hat{\mathcal{W}}\hat P_0
\end{equation}
To proceed we apply the operator expressions on both sides to a model state,
\begin{equation}
 (\mathcal E_\nu - \hat{\mathcal H}_0) \hat \Omega^{(1)} \ket{\Phi_\nu}
 = \hat Q_0 \hat{\mathcal{W}} \ket{\Phi_\nu}, \; \nu \leq N 
\end{equation} 
which leads to the explicit expression 
\begin{equation}
\label{eq:Omeg1}
\hat \Omega^{(1)} \ket{\Phi_\nu} = \frac{\hat Q_0}{\mathcal E_\nu - \hat{\mathcal{H}}_0}
\hat{\mathcal{W}}\,\ket{\Phi_\nu},\; \nu  \leq N  
\end{equation}
Obviously, this not an expression for $\hat \Omega^{(1)}$ itself but
for its application to an arbitrary model state, $\hat \Omega^{(1)} \ket{\Phi_\nu}$.
This is a characteristic feature of the Bloch equation in the case of non-degenerate model states.

For the second-order term, $m=2$, Eq.~(\ref{eq:woprec}) gives  
\begin{equation}
[\hat \Omega^{(2)}, \hat{\mathcal{H}}_0] = \hat{\mathcal{W}} \hat \Omega^{(1)} -
\hat \Omega^{(1)} \hat{\mathcal{W}}\hat P_0 - \hat P_0 \hat{\mathcal{W}} \hat \Omega^{(1)} =
\hat Q_0 \hat{\mathcal{W}} \hat \Omega^{(1)} - \hat \Omega^{(1)} \hat{\mathcal{W}}\hat P_0  
\end{equation}
Again, we have to apply the operator terms to a model state, which 
gives 
\begin{equation}
\hat \Omega^{(2)} \ket{\Phi_\nu}  =  \frac{1}{\mathcal E_\nu - \hat{\mathcal{H}}_0}
\left (\hat Q_0 \hat{\mathcal{W}} \hat{\Omega}^{(1)}\ket{\Phi_\nu}
- \hat \Omega^{(1)}\hat{\mathcal{W}}\, \ket{\Phi_\nu}\right ),\; \nu  
\leq N 
\end{equation} 
In the first term on the right-hand side, we can directly use the result~(\ref{eq:Omeg1}) for 
$\hat \Omega^{(1)} \ket{\Phi_\nu}$. In the second term, however, $\hat \Omega^{(1)}$ 
is followed by the $\hat{\mathcal{W}}$ operator. This can be handled by inserting the full
set of model states according to 
\begin{equation}
\hat \Omega^{(1)}\hat{\mathcal{W}}\, \ket{\Phi_\nu} = \hat \Omega^{(1)} \sum_{\mu = 1}^N \ket{\Phi_\mu}
\bra{\Phi_\mu}\hat{\mathcal{W}}\,\ket{\Phi_\nu} =
\sum_{\mu = 1}^N \frac{\hat Q_0}{\mathcal E_\mu - \hat{\mathcal{H}}_0}
\hat{\mathcal{W}}\,\ket{\Phi_\mu} \dirint{\Phi_\mu}{\hat{\mathcal{W}}}{\Phi_\nu}
\end{equation}
The final result reads
\begin{align}
\nonumber
\hat \Omega^{(2)} \ket{\Phi_\nu} =& 
\frac{\hat Q_0}{\mathcal E_\nu - \hat{\mathcal{H}}_0} \hat{\mathcal{W}} \frac{\hat Q_0}{\mathcal E_\nu - \hat{\mathcal{H}}_0} \hat{\mathcal{W}} \ket{\Phi_\nu} \\
&\phantom{xxx} -\sum_{\mu = 1}^N \frac{\hat Q_0}{(\mathcal E_\nu - \hat{\mathcal{H}}_0)(\mathcal E_\mu - \hat{\mathcal{H}}_0)}
\hat{\mathcal{W}}\,\ket{\Phi_\mu} \dirint{\Phi_\mu}{\hat{\mathcal{W}}}{\Phi_\nu},
 \; \nu \leq N
\end{align}
In a similar way the third- and higher-order contributions can be obtained.

The PT expansion of the wave operator can directly be applied to the  
matrix elements
\begin{equation}
\tilde{\mathcal H}'_{\mu \nu} = \dirint{\Phi_\mu}{\hat{\mathcal{H}}^{nh}_{eff}}{\Phi_\nu}, \;\; \mu,\nu \leq N 
\end{equation}
of the non-hermitian effective hamiltonian (Eq.~\ref{eq:effh}),
\begin{equation}
\hat{\mathcal{H}}^{nh}_{eff} = \hat P_0 \hat{\mathcal{H}} \hat \Omega 
= \hat P_0 \hat{\mathcal{H}}_0 \hat P_0 + \hat P_0 \hat{\mathcal{W}} \hat \Omega 
\end{equation}
The second-order expansion of $\hat \Omega$ gives rise to the third-order expansion
of $\tilde{\bs{\mathcal{H}}}'$. 
This expansion reads
\begin{align}
\nonumber
&\tilde{\mathcal{H}}'^{(0)}_{kl} + \tilde{\mathcal{H}}'^{(1)}_{kl} = \mathcal E_k \delta_{kl} + W_{kl}\\
\nonumber
&\tilde{\mathcal{H}}'^{(2)}_{kl} = \sum_a \frac{W_{ka}W_{al}}{\mathcal E_k -\mathcal E_a}
\\
&\tilde{\mathcal{H}}'^{(3)}_{kl} = 
\sum_{a,b} \frac{W_{ka} W_{ab}W_{bl} }{(\mathcal E_l -\mathcal E_a) (\mathcal E_l -\mathcal E_b)} -
\sum_{a,j} \frac{W_{ka} W_{aj}W_{jl} }{(\mathcal E_l -\mathcal E_a) (\mathcal E_j -\mathcal E_a)} 
\end{align}
Here, the "quantum chemical (QC)" notation is used in which
the indices $i,j,k,\dots$ denote model states ($\nu \leq N$), whereas $a,b,c,\dots$ refer to complementary model states ($\nu > N$).
These results can be confirmed by comparison with the not entirely explicit expressions given in Refs.~\cite{kva74:605,lin74:2441,svr87:625}.

The second- and third-order contributions to $\tilde{\bs{\mathcal{H}}}'$ are manifestly non-hermitian. Interestingly, the simple \textit{ad hoc} symmetrization reproduces the corresponding terms in the PT expansion of the hermitian effective hamiltonian 
$\tilde{\bs{\mathcal{H}}}$,
\begin{equation}
\half (\tilde{\mathcal{H}}'^{(n)}_{kl} + \tilde{\mathcal{H}}'^{(n)\,*}_{lk} ) =
\tilde{\mathcal{H}}^{(n)}_{kl}, \; n=2,3
\end{equation}
as can be seen by comparing with the hermitian results displayed in App.~E.
However, this finding will hardly apply in higher orders. In view of the strikingly different \textit{a posteriori} expressions~(\ref{eq:nheffh11}) and (\ref{eq:cfheff}) an identity of
$1/2(\tilde{\bs{\mathcal{H}}'} + \tilde{\bs{\mathcal{H}}}'^\dagger)$ and
$\tilde{\bs{\mathcal{H}}}$ is all but excluded. Indeed, discrepancies set in at fourth order~\cite{svr87:625}.

How to get a PT expansion for the hermitian effective hamiltonian associated with the symmetrized wave operator $\hat \Omega_S$ according to Eq.~(\ref{eq:uwop})?
Here it is to be noted that the Bloch equations do not apply to $\hat \Omega_S$. 
In fact, $\hat \Omega_S$ satisfies only two of the four properties~(\ref{eq:wopp1},\ref{eq:wopp2}), as $\hat \Omega_S \hat P \neq \hat P$ 
and $\hat P_0 \hat \Omega_S  \neq \hat P_0$. Moreover, $\hat \Omega_S$ lacks the projector property~(\ref{eq:wopp3}), that is, $\hat \Omega_S^2 \neq \hat \Omega_S$. 
This can be easily seen by using the \textit{a posteriori} expression~(\ref{eq:wops}) together 
with the explicit expressions~(\ref{eq:p0q0}) and (\ref{eq:tsp}) for $\hat P_0$ and $\hat P $,
respectively. This means there is no simple method to 
generate PT expansions for $\hat \Omega_S$ and the associated hermitian effective hamiltonian within the wave operator based QDPT approach. A way out has been shown by
Svr\v{c}ek and Huba\v{c}~\cite{svr87:625}. By combining the PT expansion for $\hat \Omega$
with an expansion of the inverse square root in the expression~(\ref{eq:uwop}) these authors derived the PT expansion of $\tilde{\bs{\mathcal{H}}}$ through fifth order, albeit in a not entirely explicit form.  

For a derivation of PT expansions for the hermitian effective hamiltonian within the VVPT approach the reader is referred to Kvasni\v{c}ka~\cite{kva74:605} and Shavitt and Redmon~\cite{sha80:5711}.
Pursuing the BWPT based approach Kvasni\v{c}ka~\cite{kva77:345} has derived the PT expansion for the non-hermitian effective hamiltonian through third order, and also addressed the generation of hermitian variants.

Finally, we take a look at the effective eigenvector amplitudes. These are given according to
\begin{equation}
   f'_{\kappa \mu} = \dirint{\Phi_\kappa}{\hat \Omega}{\Phi_\mu}, \,\, \kappa = 1,2,\dots ;\, \mu \leq N
\end{equation}
as matrix elements of the wave operator. In the $11$-block of $\bf f'$, there is only a zeroth order contribution,
   \begin{equation}
f'_{\mu\nu} = f'^{(0)}_{\mu\nu} = \delta_{\mu \nu}, \; \mu, \nu \leq N
\end{equation}
For the $21$-block of $\bf f'$, the expansion through second order is given by 
\begin{align}
\nonumber
f'^{(0)}_{al} &= 0\\
\nonumber
f'^{(1)}_{al} &= \frac{W_{al}}{\mathcal E_l - \mathcal E_a}\\
f'^{(2)}_{al} &=  
\sum_{b} \frac{W_{ab}W_{bl}}{(\mathcal E_l -\mathcal E_a) (\mathcal E_l -\mathcal E_b)} -
\sum_{j} \frac{W_{aj}W_{jl}}{(\mathcal E_l -\mathcal E_a) (\mathcal E_j -\mathcal E_a)}  
\end{align}   
Here, again the QC notation for the indices has been adopted.

\appendix
\renewcommand{\theequation}{E.\arabic{equation}}
\setcounter{equation}{0}
\section*{Appendix E: Explicit QDPT expansions through third order}

Here a compilation is given of the 
explicit PT expansions through third order for the hermitian effective hamiltonian 
(Eq.~\ref{eq:cfheff}) and the 
corresponding effective eigenvector amplitudes (Eq.~\ref{eq:cffeff}). 
The notations refer to the general hamiltonian
as given by Eqs.~(\ref{eq:Hgen}) and (\ref{eq:Hmunu}). 
For notational ease, the "quantum-chemical" index convention is used, where the indices $i,j,k,\dots$ and $a,b,c, \dots$ denote model states and complementary model states, respectively.\\
\\
\noindent
\textbf{Secular matrix elements}
\begin{align}
\tilde{\mathcal H}_{kl} =& \,\,\mathcal E_k \delta_{kl} +  W_{kl}
 + \tilde M^{(2)}_{kl}+ \tilde M^{(3)}_{kl} +\; O(4)\\
\label{eq:m2kl}
\tilde{M}^{(2)}_{kl} =&
  \sum_a W_{ka} W_{al} \frac{\half (\mathcal E_k + \mathcal E_l) - \mathcal E_a}
 {(\mathcal E_a - \mathcal E_k)(\mathcal E_a - \mathcal E_l )}\\
 \nonumber
 \tilde{M}^{(3)}_{kl} =&
   \half \sum_{a,b} W_{ka} W_{ab} W_{bl}\left (\frac{1}{(\mathcal E_a - \mathcal E_k )(\mathcal E_b - \mathcal E_k)} + \frac{1}{(\mathcal E_a - \mathcal E_l )(\mathcal E_b - \mathcal E_l)} \right ) \\
\label{eq:m3kl}
 & \phantom{xxxxxx}
 - \half \sum_{j,a} \frac{W_{ka}W_{aj}W_{jl}}{(\mathcal E_a - \mathcal E_l )(\mathcal E_a - \mathcal E_j )} 
 - \half \sum_{j,a} \frac{W_{kj}W_{ja}W_{al}}{(\mathcal E_a - \mathcal E_k )(\mathcal E_a - \mathcal E_j )}
\end{align}
These expressions derive from the original ADC expressions~(\ref{eq:C1pq}, \ref{eq:C2pq}) in Sec.~III and Eqs.~(\ref{eq:Cpq31},\ref{eq:Cpq32},\ref{eq:Cpq33}) in App.~A by adapting  
the notation according to Eq.~(\ref{eq:MKC}) and (\ref{eq:snotad1}),
and replacing the energies, $\epsilon_p$, and the matrix elements, 
$w_{pq}$, of the one-particle hamiltonian~(\ref{eq:H}) with the quantities $\mathcal E_p$ and $W_{pq}$, respectively, of the general hamiltonian~(\ref{eq:Hgen}). 
\\
\\
\noindent
\textbf{Effective eigenvector amplitudes}\\
\textbf{11-block}:
\begin{align}
\mathsf f^{(0)}_{kl} =&\,\, \delta_{kl}\\
\mathsf f^{(1)}_{kl} =&\,\, 0\\
\label{eq:f12}
\mathsf f^{(2)}_{kl} =& -\sum_{a} \frac{W_{ka} W_{al}}{2(\mathcal E_a-\mathcal E_k)(\mathcal E_a-\mathcal E_l)}\\
\mathsf f^{(3)}_{kl} =& \;\mathsf f^{(3,1)}_{kl} + \mathsf f^{(3,2)}_{kl} + \mathsf f^{(3,3)}_{kl} 
\end{align}
where
\begin{align}
\label{eq:f131}
\mathsf f^{(3,1)}_{kl} =& - \half \sum_{a,b}  \frac{W_{ka} W_{ab} W_{bl}}{(\mathcal E_l - \mathcal E_b) (\mathcal E_k - \mathcal E_a)} \left (\frac{1}{\mathcal E_l - \mathcal E_a} + 
\frac{1}{\mathcal E_k - \mathcal E_b} \right )\\
\label{eq:f132}
\mathsf f^{(3,2)}_{kl} =& \half \sum_{j,a} \frac{W_{ka}W_{aj}W_{jl}}{(\mathcal E_l - \mathcal E_a) (\mathcal E_k - \mathcal E_a)(\mathcal E_j - \mathcal E_a)}\\
\label{eq:f133}
\mathsf f^{(3,3)}_{kl} =&  \half \sum_{j,a} \frac{W_{kj}W_{ja}W_{al}}
{(\mathcal E_l - \mathcal E_a) (\mathcal E_k - \mathcal E_a)(\mathcal E_j - \mathcal E_a)}
\end{align}
The original ADC expressions are given by Eq.~(\ref{eq:f2pq}) (2nd order) and Eqs.~(\ref{eq:f31}), (\ref{eq:f32}),(\ref{eq:f33}) (3rd order). They have been transcribed according to Eq.~(\ref{eq:snotad2}) and by using the notations of the general hamiltonian~(\ref{eq:Hgen}) rather than those of the one-particle hamiltonian~(\ref{eq:H}). These transcription rules apply as well for the 
$21$-amplitudes below.
\\
\\
\noindent
\textbf{21-block}:
\begin{align}
\mathsf f^{(0)}_{ak}  =&\, 0\\
\label{eq:f21}
\mathsf f^{(1)}_{ak} =&\; \frac{W_{ak}}{\mathcal E_k - \mathcal E_a}\\
\label{eq:f22}
\mathsf f^{(2)}_{ak} =&
 \frac{1}{\mathcal E_a -\mathcal E_k} \sum_{j} \frac{W_{aj} W_{jk}}
{\mathcal E_j -\mathcal E_a} + \frac{1}{\mathcal E_a -\mathcal E_k}  \sum_{b} \frac{W_{ab} W_{bk}}{\mathcal E_b -\mathcal E_k}
\\ 
\mathsf f^{(3)}_{ak} =& \sum_{m=1}^6 f^{(3,m)}_{ak}
\end{align} 
where
\begin{align}
\label{eq:f231}
f^{(3,1)}_{ak} =& 
 \half \sum_{b,j}  \frac{ W_{aj} W_{jb} W_{bk}}{(\mathcal E_a-\mathcal E_j)(\mathcal E_k-\mathcal E_b)(\mathcal E_j-\mathcal E_b)}\\
\label{eq:f232}
f^{(3,2)}_{ak} =& \frac{-1}{\mathcal E_a -\mathcal E_k}
\sum_{i,j} \frac{W_{ai} W_{ij} W_{jk}}{(\mathcal E_a -\mathcal E_i)(\mathcal E_a -\mathcal E_j)} \\
\label{eq:f233}
f^{(3,3)}_{ak} =& \frac{1}{\mathcal E_a -\mathcal E_k}
\sum_{b,j} \frac{W_{ab} W_{bj} W_{jk}}{(\mathcal E_a -\mathcal E_j)(\mathcal E_b -\mathcal E_j)}\\
\label{eq:f234}
f^{(3,4)}_{ak} =& \frac{-1}{\mathcal E_a -\mathcal E_k}
\sum_{b,c} \frac{W_{ab} W_{bc} W_{ck}}{(\mathcal E_b -\mathcal E_k)(\mathcal E_c -\mathcal E_k)}\\
\label{eq:f235}
f^{(3,5)}_{ak} =& \frac{1}{\mathcal E_a -\mathcal E_k}
\sum_{b,j} \frac{W_{ab} W_{bj} W_{jk}}{(\mathcal E_b -\mathcal E_k)(\mathcal E_b -\mathcal E_j)}\\
\label{eq:f236}
f^{(3,6)}_{ak} =& \frac{1}{\mathcal E_a -\mathcal E_k}
\sum_{b,j} \frac{W_{aj} W_{jb} W_{bk}}
{(\mathcal E_a -\mathcal E_j)(\mathcal E_b -\mathcal E_k)}   
\end{align} 
Here the underlying ADC expressions are given by Eq.~(\ref{eq:f11}) in first order and Eqs.~(\ref{eq:f2pqx}), (\ref{eq:f2pqy}) in second order. 
The first third-order expression derives from Eq.~(\ref{eq:t45f}). The next four expressions,(\ref{eq:f232})-(\ref{eq:f235}), have been obtained by direct evaluation of the third-order diagrams $a9, c9, d9$, and $e9$, respectively, of Fig.~\ref{fig:1pdo3}. 
The last expression, (\ref{eq:f236}), combines the results from diagrams $b9$ and $f9$.

\end{document}